\ifdefined\CLManuscript
  \documentclass[manuscript]{clv2025}
\else
  \documentclass[final]{clv2025}
\fi

\jvol{vv}
\jnum{nn}
\jyear{2025}

\pageonefooter{Action editor: Yuki Arase. Submission received: November 14, 2025; revised version received: May 9, 2026; accepted for publication: July 1, 2026.}

\usepackage{amsmath}
\usepackage{booktabs,amsfonts,multirow,enumerate,tcolorbox,threeparttable,graphicx,array}
\tcbuselibrary{breakable}
\usepackage{subfigure}
\usepackage{subcaption}
\usepackage{float}
\usepackage{makecell}

\usepackage[final, commandnameprefix=ifneeded]{changes}
\definechangesauthor[name={JXW}, color=blue]{A}

\runningtitle{Representational Equality in Cross-country Value Simulation}
\runningauthor{Jian et al.}

\begin{document}

\title{Representational Equality in Cross-country Value Simulation: A Systematic Analysis of Large Language Models}

\author{Xiaowen Jian$^{1}$, Xinyi Mou$^{2}$, Daisong Gong$^{3}$, Chen Qian$^{4}$, Huimin Chen \thanks{Corresponding authors}$^{1}$, Maosong Sun$^{5}$}

\affilblock{
    \affil{Tsinghua University, School of Journalism and Communication\\\quad \email{jxw24@mails.tsinghua.edu.cn,huimchen@tsinghua.edu.cn}}
    \affil{Fudan University, School of Data Science\\\quad \email{xymou20@fudan.edu.cn}}
    \affil{Nankai University, College of Cryptology and Cyber Science\\\quad \email{2312325@mail.nankai.edu.cn}}
    \affil{Shanghai Jiao Tong University, School of Artificial Intelligence\\\quad \email{qianc@sjtu.edu.cn}}
    \affil{Tsinghua University, Department of Computer Science and Technology\\\quad \email{sms@tsinghua.edu.cn}}
}
\maketitle

\begin{abstract}
Traditional methods for studying human opinions often struggle to support representative and scalable research across countries. Large language models (LLMs) can serve as scalable proxies for simulating human opinions, enabling more efficient opinion analysis. However, this use of LLMs requires not only high average accuracy but also \emph{representational equality}, that is, comparable simulation accuracy across populations. Uneven simulation accuracy may reproduce or amplify societal biases in downstream applications.
This study systematically investigates country-level representational equality across 59 countries and finds substantial, systematic inequality. Populations from wealthier and more technologically advanced countries are simulated more accurately. We further compare two foundational intervention pathways, contextual adaptation and parametric modification, and show that improvements in average or target-group accuracy do not necessarily translate into greater representational equality.
For contextual adaptation, native-language prompting generally improves accuracy but remains model-dependent, whereas additional information more often improves both accuracy and equality. For parametric modification, language-specific continued post-training improves accuracy for targeted language groups but unevenly, while preference alignment yields no systematic gains in accuracy or equality. Human-annotated preference data generally preserve accuracy better than AI-annotated data. These findings highlight the need for representational equality alongside accuracy and offer guidance for more inclusive, socially responsible LLM-based simulations.\end{abstract}

\section{Introduction}

Understanding human values and perspectives in diverse social contexts is a cornerstone of social science. Traditionally, social scientists have relied on various research methods, including surveys~\citep{groves2011survey,bhattacherjee2012social, wardropper2021conducting}, observation~\citep{musante2010participant}, interviews~\citep{knott2022interviews}, and questionnaires~\citep{schensul1999essential,aithal2020development} to gather human perspectives, uncover attitudinal patterns, and guide public policy.
However, these methods typically involve high recruitment costs and lengthy research timelines to obtain broad and representative human perspectives and behaviors~\citep{Koivula2019ExaminingSD,Boas2018RecruitingLO, bonevski2014reaching,keeter2006gauging}, which limits the flexibility, efficiency, and scalability of research.

Recently, large language models (LLMs) have introduced a new paradigm for research in computational linguistics and social science. Trained on large-scale data and adapted through post-training, LLMs perform strongly on many language and reasoning tasks~\citep{radford2019language,brown2020language,wei2022chain}, and they can generate responses that resemble human discourse. Recent studies further show that contextualized inputs~\citep{durmus2023towards,santurkar2023whose,argyle2023out} and parametric modifications~\citep{ryan2024unintended,shao2023character,wang2023rolellm} can guide LLMs to produce outputs closer to the perspectives of specific individuals or groups. These findings suggest that LLMs can support social simulation and help researchers study targeted populations more efficiently~\citep{argyle2023out,park2023generative,mou2024unveiling}.

However, this promise is constrained by a foundational challenge. LLMs are trained on large web-based datasets that often contain inherent biases and provide uneven coverage of social groups~\citep{Mehrabi2019ASO,Gallegos2023BiasAF}. Their ability to simulate human populations may therefore vary substantially across groups. Unequal simulation accuracy can distort or underrepresent the perspectives of particular linguistic and cultural groups~\citep{liang2021towards,santurkar2023whose}, reinforcing disparities in downstream applications across regions and cultures. Moreover, it undermines the credibility and scalability of LLM-based simulations in real-world applications, thus limiting the broader adoption of this paradigm.

To address this challenge, we introduce and formalize the concept of \textbf{Representational Equality}, which concerns how equally well an LLM can simulate diverse populations.
This concept does not require a model to produce identical responses across groups, nor does it imply that genuine population differences should be reduced. Instead, it asks whether the model can achieve comparable simulation performance for different populations. We therefore treat representational equality as a diagnostic lens and an evaluation objective that identifies whether simulation capability is unevenly distributed across populations.
In this study, we focus on representational equality in LLM-based \textbf{value simulation}, as values are the foundation of human attitudes and behaviors~\citep{Schwartz2012AnOO}.
Specifically, we evaluate whether LLMs can approximate empirical human value response patterns across different populations, and whether this simulation accuracy is evenly distributed.
We conduct the evaluation at the country level for two reasons. First, country is a meaningful unit for cross-cultural value comparison in prior research~\citep{Hofstede2011DimensionalizingCT,obradovich2022expanding}, and it is the most discriminative grouping dimension in our analysis (Appendix~\ref{appendix:country_variation}). Second, growing evidence shows that AI systems exhibit systematic cross-country disparities tied to country-level conditions such as digital infrastructure, data visibility, and governance~\citep{manvi2024large,asiedu2024case}.

Although some studies~\citep{ryan2024unintended,durmus2023towards} have noted representation disparities across groups, they still face several limitations:
(1) \textbf{Limited analysis of equality}: Most research has focused on identifying the direction and strength of model bias (e.g., left-leaning)~\citep{Ceron2024BeyondPB,Peng2024BeyondPL} or improving simulation accuracy~\citep{Liu2025CulturalLC,cao2025specializing}. These analyses, however, do not capture how performance is distributed across populations. Representational equality evaluates this cross-group balance directly. When formalized as a quantitative index, it enables comparable measurement of equality across models and settings. Despite its importance, this dimension remains underexplored.
(2) \textbf{Limited country coverage and correlation analysis}: Most existing research focuses on value simulation for groups within the United States or is restricted to a small set of countries~\citep{Bisbee2024SyntheticRF,qu2024performance,alkhamissi2024investigating}. This leaves limited evidence about how simulation performance varies across a broader cross-country panel, including countries that may face higher risks of underrepresentation. Moreover, there is limited analysis of country-level features associated with representational inequality, such as economic and technological development as well as political and cultural factors, which constrains understanding of the structural patterns linked to such inequality.
(3) \textbf{Limited comparison of simulation strategies}: The field has identified two main intervention pathways for enhancing simulation performance, contextual adaptation and parametric modification. However, systematic comparative evidence on how these pathways affect representational equality remains scarce. A rigorous comparison is needed to understand how these approaches and their specific settings shape LLM-based value simulation.

To address these issues, this paper systematically analyzes representational equality in LLM-based value simulation across countries. First, we introduce Representational Equality as a core concept and metric for evaluating global language models, and propose a quantitative index to measure it. Second, we examine cross-country representational equality across 59 countries. For each country, simulation capability is aggregated from comparable subpopulation cells constructed from country and selected demographic attributes among gender, age, education level, and income. Beyond this foundational analysis, we investigate how different simulation strategies influence representational equality. We focus on two primary intervention pathways: (1) contextual adaptation, which modifies input prompts through native language usage and additional-information settings; and (2) parametric modification, which alters model behavior through continued post-training on native-language data and variations in alignment sources. Finally, we conduct a diagnostic analysis of how country-level simulation accuracy is associated with economic, technological, political, and cultural indicators, providing a macro-level view of the structural patterns linked to representational inequality.
This analysis helps identify systematic biases in simulation and provides evidence for more inclusive language technologies.

\section{Related Work}

\subsection{Social Simulation with LLMs}

As LLMs develop, researchers are exploring their potential for social simulation at both individual and group levels. Their ability to generate responses conditioned on social roles and context, and to support multi-agent interaction, makes them useful for approximating human decisions, attitudes, and social dynamics.

At the individual level, research has expanded from basic behavioral modeling to more nuanced social scientific investigations~\citep{lee2024can,Chu2023LanguageMT,atari2023humans,aher2023using,horton2023large,sun2024random}.
Early studies showed that LLMs can replicate patterns from classical behavioral economics experiments~\citep{horton2023large}, providing evidence about their ability to approximate human decision-making patterns. This line of work enabled broader applications in psychological research, such as simulating social psychology experiments~\citep{aher2023using}, although limitations remain in capturing collective behaviors. Recent work has introduced methods such as ``random silicon sampling'' to simulate diverse demographic perspectives~\citep{sun2024random} and support population-level analyses.

At the group level, researchers are exploring how LLMs can be used to construct virtual communities and simulate emergent group behaviors, ranging from basic social interactions to complex collective dynamics~\citep{Gao2024SimulatingHS,park2023generative}. Early studies focused on building virtual communities in which LLM agents with distinct personality traits and memory capabilities interact, revealing emergent social patterns~\citep{park2023generative}. This foundation has supported applications in simulating real-world social systems, such as social media dynamics~\citep{tornberg2023simulating}. Recent work has also combined LLM-driven agents with traditional agent-based models for large-scale social simulation~\citep{Mou2024UnveilingTT}.

\subsection{Simulation Recipes}

Research on LLM simulation recipes has primarily developed along two directions: contextual adaptation and parametric modification~\citep{mou2026individual}. These approaches aim at more accurate and targeted simulations for specific subpopulations.

For contextual adaptation, recent research explores strategies that guide LLM simulation through input design. Studies have shown that contextual information, persona details, and examples in prompts can improve targeted simulations~\citep{durmus2023towards,santurkar2023whose}. This line of work has further expanded into multimodal settings in which visual cues guide LLM behavior in social interaction~\citep{sun2024kiss}.

Parallel research on model configurations investigates how training data, model scale, and architectural choices influence simulation capabilities. Findings suggest that larger models do not always yield better performance, with smaller models occasionally matching or outperforming them on specific tasks~\citep{argyle2023out}. Recent work has introduced social science-inspired approaches, including cultural adaptation through simulated social interactions and survey-based augmentation to align models with specific cultural norms~\citep{Liu2025CulturalLC,li2024culturepark,li2024culturellm}. Alignment methods also have important side effects: while alignment can promote desirable behaviors, it may introduce or intensify cultural biases~\citep{ryan2024unintended}. Despite these findings, systematic evidence remains limited on how post-training data, alignment sources, and other training settings affect simulation accuracy and representational equality.

\subsection{Bias in LLM-based Simulations}

Bias in LLMs has long been a major area of concern. Existing research has documented harmful stereotypes and biases related to gender, race, and religion~\citep{nangia2020crows,liang2021towards,wang2023decodingtrust}, as well as nationality and other demographic factors~\citep{venkit2023nationality,ladhak2023pre,zheng2024ali}. These biases are often reflected in model-generated content, which can perpetuate social inequalities and produce errors in downstream applications~\citep{taubenfeld2024systematic}. As interest in using LLMs for global social simulation grows, it becomes important to assess how these models represent diverse cultural and demographic groups.

Recent cross-cultural evaluation studies have improved the resolution of bias detection, revealing systematic favoritism toward WEIRD populations, English-language inputs, and liberal-progressive ideologies~\citep{Helwe2025NavigatingTP,Peng2024BeyondPL}. While these studies identify whose perspectives are privileged and the direction of cultural bias, they rarely assess how evenly simulation accuracy is distributed across populations. A formal framework for representational equality is therefore needed to quantify cross-group disparities and enable systematic comparisons.

\section{Methodology}

\subsection{Overview}
\begin{figure*}
    \setlength{\belowcaptionskip}{-0.4cm}
  \centering
    {\includegraphics[width=\linewidth]{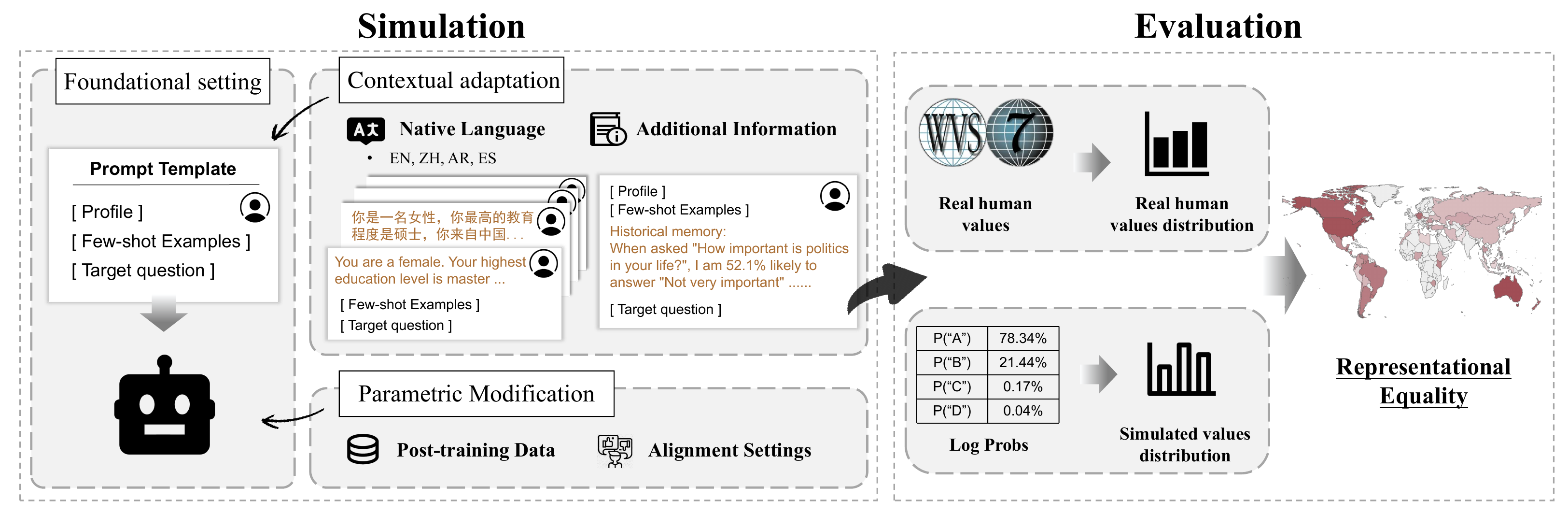}}
  \caption{Illustration of simulation and equality evaluation. We use LLMs to simulate specific populations under different settings and systematically analyze simulation accuracy and representational equality.}
  \label{fig:structure}
\end{figure*}

In this section, we present the methodology of our study. Our framework, illustrated in Figure~\ref{fig:structure}, systematically evaluates the simulation capabilities of LLMs, with a core focus on representational equality. We first define how \textbf{representational equality} is computed and outline the \textbf{analytical framework} for evaluation (right part of Figure~\ref{fig:structure}). We then describe the simulation process through which model outputs are generated under different settings (left part of Figure~\ref{fig:structure}). We establish a \textbf{foundational simulation setup} in which LLMs are prompted with demographic profiles to generate distributions of simulated values. Building on this setup, we systematically test two primary intervention strategies: (1) \textbf{contextual adaptation}, which modifies prompts with native language (English, Chinese, Arabic, or Spanish) or additional information, and (2) \textbf{parametric modification}, which alters model behavior through continued training or alignment.

\subsection{Evaluation metrics for representational equality}

Our evaluation framework measures Representational Equality through two main components: response distribution extraction and index computation. First, we adopt a first-token probability method to simulate the distribution of values within a subpopulation. Second, we use these simulated distributions to compute our primary metric, the Representational Equality Index ($\mathrm{Eq}$).

\paragraph{Extracting simulated value distribution}

Real human populations are not monolithic, and they hold a diverse spectrum of views. A meaningful social simulator must capture this diversity. In this study, we capture these viewpoints by simulating how people would respond to survey questions.
Existing research commonly uses two main approaches to extract LLM responses in survey-style evaluation: first-token probabilities over answer options, or text-based outputs~\citep{ma2024potential}. We adopt the first-token probability method because it provides a direct and computationally efficient way to approximate the full response distribution under a constrained answer space. To verify its validity, we conduct a comparison experiment showing high consistency with text-based methods in our setting (Appendix~\ref{appendix:token_consistency}).
The first-token probability method also avoids repeated generation, which is important for our large-scale evaluation over 2,420 demographic subpopulations across 59 countries for each model and intervention setting.

Following previous studies that extract first-token log probabilities over discrete answer tokens~\citep{chen2025mova,santurkar2023whose}, we request the log probabilities of the top 20 first-position tokens for each multiple-choice question. We retain tokens that correspond to valid answer-option letters (e.g., ``A'', ``B'', and ``C''), and normalize their probabilities. This yields the model-predicted response distribution $D(M,H,q)$, where $M$ denotes the evaluated model, $H$ the target subpopulation, and $q$ the survey question. Tokens outside the valid option set are excluded. Missing answer-option tokens are handled using a capped residual-mass approximation following~\citet{santurkar2023whose}. The complete extraction procedure, including top-20 token handling, option filtering, and the missing-option treatment, is detailed in Appendix~\ref{appendix:first_token_details}.

\paragraph{Calculating the representational equality index}

The primary metric of this study is the country-level Representational Equality Index, which quantifies the balance of simulation performance across countries. Our approach is informed by the concept of accuracy equality from the machine learning and statistics literature, which emphasizes comparable model performance across demographic groups~\citep{Berk2017FairnessIC,Verma2018FairnessDE}. Although value-distribution simulation differs from conventional classification, the core principle of comparable performance across groups remains directly relevant.

Representational equality measures the evenness of simulation accuracy across countries, with the empirical human distribution serving as the simulation target. It asks whether simulation capability is evenly distributed across populations, and is used as a diagnostic lens and an evaluation objective that identifies where and how simulation performance diverges across countries. Because a model that is uniformly poor could appear equal but should not be considered strong overall, we report representational equality alongside average accuracy and further introduce a composite metric that summarizes both dimensions jointly. The computation proceeds in three stages.

\textbf{Stage 1: Country-level simulation accuracy.}
Following previous studies~\citep{durmus2023towards,ryan2024unintended,cao2025specializing}, we define simulation accuracy $\mathrm{Acc}(H, M, Q)$ as the degree of similarity between the model-predicted and human response distributions.
For each subpopulation $H$ and question $q$, we obtain the human response distribution $D(H, q)$ from the World Values Survey (WVS) data (Eq.~\ref{eq:question_human_distribution}) and the model-predicted distribution $D(M,H,q)$ from first-token probabilities. We compare them with Jensen--Shannon divergence (JSD), a standard symmetric metric for comparing probability distributions. Since JSD measures dissimilarity, we define simulation accuracy as $1-\mathrm{JSD}(\cdot)$, so that higher values indicate a closer match. The accuracy for model $M$ and subpopulation $H$ on question $q$ is:
\begin{equation}
  \mathrm{Acc}(H, M, q) = 1 - \mathrm{JSD}\bigl(D(H, q), D(M,H,q)\bigr), \quad q \in Q .
\end{equation}

The overall simulation accuracy for the subpopulation, $\mathrm{Acc}(H, M, Q)$, is computed in a dimension-balanced manner. We first average question-level accuracies within each value dimension and then average across dimensions. This prevents question-dense dimensions from dominating the final score. We then average across all subpopulations within a country to obtain the country-level accuracy $A_c$.

\textbf{Stage 2: Representational equality from country accuracies.}
We use these country-level accuracies to derive the final Representational Equality Index. To capture different aspects of cross-country imbalance, we compute four indices that are widely adopted in the equality and fairness literature~\citep{Berk2017FairnessIC,Verma2018FairnessDE}. The max-min difference captures the absolute gap between the best- and worst-performing countries, while the min-max ratio captures the relative gap between them. The coefficient of variation (CV) measures relative dispersion around the cross-country mean. The Gini coefficient measures the concentration of accuracy across countries. Detailed formulas are provided in Appendix~\ref{appendix:index}.
Results in Appendix~\ref{appendix:all_equality_scores} show strong convergence across all four indices in models' representational equality and in the observed changes under intervention. For clarity and conciseness in the main text, we therefore select the CV as our primary representational equality index. We denote it as $\mathrm{Eq}_{\mathrm{CV}} = \frac{\sigma_{A_c}}{\mu_{A_c}}$, where $\mu_{A_c}$ and $\sigma_{A_c}$ are the mean and standard deviation of country-level simulation accuracy $A_c$ across all evaluated countries, respectively.

\textbf{Stage 3: Composite accuracy-equality score.}
To jointly summarize simulation accuracy and representational equality, we introduce an accuracy--equality (AE) composite metric. Specifically, we combine the average JSD with the CV-based equality index using their geometric mean:
\begin{equation}
  \mathrm{AE}(M,Q) = \sqrt{\overline{\mathrm{JSD}}(M,Q) \cdot \mathrm{Eq}_{\mathrm{CV}}(M,Q)} .
\end{equation}

Here, $\overline{\mathrm{JSD}}(M,Q)$ denotes the country-averaged JSD, which equals one minus the average simulation accuracy. Lower $\mathrm{AE}$ values indicate better joint accuracy--equality performance. This composite metric is partially compensatory as it rewards models that perform well on both accuracy and equality, while penalizing models that are strong on only one dimension. We report $\mathrm{AE}$ as a supplementary summary alongside the separate accuracy and equality measures.

\subsection{Analysis of representational inequality}
\paragraph{Selection of analytical factors}
To systematically analyze the potential correlates of representational inequality, we adapt the PEST analysis framework, a widely used tool for macro-environmental scanning from strategic management \citep{Stapenhurst2010StrategicMA}. PEST provides a structural perspective on the Political, Economic, Socio-cultural, and Technological associations of a phenomenon. We adopt this framework to identify structural patterns in simulation performance disparities of LLMs.
We operationalize this diagnostic analysis using four families of macro-level indicators: GDP per capita for economic development; Internet use and the Global Innovation Index (GII) for technological capacity; the six Worldwide Governance Indicators (WGI)~\citep{Kaufmann2010TheWG} for political governance; and Hofstede's six cultural dimensions~\citep{Hofstede2011DimensionalizingCT} for socio-cultural values. Detailed descriptions and data sources are provided in Appendix~\ref{appendix:macro_factor_details}.

\paragraph{Correlation analysis}
To explore the relationship between simulation accuracy and macro-level factors such as economic development, we obtain the macro-level indicator $f_c$ (e.g., GDP) for each country $c$. We then compute the Spearman correlation coefficient $\rho$ between simulation accuracy and the factor:
\begin{equation}
    \rho = \mathrm{Spearman}(A_c, f_c)
\end{equation}
These correlations are used as diagnostic associations rather than causal estimates. For correlation analyses involving multiple simultaneous significance tests, p-values are adjusted within the corresponding analysis family using the Benjamini--Hochberg false discovery rate (BH FDR) procedure~\citep{benjamini1995controlling}. We apply the same correction both to baseline macro-factor correlations and to the corresponding intervention-induced changes.

\subsection{Value Simulation Protocol}

Our protocol for simulating subpopulation values with LLMs is a two-stage process: first, selecting the appropriate simulation targets, and second, simulating their values through a Q\&A-based approach.

\paragraph{Subpopulation construction}

Subpopulations are formed by groups of individuals who share a common set of attributes. Demographic characteristics such as age, gender, education level, income, and country play an important role in shaping an individual's societal role, cultural context, identity, and value orientations. To ensure generalizability and consistency with WVS, we select these five indicators as the basis for subpopulation construction. The specific categories for each variable are detailed in Table~\ref{tab:demographic}.

A key methodological decision is to determine the optimal number of demographic indicators to combine when defining a subpopulation. Our goal is to balance two competing criteria: (1) statistical robustness, ensuring groups are large enough for reliable analysis, and (2) value specificity, ensuring groups are granular enough to capture meaningful cultural distinctions.
We use 10 respondents as the minimum empirical-support threshold for subgroup-level response distributions. Using this threshold, we analyzed demographic combinations of varying complexity. Combining fewer demographic indicators produced broader but less specific groups, whereas combining more indicators created fine-grained but increasingly sparse cells. Combining three demographic indicators offered the most suitable balance: it preserved substantial value variation while keeping more than 60\% of the resulting subgroup cells above the 10-respondent threshold. We therefore define each subpopulation by country and two of the four remaining attributes: gender, age, education, and income. We retain cells with more than 10 respondents and apply cross-country comparability filters. This procedure yields 2,420 country--subpopulation units in the main evaluation. Appendix~\ref{appendix:demographic} details the six attribute combinations and filtering procedure.

Furthermore, we include country in every simulated subpopulation for two reasons. First, country remains a meaningful dimension for distinguishing cultural and value variation in large-scale cross-country comparison. Prior work shows that country boundaries are a salient structuring dimension of cultural distance~\citep{obradovich2022expanding}. To examine whether country is also an important source of value variation in our empirical setting, we conducted a supplementary analysis of the WVS response distributions across the grouping variables used in this study. The results show that country produces the largest between-subgroup diversity among the demographic variables in our study (Appendix~\ref{appendix:country_variation}), with a mean JSD of 0.478 compared with 0.214 for gender, 0.272 for age, 0.288 for education, and 0.281 for income. This suggests that country is a highly discriminative macro-level grouping dimension in our study. Second,
growing evidence shows that AI systems exhibit systematic cross-country disparities linked to country-level conditions such as digital infrastructure, data visibility, and governance~\citep{manvi2024large,asiedu2024case}, making the country level a relevant scale for evaluating representational inequality. Therefore, following prior cross-country research~\citep{Hofstede2011DimensionalizingCT,licht2007culture} that compares aggregate country-level profiles, we treat country as an empirically grounded comparative unit.

\paragraph{Prompt structure}
\label{sec:simulation_process}
Given a subpopulation, we use a structured prompt to elicit simulated value responses from the LLM. The prompt consists of four components: (1) a role-playing instruction that defines the persona's profile, such as gender, age, and country. To ensure robustness, we compare five commonly used profile formats with different construction strategies, including template-based and LLM-generated profiles (Appendix~\ref{appendix:profile_format}). Since performance differences across formats are minimal, with accuracy variations below 1.3\%, we adopt the second-person format for all subsequent experiments (e.g., ``You are a female. Your age falls into the 18--24 range...''). (2) Three few-shot examples unrelated to value-based topics, which specify only the required single-letter answer format, with correct answers randomized across option positions to avoid positional bias. (3) An optional historical memory module, used only in additional information settings, which summarizes distributions for three related non-target survey questions as supplementary persona context. (4) The target survey question, where the model is asked to answer from the specified persona's perspective, with the first generated token expected to be the letter corresponding to the selected option. Appendix~\ref{appendix:prompt_structure} provides the complete prompt structure, including the system prompt, profiles, few-shot examples, optional memory, and target-question format.

\subsection{Analytical stages}
\paragraph{Foundational analysis}

To explore the initial state of current LLMs, we first evaluate the selected models under a base setting. We specify the demographic indicators to be simulated using second-person profiles, as described in Section~\ref{sec:simulation_process}. This base assessment provides the reference accuracy and Representational Equality Index of each model.

\paragraph{Comparative analysis of intervention strategies}

Previous research on enhancing value simulations has explored interventions that fall into two main strategies. The first, contextual adaptation, guides the model during inference without altering its core parameters. This category includes prompt-level methods such as specifying a user's country or language~\citep{alkhamissi2024investigating,Kwok2024EvaluatingCA}. The second pathway, parametric modification, aims to reshape the model's internal knowledge and behaviors through further training. This encompasses both knowledge infusion through culture-specific fine-tuning on curated datasets~\citep{li2024culturellm,li2024culturepark} and value alignment toward specific cultural preferences~\citep{Chakraborty2024MaxMinRLHFAW}. However, these two strategies are often studied in isolation, and their impact has rarely been assessed jointly in large-scale value simulation, especially with respect to representational equality.
Building on the foundational analysis, we therefore conduct a systematic comparative analysis of these two strategies for enhancing value simulation:

\textbf{Contextual adaptation} involves guiding the model during inference without changing its parameters. We examine two strategies: \textit{native language prompting} to probe whether the model can activate latent cultural priors through linguistic cues, and providing \textit{additional information} to test whether relevant background context can compensate for the model's knowledge gap about underrepresented populations.

\textbf{Parametric modification} involves modifying the model's internal parameters through training. We examine two key approaches: \textit{continued post-training (CPT)} to probe whether language-specific knowledge infusion can improve cultural coverage for underrepresented groups, and \textit{alignment settings} to test how preference optimization and annotation source reshape value representations across countries.

\subsection{Experimental setting}
\label{section:experimental setting}

\paragraph{Model selection}

\label{para:model-selection}
Our model selection follows a two-tiered strategy to balance broad coverage and controlled comparison. In the foundational analysis, we evaluate nine widely used open-source LLMs from institutions across different regions, spanning small-, medium-, and large-scale parameter ranges. The model set includes Llama-2-7B-Chat~\citep{Touvron2023Llama2O}, Llama-3-8B-Instruct and Llama-3-70B-Instruct~\citep{Dubey2024TheL3}, Mistral-7B-Instruct-v0.3~\citep{Jiang2023Mistral7}, ChatGLM3-6B~\citep{Zeng2022GLM130BAO}, GLM-4-9B-Chat~\citep{Zeng2024ChatGLMAF}, Qwen2.5-3B-Instruct, Qwen2.5-7B-Instruct, and Qwen2.5-72B-Instruct~\citep{Yang2024Qwen25TR}. In subsequent targeted analyses, models are selected from these families according to the specific requirements of each experimental track, such as multilingual capability or a common architectural base for controlled training. A clear overview is summarized in Table~\ref{tab:exp_summary}.

\begin{table}
\centering
\small
\setlength{\tabcolsep}{3pt}
\renewcommand{\arraystretch}{1.08}
\resizebox{\linewidth}{!}{%
\begin{threeparttable}
\caption{Summary of Experimental Design, Models, and Data.}
\label{tab:exp_summary}

\begin{tabular}{
    >{\raggedright\arraybackslash}p{2.5cm}
    >{\raggedright\arraybackslash}p{2.6cm}
    >{\raggedright\arraybackslash}p{4cm}
    >{\raggedright\arraybackslash}m{2cm}
    >{\raggedright\arraybackslash}p{3.2cm}
}
\toprule
\textbf{Analysis stage} & \textbf{Objective} & \textbf{Models} & \textbf{Coverage} & \textbf{Key data / method} \\
\midrule

\multicolumn{5}{@{}l@{}}{\textbf{Foundational analysis}}\\
- Base setting & Assess representational equality &
Full model set\tnote{a} &
59 countries &
English prompts; second-person profile \\

\midrule
\multicolumn{5}{@{}l@{}}{\textbf{Contextual adaptation}}\\
- Native language & Evaluate the impact of linguistic context &
Full model set\tnote{a} excluding Llama2-7B-chat\tnote{b} &
27 countries\tnote{c} &
Prompts in \textsc{en}/\textsc{zh}/\textsc{ar}/\textsc{es} \\

\cmidrule(lr){2-5}
- Additional Information & Evaluate the impact of explicit knowledge context &
Full model set\tnote{a} &
59 countries &
BM25, Embedding, Expansion, and Theory \\

\midrule
\multicolumn{5}{@{}l@{}}{\textbf{Parametric modification}}\\
- CPT languages & Evaluate the impact of language knowledge infusion &
\textbf{Base model:} Llama2~\citep{Touvron2023Llama2O} &
\begin{tabular}[t]{@{}l@{}}
12 countries;\\
6 languages\tnote{d}
\end{tabular} &
\textbf{CPT models}: SambaLingo~\citep{csaki2024sambalingo} (Arabic, etc.). \newline
\textbf{SFT data:} Alpaca-GPT4 dataset~\citep{peng2023instruction} \\

\cmidrule(lr){2-5}
- Alignment sources & Evaluate the impact of feedback source (human vs.\ AI) &
\textbf{Base SFT models:} Tulu-7B (Llama2)~\citep{ivison2023camels}, Tulu-8B (Llama3)~\citep{ivison2023camels}, Mistral-7B-Instruct-v0.3~\citep{Jiang2023Mistral7}, Qwen2.5-3B-Instruct, Qwen2.5-7B-Instruct~\citep{Yang2024Qwen25TR} &
59 countries &
\textbf{Alignment source:}\newline HH-RLHF (human)~\citep{Bai2022TrainingAH} vs.\ HH-RLHF-Strength-Cleaned (AI)~\citep{wang2024secrets}\newline 
\textbf{Method:} DPO~\citep{rafailov2023direct} / GRPO~\citep{shao2024deepseekmath} \\

\bottomrule
\end{tabular}

\begin{tablenotes}\footnotesize
\item[a] Full model set: ChatGLM3-6B~\citep{Zeng2022GLM130BAO}, GLM-4-9B-chat~\citep{Zeng2024ChatGLMAF}, Llama2-7B-chat~\citep{Touvron2023Llama2O}, Llama3-8B-Instruct and Llama3-70B-Instruct~\citep{Dubey2024TheL3}, Mistral-7B-Instruct-v0.3~\citep{Jiang2023Mistral7}, Qwen2.5-3B-Instruct, Qwen2.5-7B-Instruct, and Qwen2.5-72B-Instruct~\citep{Yang2024Qwen25TR}.
\item[b] Llama2-7B-chat lacks robust multilingual capability; native-language prompting not applied.
\item[c] Countries where Chinese, Arabic, Spanish, or English serve as a primary language.
\item[d] Arabic, Japanese, Russian, Serbian, Thai, and Turkish.
\end{tablenotes}

\end{threeparttable}}
\end{table}

\paragraph{Contextual adaptation settings}
Previous studies~\citep{hwang2023aligning,argyle2023out} have shown that the form and content of input can influence simulation results. We therefore examine both native-language prompting and the use of additional information at inference time.

\textbf{Native language}: Language is believed to carry implicit cultural cues that activate background information associated with a particular community within the model~\citep{Bucholtz2005IdentityAI,alkhamissi2024investigating}, improving its ability to simulate behavior for that group. Language can also reflect information produced by the group itself, rather than by an external perspective. This may help the model simulate the perspectives of underrepresented subpopulations and reduce the risk of reinforcing stereotypes. In this study, we test simulation performance using prompts in Chinese, Arabic, and Spanish on our core multilingual models. We selected these three languages because they provide broad demographic and geographic coverage within the WVS dataset and are comparatively less advantaged than English in current LLMs, making them useful cases for evaluating whether native-language prompting can improve cross-country representational equality. The multilingual prompts are constructed following the procedure described in Section~\ref{subsec:data_processing}.

\textbf{Additional information}: 
Additional information provides more detailed background knowledge about specific groups, which can help refine the model's responses. In this study, we incorporate value-based responses to non-target questions into an additional-information database for the model. We operationalize this setting as a historical-memory module built from empirical response distributions of non-target WVS questions for the same demographic subgroup. We compare four retrieval strategies: BM25 lexical retrieval (BM25), model-specific embedding retrieval (Embedding), model-specific query expansion followed by embedding retrieval (Expansion), and theory-driven retrieval based on academic keyword co-occurrence (Theory). For each target question, we include three retrieved questions with available subgroup response distributions and convert them into memory sentences reporting the subgroup's empirical answer probabilities. Appendix~\ref{appendix:theory-driven} details the retrieval procedures, and Appendix~\ref{appendix:prompt_structure} provides the prompt structure and retrieved-item examples.

\paragraph{Parametric modification settings}
The modification of a model's internal parameters represents a deeper level of intervention. We focus on two key aspects:

\textbf{CPT languages}: 
A lack of representational equality may partly reflect limited exposure to language-specific data during training. To examine this pathway without retraining models ourselves, we evaluate publicly released SambaLingo checkpoints, which apply CPT to Llama-2-7B on mixtures of English and a target language. We compare six target-language variants with a matched Llama-2-7B reference. To keep the instruction-following stage comparable, both the reference model and the CPT variants are instruction-tuned on the Alpaca-GPT4 supervised fine-tuning dataset~\citep{peng2023instruction} before evaluation. Comparing these checkpoints quantifies the target-language accuracy gains associated with CPT.

\textbf{Preference alignment}: Alignment can reshape model outputs by optimizing them toward particular preference data, and this process may affect value simulation differently across countries~\citep{perez2023discovering,ryan2024unintended}. We use this setting to examine how preference optimization and different annotation sources influence simulation accuracy and representational equality across countries. For annotation sources, we compare human-annotated and GPT-annotated preference datasets. Human annotations may provide richer human feedback but can also contain subjectivity and inconsistency~\citep{dubois2024alpacafarm}. GPT-annotated datasets are easier to scale and more consistent in format, but may inherit biases from the annotating model~\citep{wang2024secrets}. For optimization methods, we apply Direct Preference Optimization (DPO) and Group Relative Policy Optimization (GRPO). Implementation details and key hyperparameters are reported in Appendix~\ref{appendix:alignment_hyperparams}.

\section{Dataset}
\subsection{Data source}

Values are foundational to human behavior and attitudes~\citep{rokeach1973nature}. Because they are broadly shared yet vary across individuals and societies, they provide a suitable basis for evaluating social group simulations on a global scale. Evaluating the representational equality of value simulation across countries requires empirical human data that capture value differences across a broad range of countries in a standardized and comparable form. Although values may be expressed through many forms, such as natural speech, daily interactions, or scenario-based decisions, non-survey data sources currently lack the broad cross-country coverage, shared measurement structure, and respondent-level demographic information needed for our analysis. For these reasons, standardized survey data remain the most appropriate empirical basis for the present cross-country evaluation of representational equality.

Building on prior work that uses surveys to explore perspectives embedded in language models or as benchmarks for alignment~\citep{yao2023instructions,arora2022probing,santurkar2023whose}, we use questionnaire data from the World Values Survey (WVS), a global project investigating public values and their societal evolution. We use its seventh wave (2017--2022), which satisfies all three requirements above and provides multilingual questionnaire versions that support prompt interventions across language communities.
As a supplementary robustness check, Appendix~\ref{appendix:issp_benchmark} reports parallel analyses on a combined benchmark constructed from four annual International Social Survey Program (ISSP) modules administered from 2020 to 2023. ISSP provides a second standardized cross-country benchmark with different topics and confirms that the main inequality patterns are not specific to WVS.

\subsection{Data processing}
\label{subsec:data_processing}

To prepare the dataset, we extract 257 multiple-choice questions from the official multilingual WVS questionnaires, spanning 13 WVS value dimensions. Restricting the pool to 161 questions shared across all 59 countries leaves 12 dimensions because no item from \textit{Perceptions about Science and Technology} meets this criterion. We then remove \textit{Perceptions of Corruption}, which contains only one shared question, to avoid unstable dimension-level aggregation. The final benchmark contains 160 questions across 11 dimensions. Table~\ref{tab:globalvalues_stats} summarizes this cross-country shared benchmark. The questions contain 2 to 10 substantive answer options in a fixed-option format, where human respondents and models use the same answer space. For example, one item asks: ``How important is family in your life?'' with options including ``Very important,'' ``Rather important,'' ``Not very important,'' and ``Not at all important.''

For the native-language prompting experiments, we directly use the multilingual questionnaire versions provided by WVS, including Chinese, Arabic, and Spanish. These items are normalized into a unified multiple-choice format, with minor manual adjustments applied only for format consistency, such as converting Likert scales into discrete options and replacing country-specific references with generic phrasing.

\subsection{Data aggregation}
\label{sec:data-aggregation}

To obtain empirical human opinion distributions, we aggregate the survey responses by subgroup.
For each question $q$, we record its value dimension $t\in T$ and valid option set $O_q$. For subpopulation $H$, let $H_q$ denote respondents with a valid recorded answer to $q$. Missing and invalid responses are excluded from this question-specific set. For each option $o\in O_q$, the corresponding component of the human response distribution is:
\label{item:human_distribution}
\begin{equation}
  D_o(H,q)=\frac{1}{|H_q|}\sum_{h\in H_q}\mathbb{I}\bigl[A(h,q)=o\bigr], \qquad o\in O_q.
\label{eq:question_human_distribution}
\end{equation}

In the main analysis, question-level simulation accuracy is aggregated by giving each value dimension equal weight: question-level accuracies are first averaged within each retained value dimension and then averaged across the 11 retained dimensions. This choice is motivated by the uneven number of questions across WVS dimensions. Without equal weighting, question-dense domains would contribute disproportionately to the final country-level accuracy score. The resulting aggregation provides a more even summary of performance across value dimensions. Table~\ref{tab:country_coverage} reports country-level respondent counts for the shared WVS panel. Appendix~\ref{appendix:benchmark_sensitivity} shows that the main conclusions remain stable under equal question weighting and leave-one-dimension-out re-aggregation, indicating that the results are not an artifact of the weighting scheme or any single value dimension.

\section{Foundational analysis of representational equality}

\begin{figure}[t]
  \centering
  \subfigure[Overall simulation accuracy\label{fig-1:baseline_accuracy}]{%
    \includegraphics[width=0.44\linewidth]{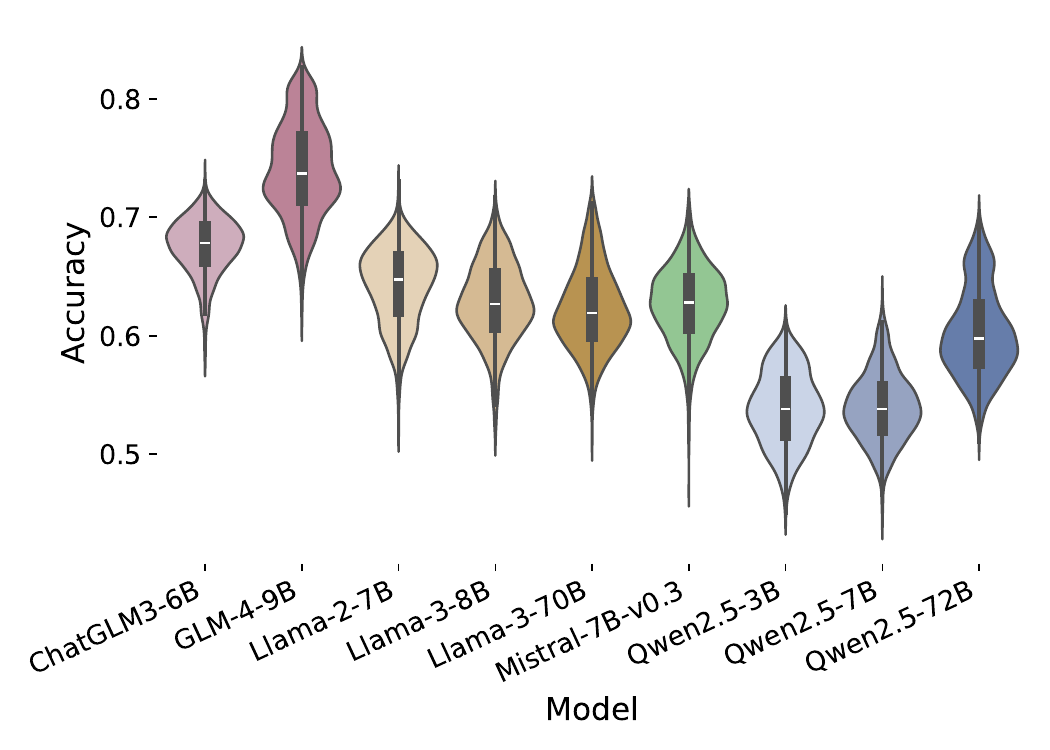}}%
  \hfill
  \subfigure[Accuracy for countries\label{fig-1:baseline-accuracy-country-mean}]{%
    \includegraphics[width=0.52\linewidth]{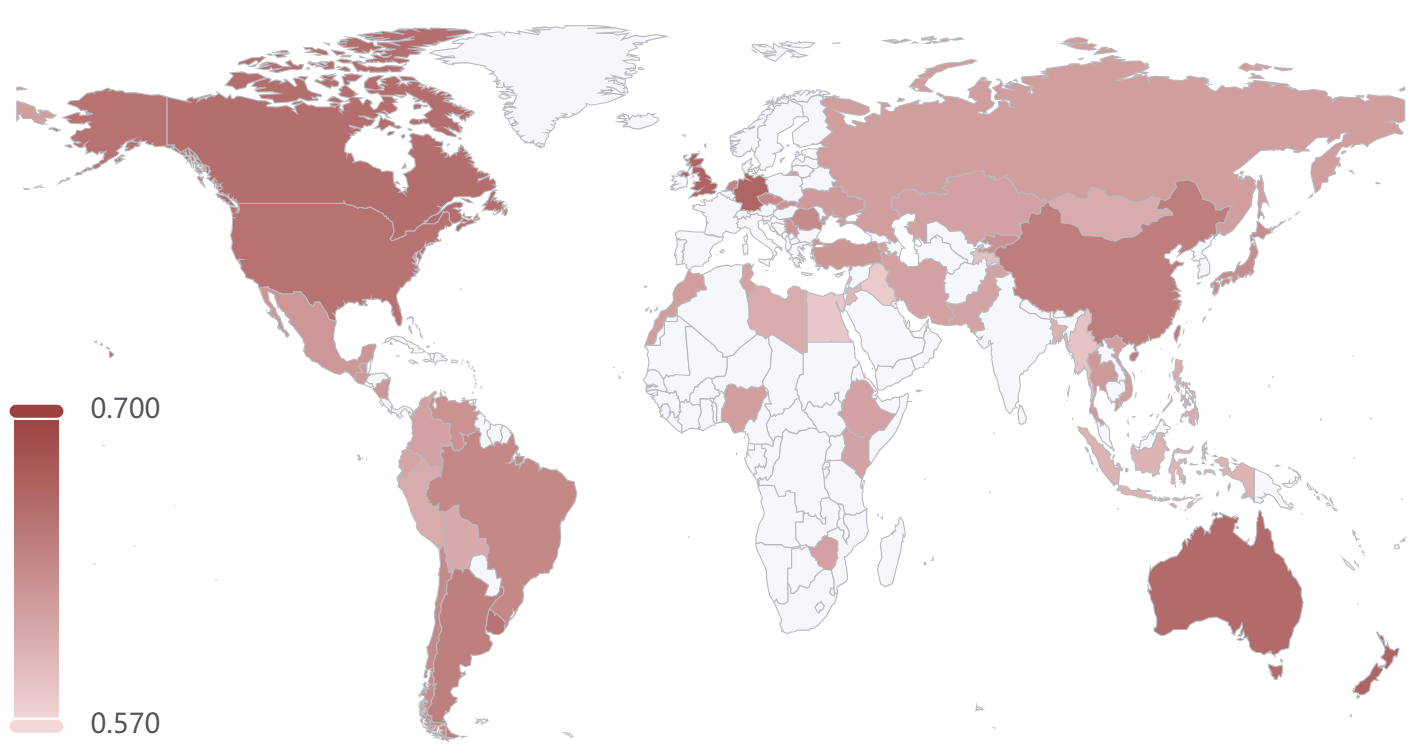}}
  \caption{Simulation accuracy for different LLMs or countries. (a) Violin plot showing the distribution of average simulation accuracy across various LLMs, with the width of each plot indicating the density of accuracy scores. (b) The map shows country-level mean accuracy, with darker red indicating higher accuracy among the countries covered by the benchmark, and white indicating countries that are not represented in the retained WVS comparison set.}
  \label{fig-1:baseline-accuracy}
\end{figure}

In this section, we conduct a foundational analysis of the representational equality of LLMs in simulating human values across different subpopulations. We first assess the overall performance of each model, focusing on both its average simulation accuracy and its representational equality. We then perform a diagnostic analysis of how country-level simulation accuracy is associated with economic, technological, political, and cultural characteristics.
This section covers all 59 countries and evaluates nine open-source LLMs spanning small-, medium-, and large-scale parameter ranges.

\subsection{Overall representational equality}
 \label{sec:Overall simulation accuracy}

\begin{figure*}[t]
  \centering
  \subfigure[Economic factor\label{fig:baseline-economy}]{%
    \includegraphics[width=0.225\textwidth]{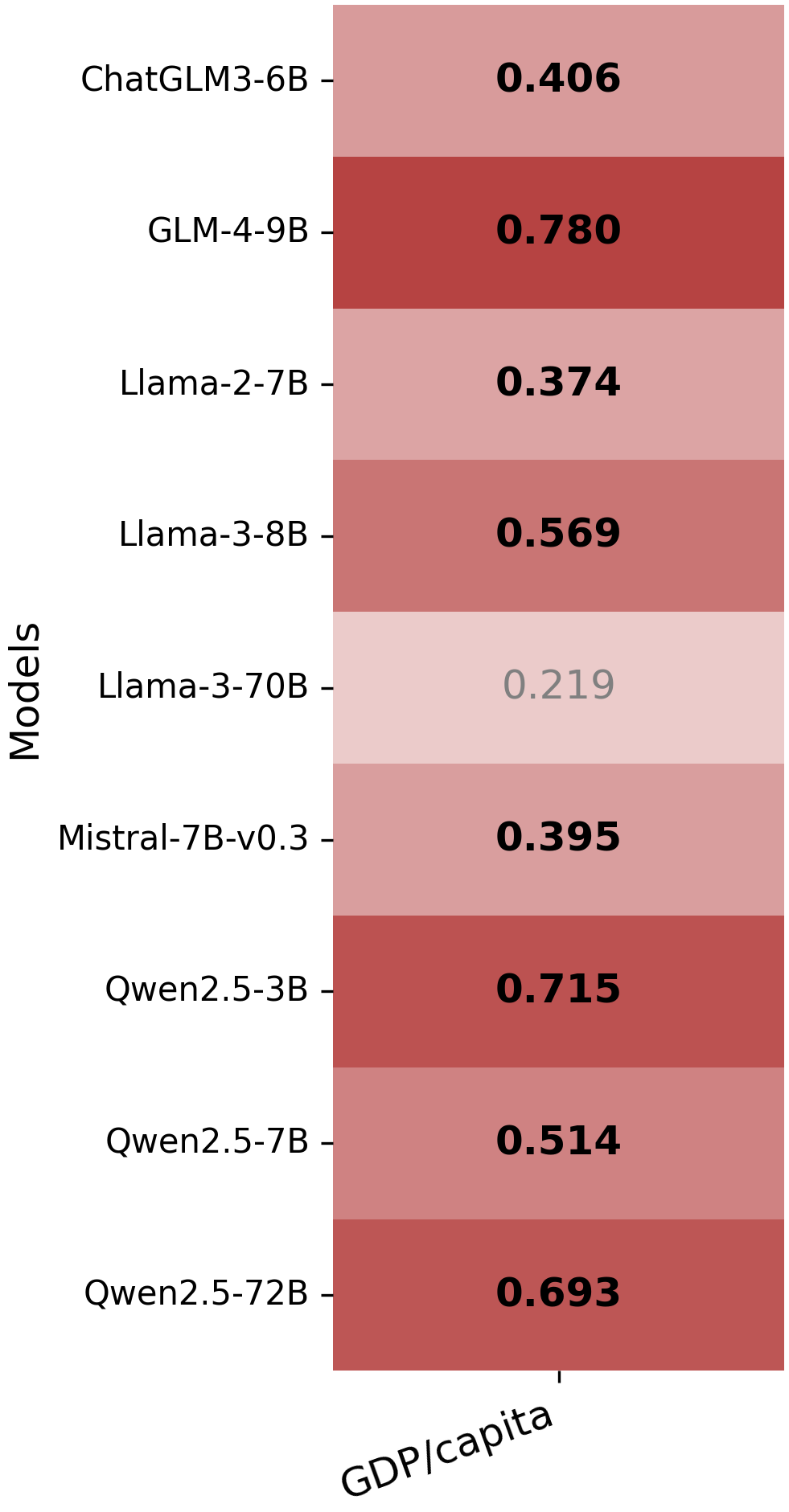}}%
  \hspace{8mm}
  \subfigure[Technological factor\label{fig:baseline-technology}]{%
    \includegraphics[width=0.224\textwidth]{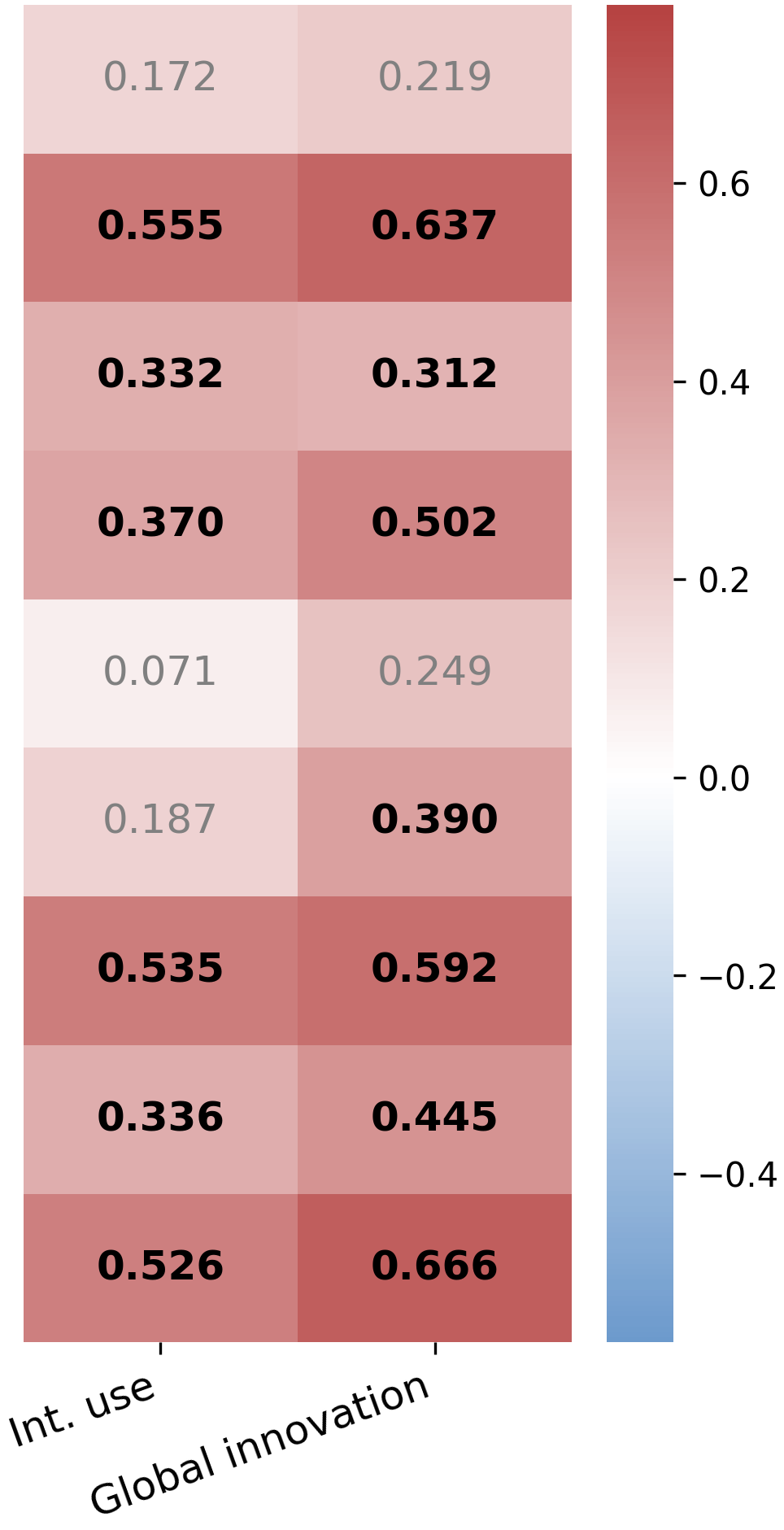}}

  \vspace{2mm}

  \subfigure[Political factor\label{fig:baseline-politics}]{%
    \includegraphics[width=0.54\textwidth]{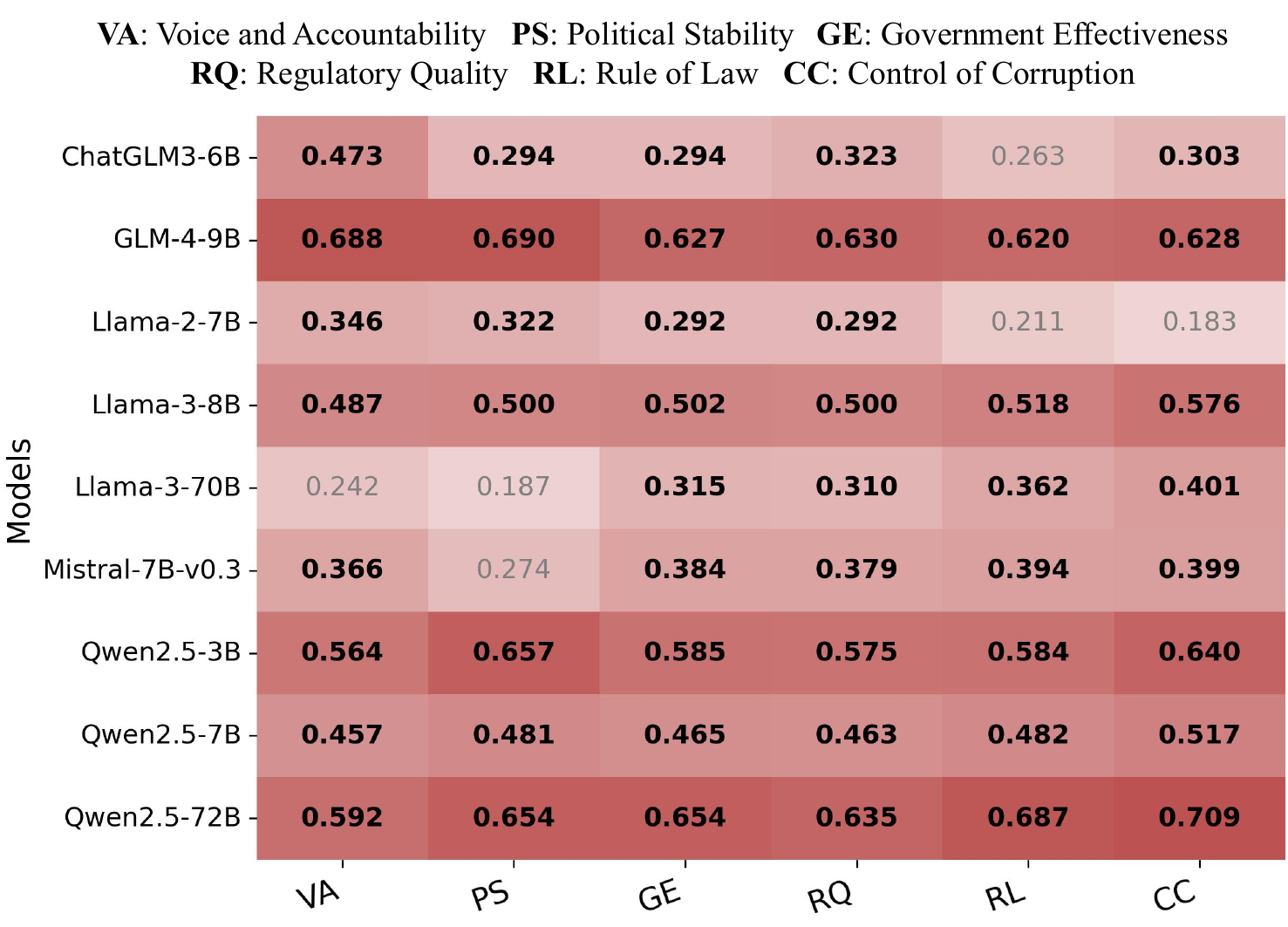}}%
  \hspace{2mm}
  \subfigure[Cultural factor\label{fig:baseline-culture}]{%
    \includegraphics[width=0.432\textwidth]{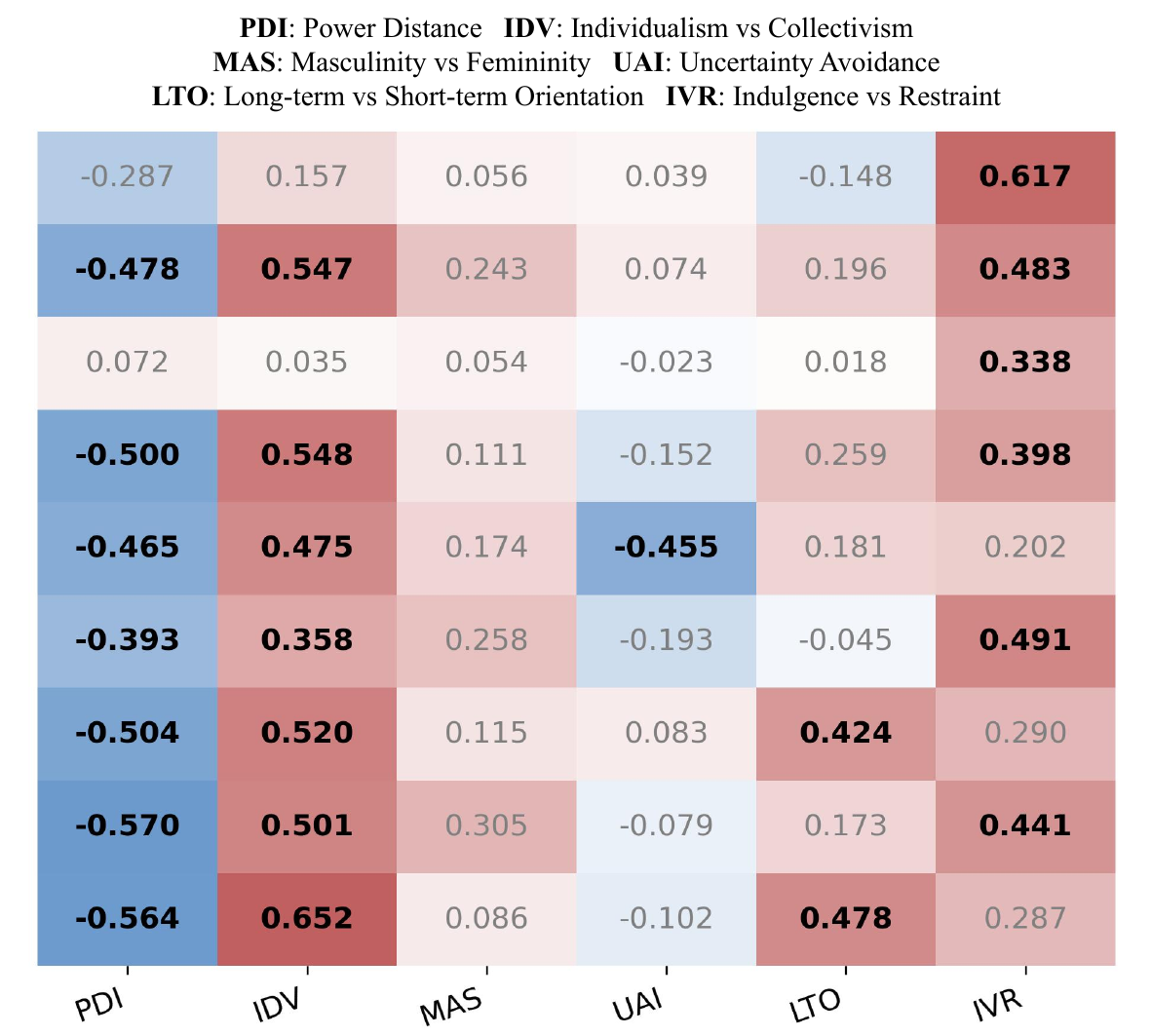}}

  \caption{Country-level correlates of representational inequality across four macro-level factor families. Cells report Spearman correlations between country-level simulation accuracy and macro-level indicators. Red indicates a positive correlation, while blue indicates a negative correlation, with darker shades reflecting a stronger absolute correlation. Bolded values denote Spearman correlations that remain statistically significant after BH-FDR correction at $q<0.05$ across all displayed model--factor tests.}
  \label{fig-1:baseline_country_similarity}
\end{figure*}

Overall, the evaluated models exhibit varying levels of capability in replicating human value judgments. As summarized in Figure~\ref{fig-1:baseline_accuracy}, mean country-level accuracy ranges from 0.538 to 0.739. GLM-4-9B achieves the highest average accuracy, while the Llama family models perform moderately lower yet remain closely clustered across different versions and scales. In contrast, the two small-scale Qwen models consistently rank at the lower end of the accuracy spectrum.

However, average accuracy alone does not capture whether a model performs evenly across different countries. To assess this balance, we further examine each model's representational equality.
As the primary equality metric, a lower $\mathrm{Eq}_{\text{CV}}$ value indicates more equal cross-country simulation performance. The results reveal clear disparities, with $\mathrm{Eq}_{\text{CV}}$ values ranging from 0.0286 to 0.0546. ChatGLM3-6B demonstrates the highest equality ($\mathrm{Eq}_{\text{CV}}=0.0286$). In contrast, Qwen2.5-72B-Instruct shows the weakest equality ($\mathrm{Eq}_{\text{CV}}=0.0546$). Notably, GLM-4-9B remains the accuracy leader but still has relatively high dispersion across countries ($\mathrm{Eq}_{\text{CV}}=0.0486$). This pattern highlights a key insight: \textbf{higher average accuracy does not necessarily imply more evenly distributed simulation performance across countries.}

To jointly summarize accuracy and representational equality, we additionally report the $\mathrm{AE}$ composite metric in Table~\ref{tab:equality_summary_exp1}, which integrates average simulation accuracy with representational equality. Lower values indicate better joint accuracy--equality performance. The results diverge from accuracy-only or equality-only rankings. ChatGLM3-6B achieves the strongest overall composite performance ($\mathrm{AE}=0.096$), followed by GLM-4-9B ($\mathrm{AE}=0.113$) and Mistral-7B-v0.3 ($\mathrm{AE}=0.116$). Notably, while GLM-4-9B attains the highest average accuracy, its pronounced cross-country inequality diminishes its overall standing, whereas Mistral-7B-v0.3 obtains a competitive composite score through a well-balanced profile without leading on either metric alone.

\subsection{Analysis of representational inequality}

The overall analysis reveals substantial representational inequality in subpopulation simulation performance across countries, as shown in Figure~\ref{fig-1:baseline-accuracy-country-mean}. LLMs demonstrate higher accuracy in simulating populations from Australia, North America, and Europe, while countries from regions such as the Middle East, including Egypt, Jordan, and Iraq, exhibit lower simulation accuracy, indicating a regional bias.
To further characterize these structural inequalities, we perform a diagnostic analysis
of the correlations between each country's simulation accuracy and its economic, technological, cultural, and political factors, using Spearman correlation coefficients. As shown in Figure~\ref{fig-1:baseline_country_similarity}, these coefficients range from \text{-}1 to 1 and measure association strength and direction.

\textbf{Economic development correlates positively with simulation accuracy via GDP per capita.} GDP per capita shows significant positive correlations with simulation accuracy for most model families, particularly for GLM-4-9B, indicating that populations from economically advantaged countries tend to be simulated more accurately.

\textbf{Digital infrastructure and innovation intensity are linked to accuracy.} Both Internet use and the Global Innovation Index (GII) exhibit significant positive correlations with accuracy for most models, suggesting that stronger digital infrastructure and innovation capacity are mirrored by higher simulation accuracy.

\textbf{Better governance quality is associated with higher accuracy.} Analysis through the WGI demonstrates significant positive correlations between governance quality and simulation accuracy across almost all models, suggesting that institutional advantages are reproduced as representational advantages in AI.

\textbf{Systematic cultural biases emerge, favoring lower power distance, higher individualism, and higher indulgence.} Most models show significant negative correlations with Power Distance Index (PDI), while Individualism (IDV) and Indulgence (IVR) correlate positively with accuracy. Other cultural dimensions generally exhibit weaker correlations. These patterns reflect a structural bias in LLMs that aligns more closely with specific cultural archetypes.

The observed patterns indicate that disparities in LLM simulation accuracy are associated with country-level characteristics across four dimensions. This finding shows how societal inequalities are systematically reproduced within LLM representations and highlights the significant challenge of achieving equality in global contexts.

\subsection{Case study}

To investigate how macro-level inequality materializes in specific value simulations, we compare the simulated value distributions of a country with high simulation accuracy and one with low simulation accuracy in our evaluation. We select the United States and Iraq as a representative pair because they occupy contrasting positions in country-level simulation accuracy in our benchmark and exhibit distinct empirical value distributions in the WVS. We use GLM-4-9B as the primary lens because it achieves the strongest overall simulation accuracy in our evaluation while still exhibiting pronounced representational inequality, making it a suitable case for examining how high aggregate accuracy can coexist with uneven cross-country simulation performance.

\begin{figure}[t]
  \centering
  \subfigure[Distributional comparison\label{fig-1:case-bar}]{%
    \includegraphics[width=0.53\linewidth]{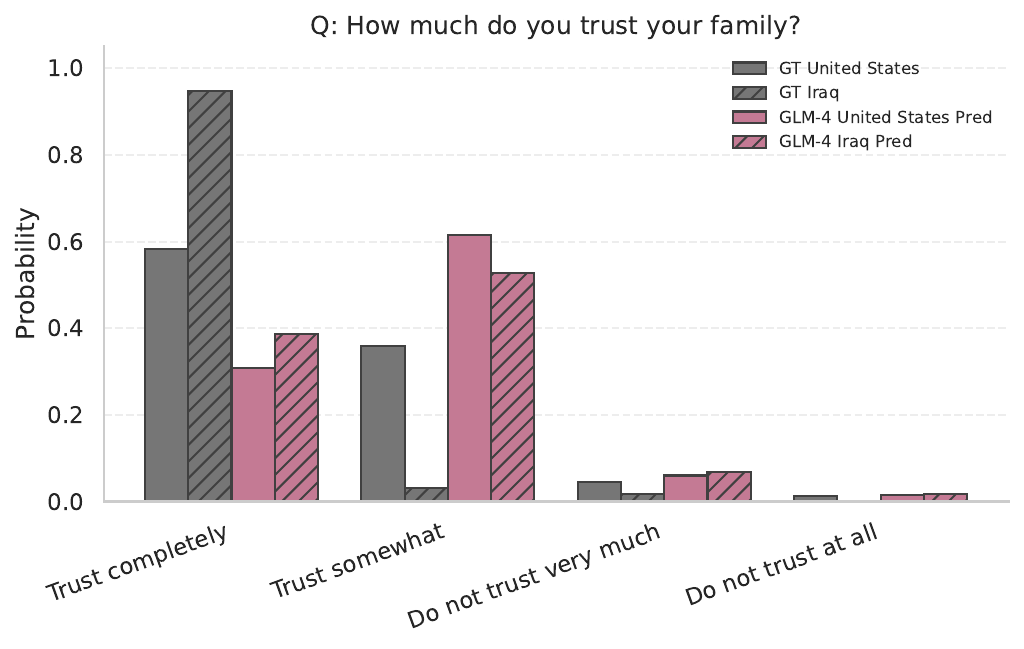}}%
  \hfill
  \subfigure[Cross-question shift pattern\label{fig-1:case-scatter}]{%
    \includegraphics[width=0.43\linewidth]{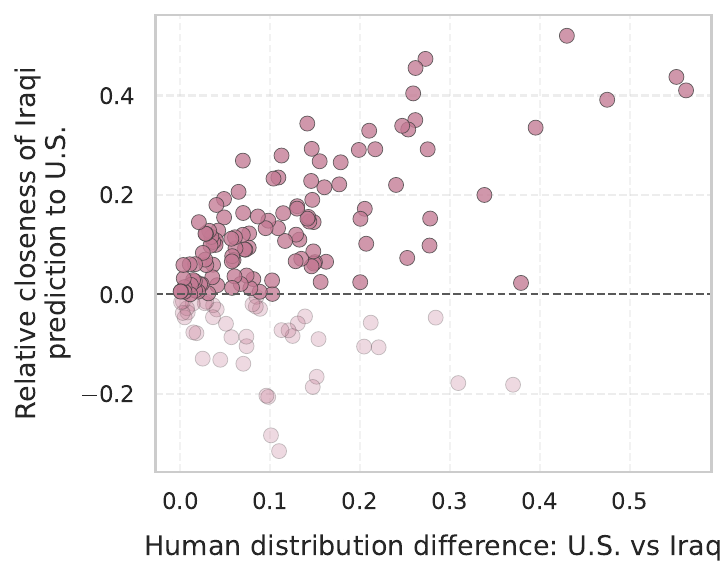}}
  \caption{Simulated value convergence toward the better-simulated country's distribution in GLM-4-9B.
  (a) Human and model-predicted response distributions for the United States and Iraq on WVS Q58 about how much respondents trust their family.
  (b) Cross-question comparison between empirical cross-country differences and model prediction shifts. The horizontal axis reports the Jensen--Shannon divergence between the U.S. and Iraqi human response distributions. The vertical axis is the divergence of the Iraqi-profile prediction from the Iraqi human distribution minus its divergence from the U.S. human distribution. Positive values therefore indicate that the prediction is closer to the U.S. human distribution.}
  \label{fig-1:case-study}
\end{figure}

When simulating Iraq, GLM-4-9B often shifts the predicted distribution closer to the distribution of the United States. Figure~\ref{fig-1:case-bar} illustrates this on WVS Q58 about how much respondents trust their family. In the human data, approximately 60\% of U.S. respondents answer ``Trust completely'', while over 95\% of Iraqi respondents do so. However, GLM-4-9B predicts a similar distribution of around 40\% for both countries, indicating that the model fails to preserve the near-unanimous Iraqi response pattern and instead anchors its output toward the United States.
Figure~\ref{fig-1:case-scatter} further shows that this convergence pattern extends across the full question set. The relative closeness scores for more than half of the questions fall above zero, indicating that GLM-4-9B's simulated Iraqi distribution is more often closer to the empirical U.S. distribution than to the empirical Iraqi distribution. This tendency is especially visible when the two countries exhibit large empirical differences.

In this country pair, representational inequality is accompanied by directional convergence toward the U.S. value distribution. A plausible account is that large-scale web corpora overrepresent certain viewpoints~\citep{bender2021dangers,navigli2023biases}, and models are often more accurate on high-frequency patterns~\citep{razeghi2022impact}. These conditions may create a latent prior toward the cultural responses of better-represented countries~\citep{manvi2024large}. Consequently, even when an underrepresented country is explicitly specified, the model may anchor its output toward the better-represented country's profile and fail to preserve distinctive value distributions of underrepresented countries, even when those distributions are highly homogeneous within the target population. Mitigating this pattern may require richer culture-specific contextual signals at inference time or targeted reshaping of the model's internal value distribution through training. The following sections investigate these two intervention pathways: contextual adaptation and parametric modification.

\section{Impact of contextual adaptation}

In this section, we investigate how two key aspects of contextual adaptation, native language prompting and additional-information settings, influence LLMs' simulation performance for different subpopulations. Through this comparison, we assess whether inference-time interventions can improve representational equality in value simulation while maintaining accuracy across diverse populations.
The native-language prompting analysis covers a subset of 27 countries where Chinese, Arabic, Spanish, or English serve as primary languages, and is conducted on eight core multilingual models. The additional-information analysis covers all 59 countries of the shared evaluation panel and is evaluated on nine models.

\subsection{Native language}

\begin{figure}[t]
  \centering
    {\includegraphics[width=\linewidth]{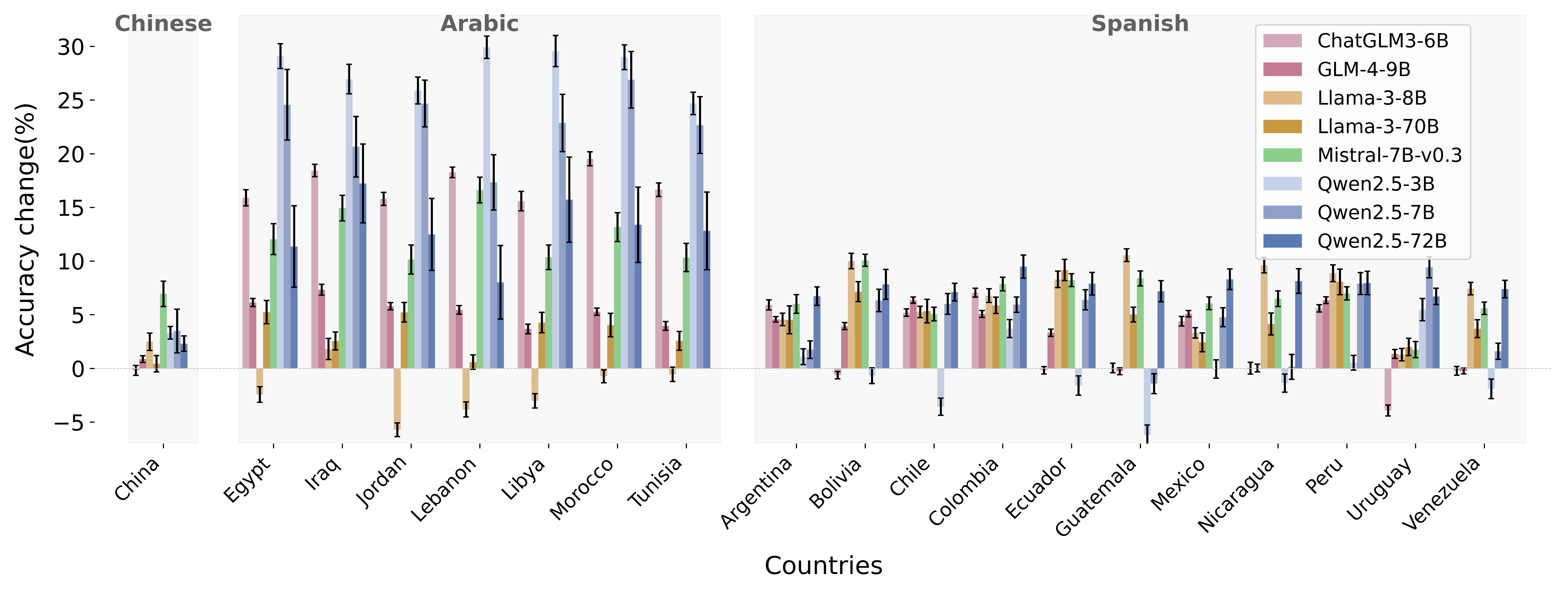}}
  \caption{Simulation accuracy changes (\%) of LLMs under native-language prompting. Bars show the mean relative change from English prompting for each country. Error bars denote two-sided 95\% $t$ confidence intervals over aligned demographic subgroups.}
  \label{fig-2:language-accuracy}
\end{figure}

Native-language prompting may serve as a linguistic cue that shifts model predictions toward the local response distribution. Such shifts may move predictions away from an English-default pattern and closer to the target population's empirical distribution.
We examine Chinese, Arabic, and Spanish prompting for populations in countries where these languages are primarily spoken. These three languages are selected for their broad global coverage in terms of number of speakers\footnote{Language coverage was considered with reference to Ethnologue's summary of the world's most widely spoken languages (\url{https://www.ethnologue.com/insights/ethnologue200/}).}, the availability of corresponding multilingual questionnaire support in our WVS-based evaluation setting, and their comparatively weaker advantage relative to English in current LLMs. Together, these properties make them informative cases for examining this intervention.

As shown in Figure~\ref{fig-2:language-accuracy} and Table~\ref{tab:equality_summary_exp2}, native language prompts generally enhance average simulation accuracy, but this effect varies across languages, populations, and models. For Chinese, the effect is modest but generally positive. For Arabic and Spanish, outcomes are largely beneficial, while a few models exhibit declines, such as the Llama-3 family in Arabic-speaking regions and Qwen2.5-3B in Spanish-speaking regions. These deviations indicate that although native language can act as a strong cultural cue, its effectiveness remains highly contingent on multilingual capacity and alignment imbalances within individual model ecosystems.

We next evaluate representational equality on the same 27-country subset.
The results reveal two patterns, depending on whether accuracy gains are evenly distributed across language groups. For Llama-3-70B, Qwen2.5-72B, GLM-4-9B, and Mistral-7B-v0.3, improvements are relatively balanced across Chinese, Arabic, and Spanish communities, so cross-country variance is reduced or stable. For Qwen2.5-7B, Llama-3-8B, Qwen2.5-3B, and ChatGLM3-6B, gains concentrate in specific language groups. This creates a new dominant group in place of the previously advantaged English-speaking countries, relocating rather than reducing inequality. The $\mathrm{AE}$ composite metric in Table~\ref{tab:equality_summary_exp2} supports this reading. The first group improves on the joint accuracy--equality score, whereas the second deteriorates even when mean accuracy rises.

\begin{figure}[t]
  \centering
    {\includegraphics[width=0.55\linewidth]{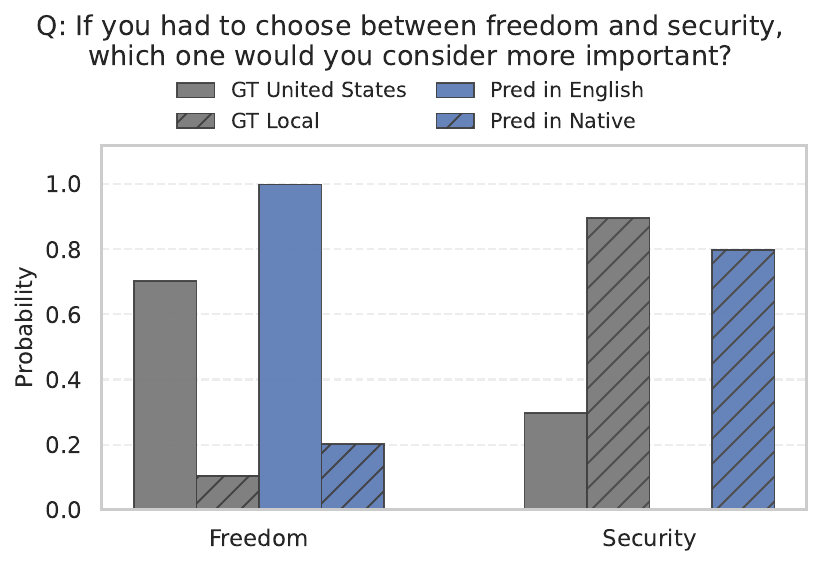}}
  \caption{Simulation results for Jordan on WVS Q150 (forced choice between freedom and security) under English and Arabic prompts. Gray bars show the ground-truth (GT) human distributions for Jordan (Local) and the United States. Blue bars show Qwen2.5-72B predictions (Pred) under English and Arabic (Native) prompting.}
  \label{fig-2:case-qwen}
\end{figure}

\paragraph{Case study}
To illustrate this distributional shift, we examine Qwen2.5-72B on Jordan under WVS Q150. The item asks respondents which they would prioritize if forced to choose between freedom and security, two values that are both widely endorsed. We select Jordan because its empirical distribution differs sharply from that of the United States. Specifically, 70.0\% of U.S.\ respondents prioritize freedom, whereas 90.0\% of Jordanian respondents favor security. This contrast provides a clear case for examining whether a language switch shifts predictions toward the local distribution.
As shown in Figure~\ref{fig-2:case-qwen}, under English prompting Qwen2.5-72B assigns 100.0\% probability to freedom, matching the U.S.\ pattern rather than Jordan's near-unanimous security preference, even though the target country is specified. With an Arabic prompt, the prediction shifts to a 79.2\% preference for security, closely approximating the Jordanian distribution. This case indicates that, for models with sufficient multilingual capacity, native language can serve as an effective linguistic cue that activates country-specific representations and moves the simulation output closer to the target population's distribution.

\paragraph{Summary and implications}
\textbf{Native language prompting can improve simulation accuracy for the corresponding language communities, but the gains are model-dependent and uneven.} Across the 27-country evaluation, accuracy rises for many language--model pairs, yet the gains remain uneven and some pairs decline. When improvements concentrate in particular language groups, cross-country inequality may be relocated rather than reduced, as reflected in the $\mathrm{AE}$ split between balanced and unbalanced models. Native-language prompting is therefore a useful inference-time lever when multilingual capacity is already present, but only a partial remedy. More even global equality requires more balanced multilingual pretraining beyond the relatively high-resource languages studied here.

\subsection{Additional information}

\begin{figure}[t]
  \centering
    {\includegraphics[width=\linewidth]{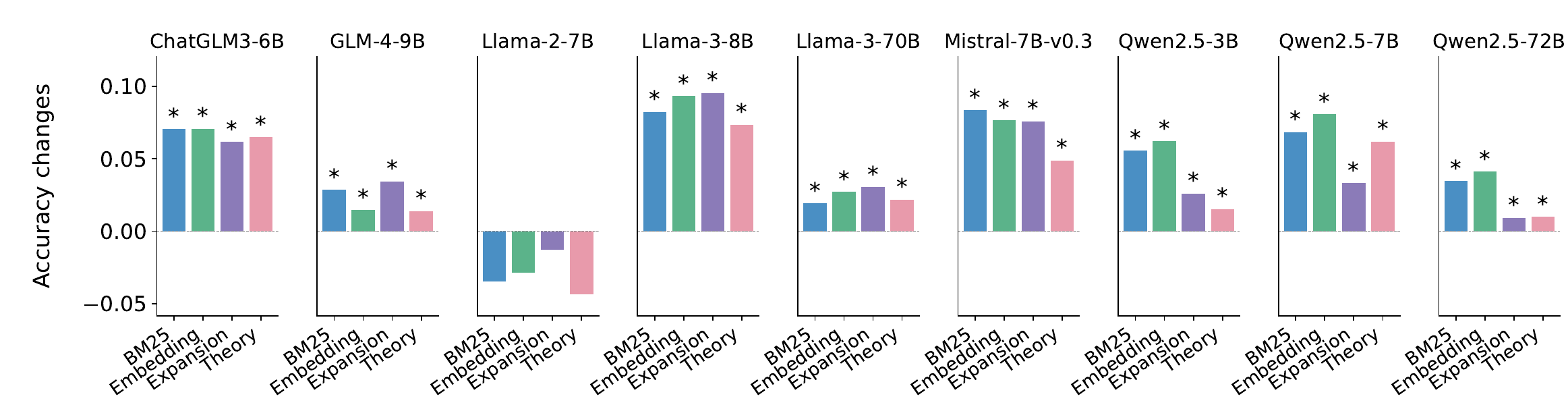}}
  \caption{Simulation accuracy changes across LLMs under different additional-information settings. Bars show mean country-level paired changes over 59 countries. Asterisks mark statistically significant changes in two-sided paired $t$-tests over countries after BH-FDR correction at $q<0.05$ across all comparisons.}
  \label{fig-2:additional}
\end{figure}

Compared with using only demographic profiles to specify simulation identities, incorporating additional information gives the model more concrete background about the target population. Demographic profiles alone may be too sparse, causing the model to rely on broad cues such as generic stereotypes about a population. By providing relevant information from related WVS questions, this setting examines whether more specific context can help the model answer the target survey question for the target population, rather than relying mainly on coarse country-level associations.

\paragraph{Overall representational equality}
As shown in Figure~\ref{fig-2:additional}, incorporating additional information generally improves mean country-level simulation accuracy, with scores ranging from 0.553 to 0.774 across models. The main accuracy exception is Llama-2-7B, whose scores do not rise relative to its baseline; this weaker response may reflect limited ability to integrate additional data and follow the instructed format.

A notable finding is the improvement in representational equality across models, as shown in Table~\ref{tab:equality_summary_exp2}. Providing additional information improves equality scores for almost every model across the retrieval strategies we evaluate. For example, GLM-4-9B, which had relatively high baseline inequality, saw its $\mathrm{Eq}_{\text{CV}}$ drop from 0.0486 to 0.0255 under BM25.
The $\mathrm{AE}$ composite metric further supports this pattern. $\mathrm{AE}$ scores decrease relative to the corresponding baselines across almost all models. The main exception is Qwen2.5-7B under Expansion, which already indicates that retrieval design, especially expansion-style reformulation, can blunt the joint gain. Overall, additional information is a comparatively effective contextual intervention for accuracy and representational equality in our setting, but the gains remain retrieval-dependent.

\paragraph{Analysis of representational inequality}

\begin{figure*}[t]
  \centering
    {\includegraphics[width=\linewidth]{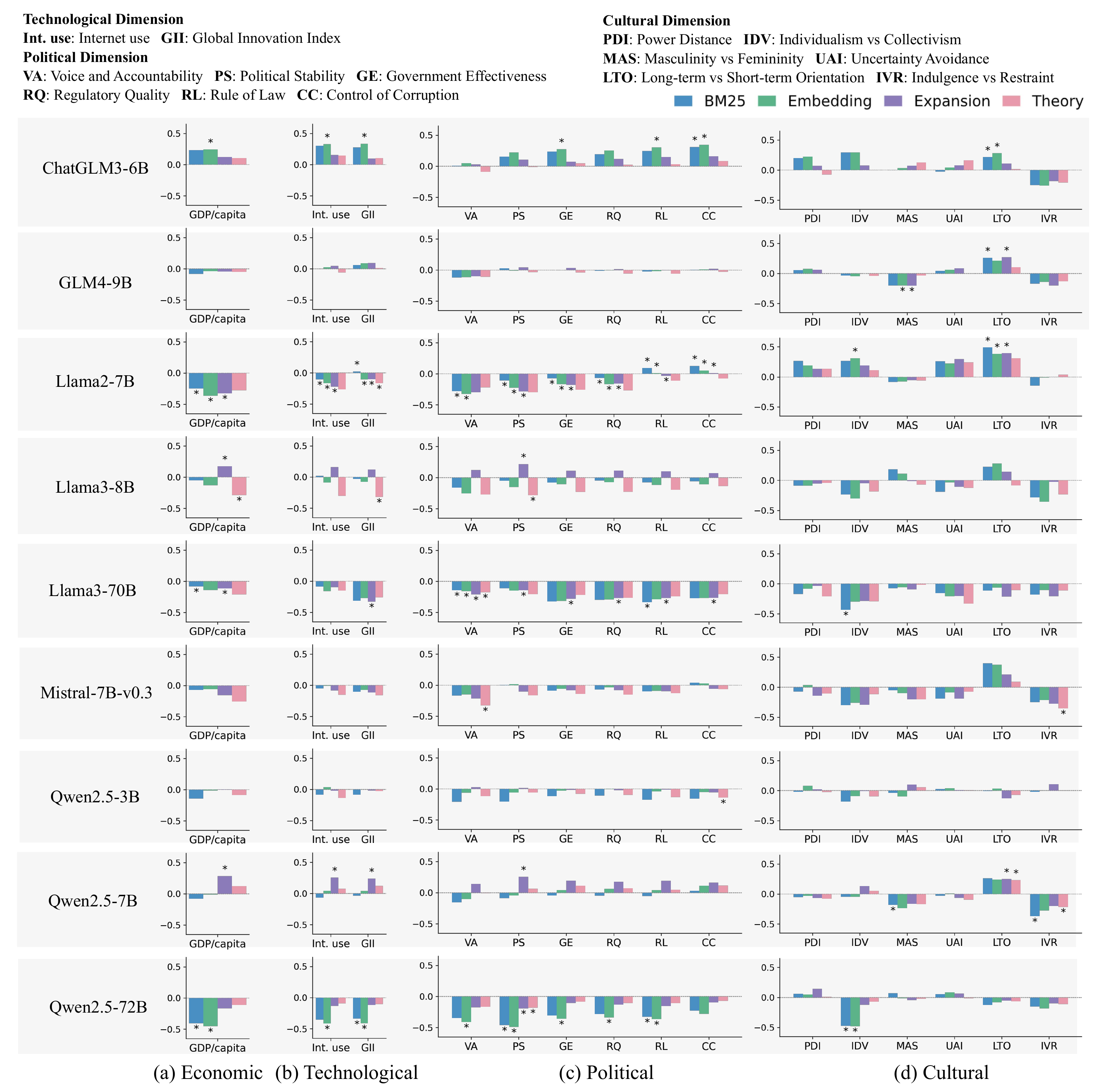}}
  \caption{Changes in the correlations between simulation accuracy and economic, technological, cultural, and political factors under different additional-information strategies. Bars represent the change in absolute Spearman correlation relative to the baseline, with positive values indicating an increase and negative values indicating a decrease. Asterisks indicate statistically significant differences in two-sided Steiger tests~\citep{steiger1980tests} after BH-FDR correction at $q<0.05$ across all displayed comparisons.}
  \label{fig:additional-correlation-change}
\end{figure*}

Lower $\mathrm{Eq}_{\text{CV}}$ values indicate a more even distribution of simulation accuracy across countries, but they do not by themselves establish weaker associations with macro-level conditions. Figure~\ref{fig:additional-correlation-change} therefore reports changes in the absolute Spearman correlations between country-level accuracy and economic, technological, political, and cultural factors. Negative bars indicate weaker absolute associations than at baseline. Descriptively, retrieval reduces these associations in many settings, particularly for Qwen2.5-72B and the Llama models on economic, technological, and political factors. Conversely, ChatGLM3-6B shows stronger associations across retrieval strategies, while Expansion strengthens selected associations for Qwen2.5-7B and Llama-3-8B. Additional information can thus narrow cross-country accuracy gaps while either weakening or strengthening structural associations, depending on the model, retrieval strategy, and factor.

\begin{figure*}[t]
  \centering
  \subfigure[GDP-Accuracy Association\label{fig-2:supplement-case-gdp}]{%
    \includegraphics[width=0.46\linewidth]{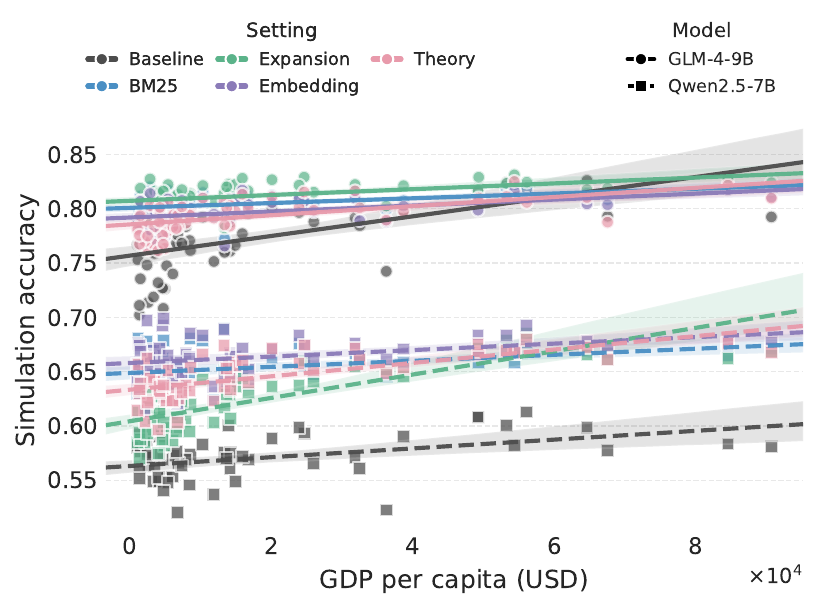}}%
  \hfill
  \subfigure[Distribution Shift\label{fig-2:supplement-case-dist}]{%
    \includegraphics[width=0.50\linewidth]{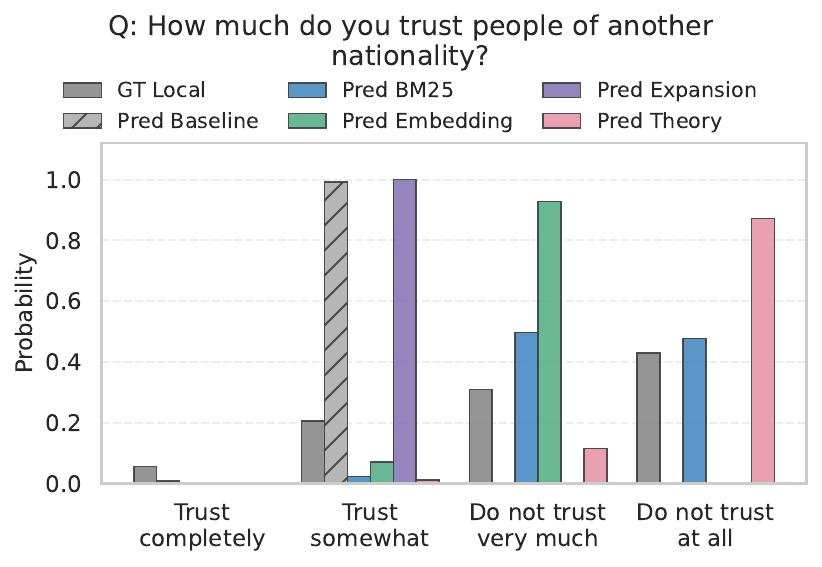}}
  \caption{Effect of additional information on GDP-accuracy association and country-specific response distributions. (a) Scatter plot with fitted regression lines showing the relationship between country GDP per capita and simulation accuracy under different retrieval settings for GLM-4-9B and Qwen2.5-7B. (b) Human ground-truth (gray bars) and average model-predicted (colored bars) response distributions when Qwen2.5-7B simulates the Mexican subpopulation for WVS Q63 (trust people of another nationality) under various retrieval strategies.}
  \label{fig-supplement-case-study}
\end{figure*}

\paragraph{Case study}
To concretely illustrate how different retrieval strategies reshape these structural inequalities, we examine both country-level trends across the economic dimension and a specific simulation case. 
Figure~\ref{fig-2:supplement-case-gdp} shows that in the baseline setting, accuracy is strongly positively associated with country GDP. Under several retrieval settings for GLM-4-9B, the fitted line flattens and lower-income countries improve more, consistent with a descriptive decrease in absolute GDP--accuracy association. For Qwen2.5-7B, Expansion instead reverses this pattern and yields a steeper slope.

Figure~\ref{fig-2:supplement-case-dist} further illustrates the retrieval-dependent pattern for Qwen2.5-7B when simulating the Mexican population on trust in people of another nationality. In the baseline, the model concentrates predictions on ``Trust somewhat.'' In this case, BM25 moves the prediction closer to the local distribution, whereas Expansion shifts it back toward the baseline pattern. This result suggests that model-generated expansions may reinforce an existing directional bias. Reformulating the query through the model's own assumptions may retrieve context that confirms the baseline pattern and weakens the equalizing effect of additional information.

\paragraph{Summary and implications}
\textbf{Additional information is a comparatively effective contextual intervention for improving both simulation accuracy and representational equality.} Joint gains are attainable when missing local context is supplied at inference. Direct retrieval such as BM25 most clearly improves performance for lower-income countries and corrects biased defaults, whereas Expansion can reintroduce model priors and reverse equalizing trends for models such as Qwen2.5-7B. Retrieval can also attenuate selected associations, as for Qwen2.5-72B under BM25 and Embedding, but the direction and statistical support vary by model, strategy, and factor. Future designs should therefore prioritize careful evidence selection and integration, and treat expansion-style reformulation with caution when the goal is equitable cross-country simulation.

\section{Impact of parametric modification}

In this section, we investigate how two aspects of parametric modification, CPT on target-language corpora and preference alignment, influence the cross-country equality of value simulation. The goal is to assess whether training-time interventions can promote more balanced performance.
The CPT analysis covers a subset of 12 countries where the six target languages are primary national languages. The alignment analysis covers all 59 countries using five base SFT models and four configurations. Table~\ref{tab:exp_summary} summarizes these settings.

\subsection{CPT languages}

\begin{figure}[t]
  \centering
  \subfigure[Simulation accuracy change\label{fig:acc-cpt}]{%
    \label{fig-3:train-accuracy}
    \includegraphics[width=0.44\linewidth]{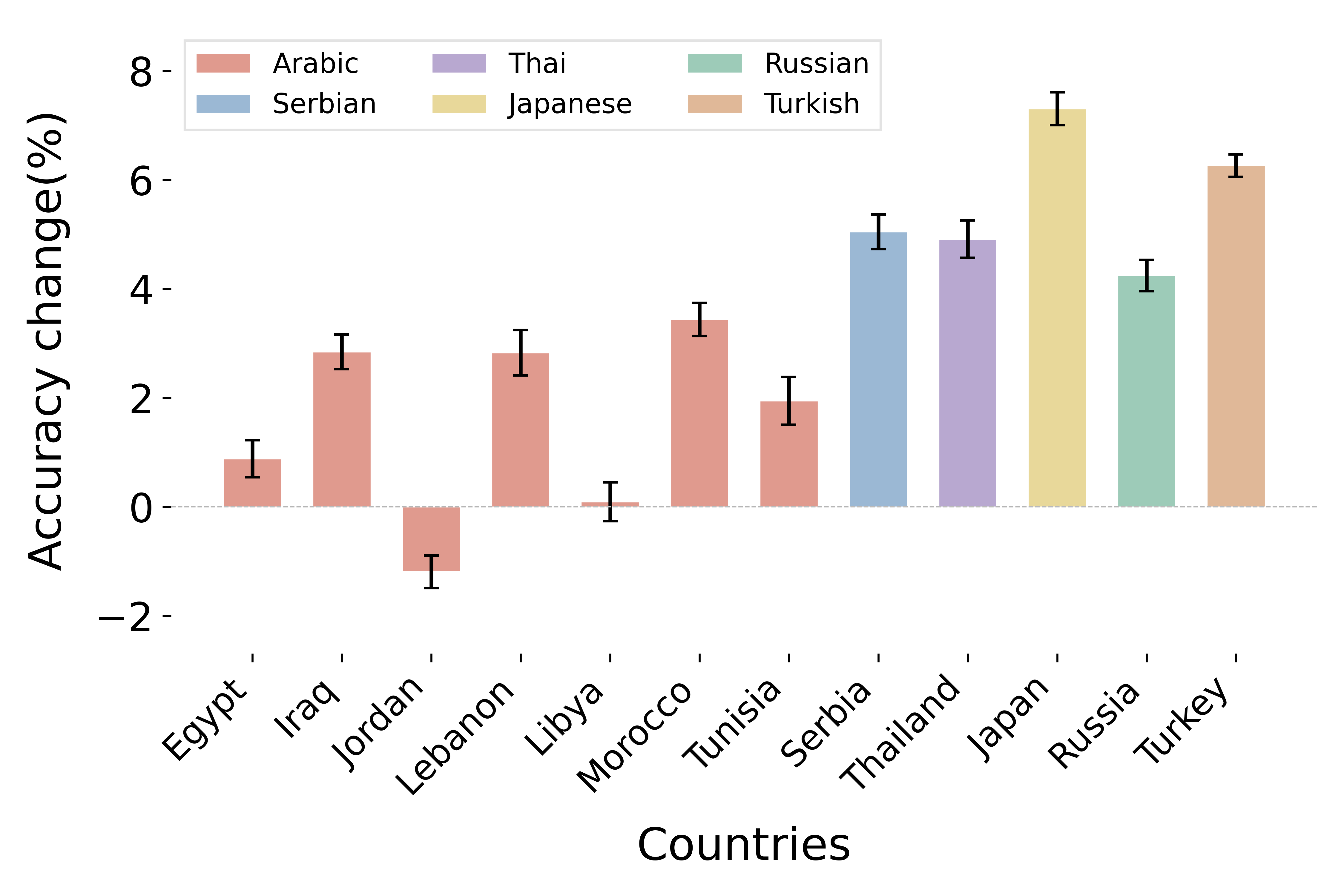}}%
  \hfill
  \subfigure[Directional gains of CPT\label{fig:cpt-directional-gains}]{%
   \label{fig-3:cpt-directional-case}
    \includegraphics[width=0.52\linewidth]{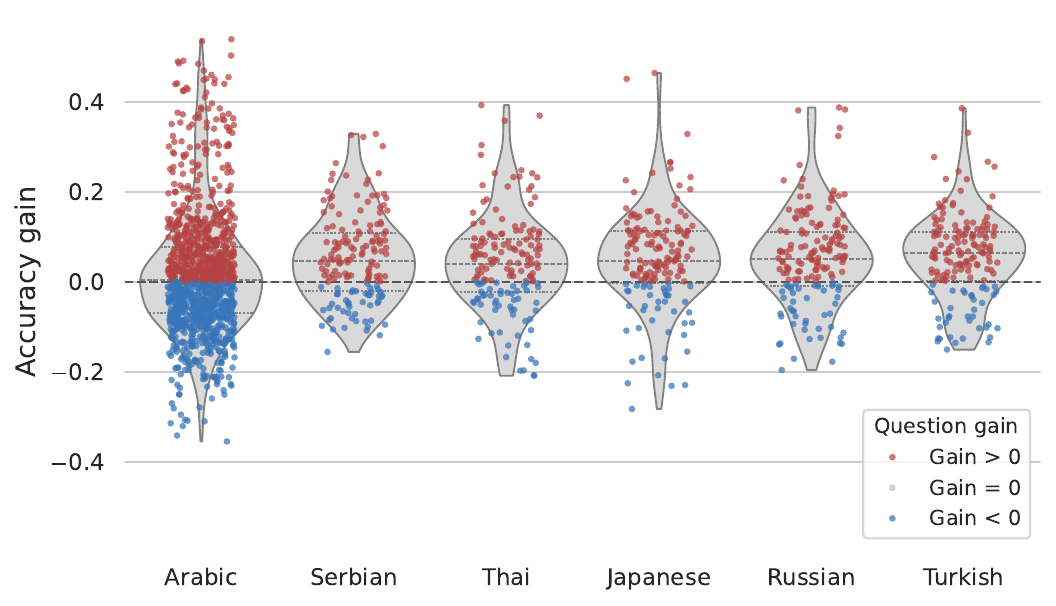}}

  \caption{Effects of multilingual CPT on simulation accuracy.
  (a)~\textbf{Country-level} relative accuracy changes after multilingual CPT, averaged over demographic subgroups within each country; error bars show 95\% confidence intervals over subgroups.
  Bars are grouped by the CPT target language (six languages; Arabic covers multiple countries).
  (b)~Question-level accuracy gain distributions by language.
  Red points indicate positive gains, blue points indicate negative gains, and gray points indicate no change.}
  \label{fig:accuracy-combined}
\end{figure}

One plausible source of simulation disparities for underrepresented subpopulations is their limited exposure to relevant language data during post-training. We therefore hypothesize that CPT on target-language corpora can improve simulation accuracy for the corresponding language groups by increasing the model's familiarity with their linguistic and cultural context.

To explore this, we conduct CPT experiments with six target languages: Arabic, Japanese, Russian, Serbian, Thai, and Turkish. Figure~\ref{fig-3:train-accuracy} reports country-level relative changes in mean simulation accuracy after CPT, $(\mathrm{Acc}_{\mathrm{CPT}}-\mathrm{Acc}_{\mathrm{base}})/\mathrm{Acc}_{\mathrm{base}}$. Scores are averaged over demographic subpopulations within each country, and error bars denote 95\% confidence intervals over subgroups. CPT improves mean accuracy for all six language-level groups. Arabic shows the greatest cross-country heterogeneity: its country-level changes range from approximately $-1.2\%$ in Jordan to 3.4\% in Morocco, while Egypt improves by 0.9\%. Its language-average gain is approximately 1.6\%.

The gains are also uneven across survey items. Figure~\ref{fig-3:cpt-directional-case} shows limited or negative changes for some questions, including within language groups with positive average gains. Thus, language-specific continued post-training improves aggregate accuracy more consistently than it improves every country or value item. The Arabic pattern may be especially heterogeneous because Arabic is used across countries with diverse cultural contexts. A shared Arabic CPT corpus may therefore provide unevenly relevant regional and value-domain coverage across these populations.
\paragraph{Summary and implications}
\textbf{CPT improves average simulation accuracy across all six target-language groups, although the gains are not uniform across all value-related questions or language groups.} Language coverage alone is therefore insufficient to ensure representational equality. More targeted corpus curation and controlled training comparisons are needed to determine how language-specific data can yield more even improvements across populations.

\subsection{Preference alignment}

\begin{figure}[t]
  \centering
  \includegraphics[width=\linewidth]{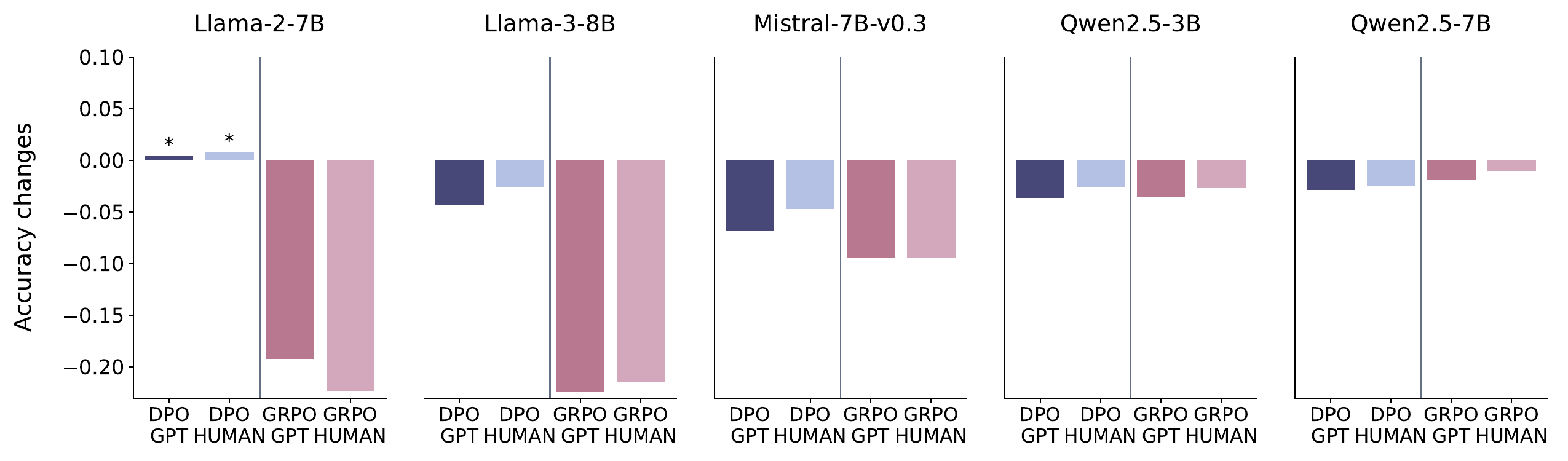}
  \caption{Simulation accuracy changes across LLMs under different alignment settings. Bars show mean country-level paired changes over 59 countries. Asterisks mark statistically significant paired changes in two-sided paired $t$-tests over countries after BH-FDR correction at $q<0.05$ across all comparisons.}
   \label{fig-3:dpo_accuracy}
\end{figure}

Preference alignment can reshape model outputs by optimizing them toward particular preference data, which may affect both simulation accuracy and representational equality. To examine this, we apply two preference optimization methods (DPO and GRPO) using two preference datasets: one annotated by humans and another by AI. This setting allows us to assess whether preference optimization improves value simulation and whether different annotation sources lead to different accuracy--equality outcomes.

\paragraph{Overall representational equality}
As shown in Figure~\ref{fig-3:dpo_accuracy} and Table~\ref{tab:equality_summary_exp3}, preference alignment often does not improve value simulation. In terms of average accuracy, aligned models mostly do not outperform the corresponding base settings. Preference optimization can therefore substantially change simulation behavior, but these changes are not systematically beneficial for distributional value simulation.
The equality results reveal a similar limitation. Across the majority of evaluated configurations, alignment either worsens or fails to improve cross-country representational balance. Only a limited number of settings reduce $\mathrm{Eq}_{\text{CV}}$ relative to their corresponding base models, while many settings increase $\mathrm{Eq}_{\text{CV}}$ or leave it close to the base level.
The composite $\mathrm{AE}$ metric reinforces this pattern. Most aligned settings either increase $\mathrm{AE}$ or leave it largely unchanged, indicating that alignment rarely improves joint accuracy--equality performance. In other words, preference alignment does not generally produce simultaneous gains in simulation accuracy and cross-country balance.

\paragraph{Effect of annotation source}
The annotation source further shapes alignment outcomes, especially for average accuracy. Across paired aligned settings, human-annotated preference data more consistently preserve simulation accuracy than GPT-annotated data. In most model--setting pairs, the human-annotated variant achieves equal or higher accuracy than its GPT-annotated counterpart. This pattern suggests that human feedback provides a more stable signal for maintaining distributional similarity to empirical human value responses.
However, the advantage of human annotations is less consistent for representational equality. Human-annotated alignment often yields lower $\mathrm{Eq}_{\text{CV}}$ than GPT-annotated alignment, but this pattern does not hold across all settings. In some cases, GPT-annotated alignment produces lower $\mathrm{Eq}_{\text{CV}}$.

Thus, human annotations are more favorable for preserving simulation accuracy, while their equality advantage is not systematic. GPT annotations can sometimes yield lower cross-country accuracy dispersion, but this result should be interpreted together with average accuracy and the composite $\mathrm{AE}$ metric. The annotation source therefore matters, but it does not overturn the broader conclusion that preference alignment yields no systematic gains in representational equality.

\paragraph{Summary and implications}
\textbf{Preference alignment yields no systematic gains in simulation accuracy or representational equality.} Aligned models mostly do not outperform their base settings on average accuracy, and most configurations either worsen $\mathrm{Eq}_{\text{CV}}$ or leave it close to the baseline. Human-annotated preference data more consistently preserve simulation accuracy than GPT-annotated data, but this advantage does not systematically extend to equality. These patterns appear under both DPO and GRPO, so the conclusion is not tied to one optimization method.

\section{Conclusion and limitations}

This study provides a systematic analysis of representational equality in LLM-based human value simulation. Our findings can be summarized as follows:

\begin{itemize}
    \item \textbf{Cross-country representational inequality} is substantial and systematic in LLM-based value simulation. Models tend to simulate countries with stronger economies, more advanced technological infrastructure, higher governance quality, lower power distance, and greater individualism more accurately. These patterns appear across model families and scales, suggesting a systemic limitation of current LLMs.
    \item \textbf{Native-language prompting} can improve simulation accuracy for target populations by activating culturally grounded representations, but its effects vary across models and language communities and do not consistently improve cross-country representational equality.
    \item \textbf{Additional-information settings} are a comparatively effective and robust contextual intervention in our evaluation. They improve simulation accuracy and reduce cross-country inequality across most models.
    \item \textbf{Language-specific continued post-training} improves average simulation accuracy for the targeted language groups, but the gains are uneven across languages, countries, and questions. More targeted data construction is therefore needed to achieve broader representational gains.
    \item \textbf{Preference alignment} yields no systematic gains in simulation accuracy or representational equality. Human-annotated preference data generally preserve simulation accuracy better than GPT-annotated data, suggesting that annotation source matters for alignment in this setting.
\end{itemize}

These results have empirical and methodological implications for LLM-based social simulation. In the WVS analysis, simulation performance is systematically associated with macro-level economic, technological, political, and cultural conditions across model scales and the intervention settings examined here. The ISSP analysis provides supplementary evidence that cross-country accuracy differences and the relative effects of the evaluated strategies are not specific to a single survey instrument. Methodologically, the supplementary $\mathrm{AE}$ composite metric shows why accuracy and representational equality should be evaluated jointly: models with the highest average accuracy do not necessarily achieve the best joint accuracy--equality performance, and additional-information strategies provide the most consistent improvement on both dimensions. For practitioners, these findings point to the importance of inclusive model design, diversified data construction, and careful use of alignment when deploying LLMs across diverse cultural contexts.

While these findings provide useful insights, several limitations warrant attention in future research. First, our evaluation relies on standardized, fixed-option cross-country surveys. This design enables the comparable country-level evaluation required by our research question, but the findings may not generalize to other forms of value expression, including open-ended discourse, moral reasoning, or behavioral enactments beyond structured questionnaires. Second, although country-level analysis is empirically supported as the most discriminative demographic dimension in our setting and is consistent with established cross-country research practice, it inevitably masks within-country subpopulation heterogeneity and cannot fully represent linguistic minorities or languages that cut across national boundaries. Third, while our robustness analysis shows no systematic association between within-country response heterogeneity and simulation accuracy in the current setting (Appendix~\ref{appendix:heterogeneity}), response heterogeneity remains a conceptually important factor that may affect the comparability of simulation equality scores. Fourth, the retained WVS dimensions differ in their substantive focus, level of abstraction, cultural sensitivity, and susceptibility to translation or social-desirability effects. Although our main analysis reports an overall dimension-balanced result and robustness checks suggest stable aggregate patterns, individual value dimensions may still show heterogeneous patterns obscured by the overall score.

Our findings also suggest several directions for future research. A central source of cross-country inequality is likely data scarcity: underrepresented countries often have weaker digital infrastructure and sparser linguistic representation in pretraining corpora, which limits the model's capacity to simulate their populations. Our intervention experiments suggest that addressing this scarcity is a promising route to improving simulation accuracy for underrepresented countries. Additional information at inference time compensates for missing contextual knowledge and most consistently improves joint accuracy--equality performance. Our CPT comparison further indicates that low-resource-language post-training can enrich parametric knowledge, although its effects remain uneven. Future work can examine broader data-centric strategies, such as monolingual corpus augmentation, cross-lingual back-translation~\citep{mcgiff2025overcoming,yirmibecsouglu2023morphologically}, and structured cross-lingual transfer learning~\citep{chang2024multilinguality,downey2024targeted}, and evaluate them jointly in terms of simulation accuracy and representational equality. Future work should also construct richer cross-country benchmarks, pursue finer-grained sub-country, multilingual, and dimension-level analyses, and further examine cross-cultural measurement differences in cross-country comparisons.

Overall, this study shows that building reliable LLM-based social simulators requires more than improving average accuracy. Models must also provide balanced simulation capability across countries and avoid reproducing structural inequalities embedded in global data ecosystems. Progress on this goal requires broader demographic coverage, finer-grained multilingual evaluation, and a clearer understanding of how model architectures, training processes, and data composition shape representational equality. Addressing these challenges can support more inclusive and representative AI systems for diverse global populations.

\appendix
\renewcommand{\thesection}{\Alph{section}}

\renewcommand{\thesubsection}{\thesection.\arabic{subsection}}

\renewcommand{\theHsection}{appendix.\Alph{section}}
\renewcommand{\theHsubsection}{appendix.\Alph{section}.\arabic{subsection}}
\renewcommand{\theHfigure}{appendix.\Alph{section}.\arabic{figure}}
\renewcommand{\theHtable}{appendix.\Alph{section}.\arabic{table}}
\renewcommand{\theHequation}{appendix.\Alph{section}.\arabic{equation}}

\apptocmd{\appendixsection}{%
  \setcounter{subsection}{0}%
}{}{}

\section{Methodological Details}

\subsection{Representational equality index definitions}
\label{appendix:index}

Based on the 59 country-level accuracy scores, we compute four established indices to measure representational equality. For each country $c$, $A_c$ denotes its country-level simulation accuracy, obtained by aggregating the simulation accuracy of all retained subpopulations within that country. Let $N=|\mathcal{C}|$ denote the number of evaluated countries. These indices therefore measure cross-country dispersion in simulation performance.

\begin{itemize}
\item \textbf{Max-Min Difference.} This index captures the worst-case inequality by measuring the absolute gap between the best- and worst-performing countries. A lower value indicates higher equality.
\begin{equation}
    Eq_{\text{diff}} = \max_{c \in \mathcal{C}} A_c - \min_{c \in \mathcal{C}} A_c
\end{equation}

\item \textbf{Min-Max Ratio.} This index measures the relative performance disparity between the lowest and highest country-level simulation accuracy~\citep{Ghosh2021CharacterizingIG}; values closer to 1 indicate higher equality.
\begin{equation}
    Eq_{\text{ratio}} = \frac{\min_{c \in \mathcal{C}} A_c}{\max_{c \in \mathcal{C}} A_c}
\end{equation}

\item \textbf{Coefficient of Variation (CV).} As a standard statistical measure of relative dispersion, CV quantifies the extent of variability relative to the mean of country-level accuracy scores~\citep{Hooker2021AGT}. A lower value indicates higher equality.
\begin{equation}
  Eq_{\text{CV}} = \frac{\sigma_{A_c}}{\mu_{A_c}}
\end{equation}
where $\mu_{A_c}$ and $\sigma_{A_c}$ denote the mean and population standard deviation of $\{A_c\}_{c \in \mathcal{C}}$, respectively.

\item \textbf{Gini Coefficient.} Borrowed from economics, the Gini Coefficient measures the concentration of the ``resource'' of accuracy across countries~\citep{Murakami1976GiniCA}. It provides a holistic assessment of the full distribution of country-level accuracy scores (0 denotes perfect equality).
\begin{equation}
    Eq_\text{Gini} = \frac{\sum_{c \in \mathcal{C}} \sum_{c' \in \mathcal{C}} |A_c - A_{c'}|}{2N \sum_{c \in \mathcal{C}} A_c}
\end{equation}
\end{itemize}

\subsection{Details of first-token probability extraction}
\label{appendix:first_token_details}

For each multiple-choice question, we request the logarithmic probabilities of the top 20 first-position tokens generated by the model, a widely adopted convention in LLM APIs for obtaining token-level probability estimates. We then retain only the subset of returned tokens that correspond to valid answer-option letters (e.g., ``A'', ``B'', and ``C''). Tokens among the top 20 that do not correspond to valid answer options are treated as outside the option set and are not reassigned to any survey choice. The retained logarithmic probabilities are exponentiated and normalized to form the final probability distribution $D$ for the model's simulated answers.

Let $O_q$ be the valid option set for question $q$, $I$ the returned top-20 token set, and $V\subseteq O_q$ the valid options observed in $I$. The missing valid options are $U=O_q\setminus V$, with $K=|U|$. Following~\citet{santurkar2023whose}, we compute the residual probability mass $r=\max(0,1-\sum_{i\in I}p_i)$. When $K>0$ and $V$ is nonempty, each missing option receives the provisional value
\begin{equation}
  \widetilde{p}_u=\min\left(\frac{r}{K},p_{\mathrm{cap}}\right), \qquad
  p_{\mathrm{cap}}=\min_{v\in V}p_v, \quad u\in U.
\end{equation}
We then renormalize the provisional probabilities over all valid options. If no usable valid-option probability is available, we use a uniform distribution over $O_q$. We refer to this procedure as a capped residual-mass approximation. The cap constrains the provisional allocation from above. For multiple missing options, $r/K$ equally allocates residual mass without bounding any option's unknown true probability.

\subsection{Macro-level indicators for diagnostic analysis}
\label{appendix:macro_factor_details}

This section provides detailed descriptions and data sources for the PEST indicators used in the diagnostic analysis.

\textbf{Economic factor.} GDP per capita is taken from the World Bank indicator database, using the 2024 country values\footnote{\url{https://data.worldbank.org/indicator}}. This indicator reflects a country's overall economic capacity to invest in education, digital literacy, and technological infrastructure, which may relate to how large, visible, and diverse the available digital traces are.

\textbf{Technological factor.} Internet use is taken from the ITU statistics portal using the latest available country value up to 2024.\footnote{\url{https://www.itu.int/en/ITU-D/Statistics/pages/stat/default.aspx}} The Global Innovation Index (GII) is taken from the WIPO Global Innovation Index 2024.\footnote{\url{https://wipo.int/web-publications/global-innovation-index-2024/en/gii-2024-results.html}} Internet use captures population participation in online spaces and potential data generation, while GII reflects a country's broader innovation capability and technological ecosystem.

\textbf{Political factor.} The six Worldwide Governance Indicators (WGI)~\citep{Kaufmann2010TheWG}, including Voice and Accountability (VA), Political Stability (PS), Government Effectiveness (GE), Regulatory Quality (RQ), Rule of Law (RL), and Control of Corruption (CC), are taken from the World Bank's 2024 release.\footnote{\url{https://www.worldbank.org/en/publication/worldwide-governance-indicators}} These indicators jointly characterize institutional quality and governance effectiveness.

\textbf{Socio-cultural factor.} Hofstede's cultural dimensions~\citep{Hofstede2011DimensionalizingCT}, including Power Distance (PDI), Individualism versus Collectivism (IDV), Masculinity versus Femininity (MAS), Uncertainty Avoidance (UAI), Long-term versus Short-term orientation (LTO), and Indulgence versus Restraint (IVR), are used based on the 2015 release. These dimensions capture cross-cultural variation in values and norms that may influence digital behavior and content creation patterns across countries.

\subsection{Additional-information retrieval strategies}
\label{appendix:theory-driven}

We compare four strategies for retrieving non-target WVS questions. The target question itself is excluded from every candidate ranking. From each ranking, we select the first three questions with empirical response distributions for the target subpopulation.

\textbf{BM25 retrieval (BM25).} We lowercase and whitespace-tokenize each question, then rank candidates by BM25 lexical similarity using the \texttt{rank\_bm25} implementation.

\textbf{Embedding retrieval (Embedding).} For each evaluated model $M$, we encode each survey question with $M$'s native tokenizer and checkpoint in evaluation mode. We mean-pool final-layer hidden states over non-padding tokens and rank non-target candidates by cosine similarity. Retrieval is conducted independently within each model's representation space; scores are not compared across models.

\textbf{Query-expansion retrieval (Expansion).} Each evaluated model $M$ generates a short, question-specific expansion by reflecting on the target item and its response options. We concatenate this expansion with the original target question and encode the resulting query with $M$ using the same procedure as in Embedding. Candidate questions are encoded in their original form with $M$ and ranked by cosine similarity to the expanded-query representation.

\textbf{Theory-driven retrieval (Theory).} We use Scopus to measure academic similarity between questions. We first extract key concepts from each question using GPT-4o, then calculate their co-occurrence frequency within Scopus publications. The resulting ranking identifies questions connected through academic concept co-occurrence.

\subsection{Alignment training hyperparameters}
\label{appendix:alignment_hyperparams}

This appendix reports the key hyperparameters used in the alignment experiments.
The DPO experiments are implemented with the LLaMA-Factory framework~\citep{zheng2024llamafactory} using Low-Rank Adaptation (LoRA) fine-tuning.
The GRPO experiments are conducted with the VERL reinforcement-learning framework~\citep{sheng2025hybridflow} for response generation and a reward model for preference scoring.
For each method, the same hyperparameter configuration is used for the corresponding human-annotated and GPT-annotated preference data.

\begin{table}[htbp]
\centering
\caption{Key hyperparameters for DPO and GRPO alignment experiments.}
\label{tab:alignment_hyperparams}
\begin{tabular}{lcc}
\toprule
\textbf{Hyperparameter} & \textbf{DPO} & \textbf{GRPO} \\
\midrule
Fine-tuning method & LoRA & LoRA \\
LoRA rank & 8 & 32 \\
Learning rate & $1 \times 10^{-5}$ & $5 \times 10^{-6}$ \\
Number of epochs & 2 & 2 \\
Effective batch size & 16 & 512 \\
Maximum prompt length & 1024 & 1024 \\
DPO beta ($\beta$) & 0.1 & -- \\
Rollout samples per prompt & -- & 16 \\
KL coefficient & -- & 0.001 \\
\bottomrule
\end{tabular}
\end{table}

For DPO, we use the sigmoid preference loss with $\beta=0.1$, train LoRA adapters for two epochs, and set the effective batch size to 16.
For GRPO, we train for two epochs with an effective batch size of 512 and sample 16 responses per prompt during rollout.
A KL regularization coefficient of 0.001 is used in GRPO.
These settings are reported to make the alignment experiments more transparent and reproducible.

\section{Benchmark and Data}

\subsection{Demographic indicators}
\label{appendix:demographic}

\begin{table}[h]
  \centering
  \resizebox{1\linewidth}{!}{%
  \begin{threeparttable}
    \caption{Demographic indicators and their values}
    \label{tab:demographic}
    \begin{tabular}{ll}
      \toprule
      Demographics & Values                         \\
      \midrule
      Gender   & Male, Female \\
      Age   & 18-24, 25-44, 45-59, 60-75, 75+                    \\
      Country & China, USA, Russia, and other countries (59 countries total)\tnote{*}    \\
      Income Level & Levels 1-10   \\
      Education Level & No education, Primary school, Middle school, High school,\\&  Post-secondary non-tertiary education, Short-cycle tertiary education,\\&  Bachelor's degree, Master's degree, Doctorate or above \\
      \bottomrule
    \end{tabular}
    \begin{tablenotes}
      \footnotesize
      \item[*] The World Values Survey (WVS) dataset originally contains data from 64 different countries and regions. Since our study focuses on country-level indicators and cross-national comparisons, we excluded sub-national regions such as Northern Ireland from the analysis. This ensures that all measurement variables are consistently applied at the national level, resulting in a final dataset of 59 countries.
    \end{tablenotes}
  \end{threeparttable}}
\end{table}

We construct six three-attribute subpopulation families: country--gender--age, country--gender--income, country--gender--education, country--age--income, country--age--education, and country--income--education. Cells are retained when they contain more than 10 respondents. For families without education, retained raw-category patterns meet this threshold in all 59 countries. For families containing education, we map the nine education categories to three broader International Standard Classification of Education (ISCED) bands when applying the cross-country presence filter. This harmonization accommodates cross-national differences in education-system structure and stage boundaries, which otherwise make fine-grained categories unevenly represented across countries. The broader mapping is used only to determine cross-country subgroup presence, and the original education categories are retained when computing response distributions. The resulting evaluation set contains 2,420 country--subpopulation units, with 34--47 units per country.

\subsection{Dataset}
\label{appendix:dataset}

Table~\ref{tab:globalvalues_stats} summarizes the number of questions, average options per question, and average question length across value dimensions. Table~\ref{tab:country_coverage} reports
  the country-level respondent counts for the 59 countries in the shared WVS evaluation panel.

\begin{table}[h]
\centering
\resizebox{\linewidth}{!}{
\begin{tabular}{lccc}
\toprule
Values Dimension & \# Questions & Avg. Options & Avg. Question Length \\
\midrule
Economic Values & 6 & 8.8 & 38.5\\
Ethical Values & 12 & 10.0 & 19.4\\
Happiness and Wellbeing & 11 & 5.6 & 21.1\\
Index of Postmaterialism & 6 & 4.0 & 13.2\\
Perceptions of Migration & 8 & 3.2 & 11.6\\
Perceptions of Security & 5 & 3.2 & 14.8\\
Political Culture and Political Regimes & 19 & 7.6 & 30.8\\
Political Interest and Political Participation & 16 & 3.9 & 19.2\\
Religious Values & 4 & 5.5 & 15.2\\
Social Capital, Trust and Organizational Membership & 31 & 3.6 & 13.0\\
Social Values, Norms, Stereotypes & 42 & 3.2 & 15.8\\
\bottomrule
\end{tabular}}
\caption{Statistical summary of the retained common-question benchmark used in the main analysis. The final panel contains 160 questions across 11 value dimensions after restricting the WVS item pool to questions shared by all 59 countries and removing the singleton \textit{Perceptions of Corruption} dimension.}
\label{tab:globalvalues_stats}
\end{table}

\begin{table*}[t]
\centering
\scriptsize
\setlength{\tabcolsep}{4pt}
\resizebox{\textwidth}{!}{
\begin{tabular}{lrlrlr}
\toprule
Country & Respondents & Country & Respondents & Country & Respondents \\
\midrule
Chile & 1000 & Cyprus & 1000 & Uruguay & 1000 \\
Argentina & 1003 & Andorra & 1004 & Maldives & 1039 \\
Serbia & 1046 & New Zealand & 1053 & Venezuela & 1190 \\
Libya & 1196 & Bangladesh & 1200 & Czechia & 1200 \\
Ecuador & 1200 & Egypt & 1200 & Greece & 1200 \\
Iraq & 1200 & Kyrgyzstan & 1200 & Lebanon & 1200 \\
Morocco & 1200 & Myanmar & 1200 & Nicaragua & 1200 \\
Philippines & 1200 & Slovakia & 1200 & Tajikistan & 1200 \\
Vietnam & 1200 & Jordan & 1203 & Tunisia & 1206 \\
Zimbabwe & 1215 & Armenia & 1223 & Guatemala & 1229 \\
Ethiopia & 1230 & Nigeria & 1237 & South Korea & 1245 \\
Romania & 1257 & Kenya & 1266 & Kazakhstan & 1276 \\
Ukraine & 1289 & Malaysia & 1313 & Japan & 1353 \\
Peru & 1400 & Iran & 1499 & Thailand & 1500 \\
Colombia & 1520 & Germany & 1528 & Mongolia & 1638 \\
Mexico & 1741 & Brazil & 1762 & Australia & 1807 \\
Russia & 1810 & Pakistan & 1995 & Singapore & 2012 \\
Bolivia & 2067 & Netherlands & 2145 & Turkey & 2415 \\
United States & 2596 & Great Britain & 2600 & China & 3036 \\
Indonesia & 3200 & Canada & 4018 &  &  \\
\bottomrule
\end{tabular}
}
\caption{Country-level respondent counts for the 59 countries in the shared WVS evaluation panel. Each country is evaluated on the same 160 shared questions after common-question filtering. Subgroup-level human response distributions are computed from valid (non-missing) responses only.}
\label{tab:country_coverage}
\end{table*}

\subsection{Benchmark construction and sensitivity analysis}
\label{appendix:benchmark_sensitivity}

The main analysis uses a cross-country shared panel constructed in two steps. Starting from 257 multilingual WVS questions spanning 13 dimensions, we retain the 161 questions asked in all 59 countries. These shared questions span 12 dimensions because no item from \textit{Perceptions about Science and Technology} is common to the full country set. We then remove the singleton \textit{Perceptions of Corruption} dimension, leaving 160 questions across 11 retained dimensions. In this panel, question-level accuracies are first averaged within each dimension and then averaged across dimensions. This procedure gives each of the 11 value dimensions equal weight regardless of how many questions it contains, preventing question-dense domains from dominating the aggregate while maintaining cross-country question comparability.

To test whether the conclusions depend on this benchmark construction, we conduct two sensitivity analyses. The first is a leave-one-dimension-out (LODO) analysis, which tests whether the results are driven by any single value dimension. The second compares the main aggregation (equal weight per dimension) with equal question weighting, to test whether the weighting rule changes the main findings.

\begin{figure*}
  \centering
  \subfigure[Equal dimension weighting vs.\ equal question weighting\label{fig:benchmark_accuracy_dumbbell}]{
    \includegraphics[width=0.46\linewidth]{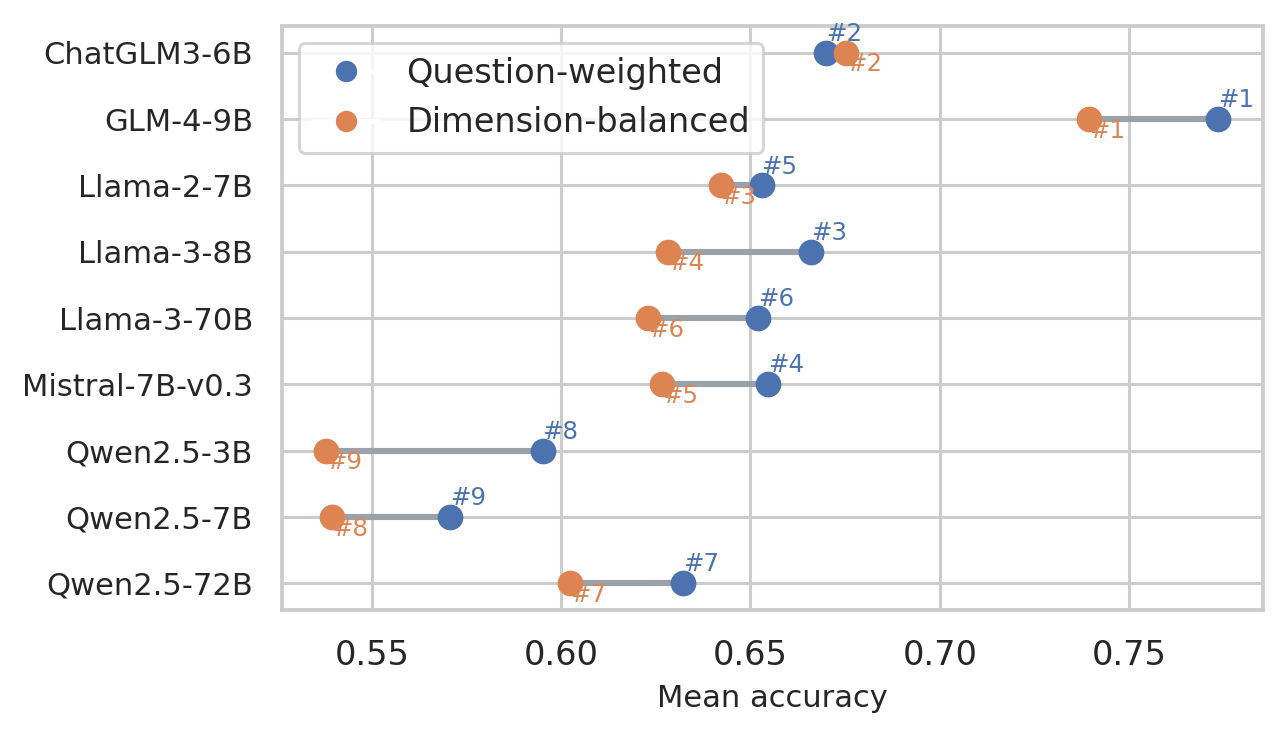}}
    \hspace{2mm}
  \subfigure[LODO stability of the $AE$ composite metric\label{fig:benchmark_lodo_ae}]{
    \includegraphics[width=0.46\linewidth]{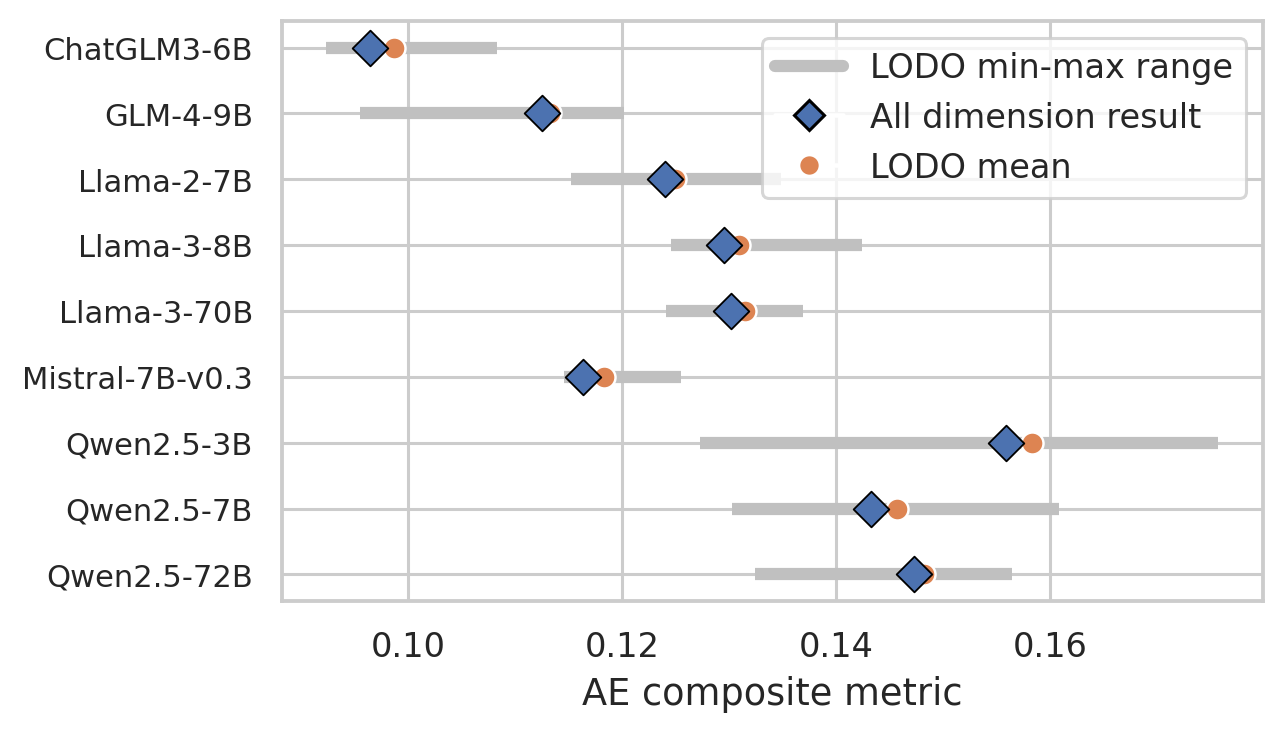}}
  \caption{Sensitivity of the main results to benchmark construction. Panel (a) compares foundational average accuracy under equal dimension and equal question weighting on the cross-country shared panel. Rank annotations indicate the model's ranking under each aggregation scheme. Panel (b) summarizes the stability of the $AE$ composite metric under leave-one-dimension-out (LODO) re-aggregation. The horizontal range shows the minimum and maximum $AE$ values across the 11 LODO runs, the circular marker shows the mean LODO value, and the diamond marker shows the full-panel baseline. Lower $AE$ values indicate a better joint accuracy-equality profile.}
  \label{fig:benchmark_sensitivity}
\end{figure*}

The LODO analysis removes one of the 11 retained value dimensions at a time and re-aggregates the remaining 10 dimensions using the same equal dimension weighting as in the main analysis. For each omitted dimension, we recompute average accuracy, representational equality, the $AE$ composite metric, country-level accuracy rankings, and correlations between country-level accuracy and macro-level factors. The results indicate that the main patterns are not driven by a single value dimension. Across models, mean country-rank Spearman values range from 0.939 to 0.983, showing that country-level rankings remain close to the full-panel baseline after removing any one dimension. The sign stability of macro-factor correlations ranges from 0.903 to 1.000, indicating that the positive or negative direction of most structural associations is preserved under dimension removal. The $Eq_{CV}$ span under LODO remains limited for all models, ranging from 0.0053 to 0.0298. The $AE$ composite metric also remains stable, with LODO spans ranging from 0.0109 to 0.0484 across models. As shown in Figure~\ref{fig:benchmark_lodo_ae}, the full-panel $AE$ values generally fall close to the LODO ranges, suggesting that the joint accuracy-equality performance is not determined by a single dominant topic block.

The aggregation comparison examines whether changing the weighting rule affects the main findings. In the main analysis, each of the 11 retained value dimensions carries the same weight regardless of how many questions it contains. The alternative weights each question equally, averaging accuracy directly across all 160 retained questions, so that dimensions with more questions contribute proportionally more to the final score. We compare the two schemes at three levels: average accuracy, representational equality, and the $AE$ composite metric. We also compare country-level mean-accuracy rankings and macro-factor correlations. The results show that the main conclusions are stable under the two weighting schemes. In foundational analysis, the model-rank Spearman correlations between equal question weighting and equal dimension weighting are 0.933 for average accuracy, 0.567 for $Eq_{CV}$, and 0.750 for the $AE$ composite metric. At the country level, the mean-accuracy ranking correlation is 0.935, indicating that the overall country pattern is highly similar across the two schemes. The signs of the macro-factor correlations are also fully preserved, suggesting that the direction of the economic, technological, political, and cultural associations is not an artifact of the weighting rule. Figure~\ref{fig:benchmark_accuracy_dumbbell} illustrates this comparison for foundational average accuracy, showing that the model rankings remain largely consistent after switching from equal dimension weighting to equal question weighting.

\section{Supplementary Validation Analyses}

\subsection{First-token probability versus repeated generation}
\label{appendix:token_consistency}

There are two common ways to estimate model response distributions in multiple-choice survey simulation. The first is repeated generation, where the model is sampled multiple times and the generated answer choices are aggregated into an empirical distribution. The second is probability-based estimation, where the model's token probabilities over the fixed answer options are directly used to construct the response distribution. This study adopts the first-token probability method because it provides a direct and efficient estimate under a fixed option space. However, prior work has shown that first-token probabilities may not always match the final text answers generated by instruction-tuned models in multiple-choice settings~\citep{wang2024my}. We therefore conduct a validation analysis to examine whether, in our specific survey simulation setting, the response distributions estimated from first-token probabilities are consistent with those estimated from repeated generation.

For this validation analysis, we select representative models from several model families evaluated in this study, including GLM-4-9B, Llama-3-8B, Mistral-7B-v0.3, and Qwen2.5-7B. We use the same setup as the foundational analysis. Specifically, each model is prompted with the second-person profile format and asked to simulate responses to the WVS questions included in our dataset. The repeated-generation estimator draws 100 completions per prompt at temperature 1.0. We compare this empirical distribution with the distribution induced by first-token probabilities over the answer-option letters. For each model and intervention setting, we simulate value distributions for 2,420 distinct demographic subpopulations across 59 countries. Applying repeated generation throughout the full evaluation would substantially increase computational cost. The two estimators are highly consistent in this constrained survey setting. Across all question-level comparisons, the mean JSD-based similarity is 0.980 and the median is 0.996. At the model level, mean JSD-based similarity ranges from 0.960 to 0.992, indicating strong agreement even for the weakest model.

These results support the use of first-token probabilities as a reliable and efficient operational approximation for fixed-option survey questions in our evaluation setting. This conclusion is limited to the option-constrained questionnaire setting studied here. It does not imply that first-token probabilities are interchangeable with open-ended generation for broader forms of value expression, complex reasoning, or real-world behavior. Rather, the comparison shows that, for large-scale survey-based population simulation, first-token probabilities closely track repeated-generation estimates while greatly reducing computational cost.
\begin{figure*}
  \centering
  \includegraphics[width=0.62\linewidth]{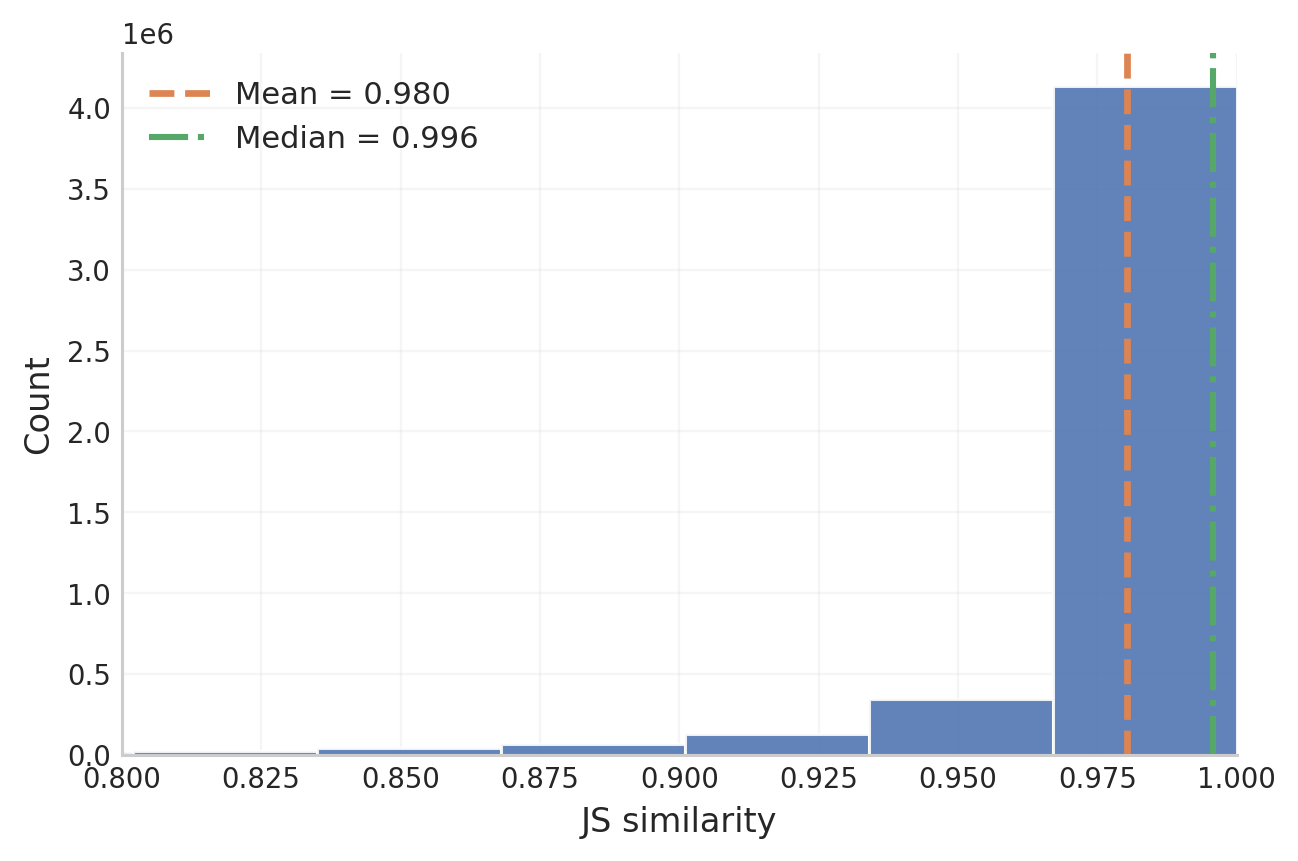}
  \caption{Distribution of question-level JSD-based similarity between the response distribution estimated from first-token probabilities and the distribution estimated from repeated generation on the same WVS multiple-choice questions.}
  \label{fig:token_repeat_js_similarity}
\end{figure*}

\subsection{Analysis on demographic variation in human values}
\label{appendix:country_variation}

To further justify the use of country as a primary aggregation dimension in our subgroup construction, we conduct an additional analysis on real human value distributions derived from the WVS data. The goal of this analysis is to compare how much the empirical human response distributions change when varying one demographic variable at a time, including country, gender, age, education, and income. This empirically supports country as the most informative dimension for cross-group value variation.

\paragraph{Analytical design}

For each demographic variable, we construct matched subgroup pairs that differ in exactly one variable while holding the remaining demographic attributes fixed. For each matched pair, we compute the Jensen--Shannon divergence (JSD) between their empirical human response distributions for each question, and then average the divergence across questions. This yields a pair-level measure of how much human value distributions change when one demographic variable is varied.

Using this procedure, we obtain five groups of divergence scores, corresponding to changes in country, gender, age, education, and income. Larger JSD values indicate larger shifts in real human value distributions and therefore greater differentiation associated with the corresponding demographic variable.

\paragraph{Results}

The results show that varying \textbf{country} produces the largest divergence in real human value distributions by a substantial margin (mean JSD $= 0.478$), clearly exceeding age ($0.272$), education ($0.288$), income ($0.281$), and gender ($0.214$). This pattern is also reflected in the median values.

\begin{figure*}
  \centering
    {\includegraphics[width=0.55\linewidth]{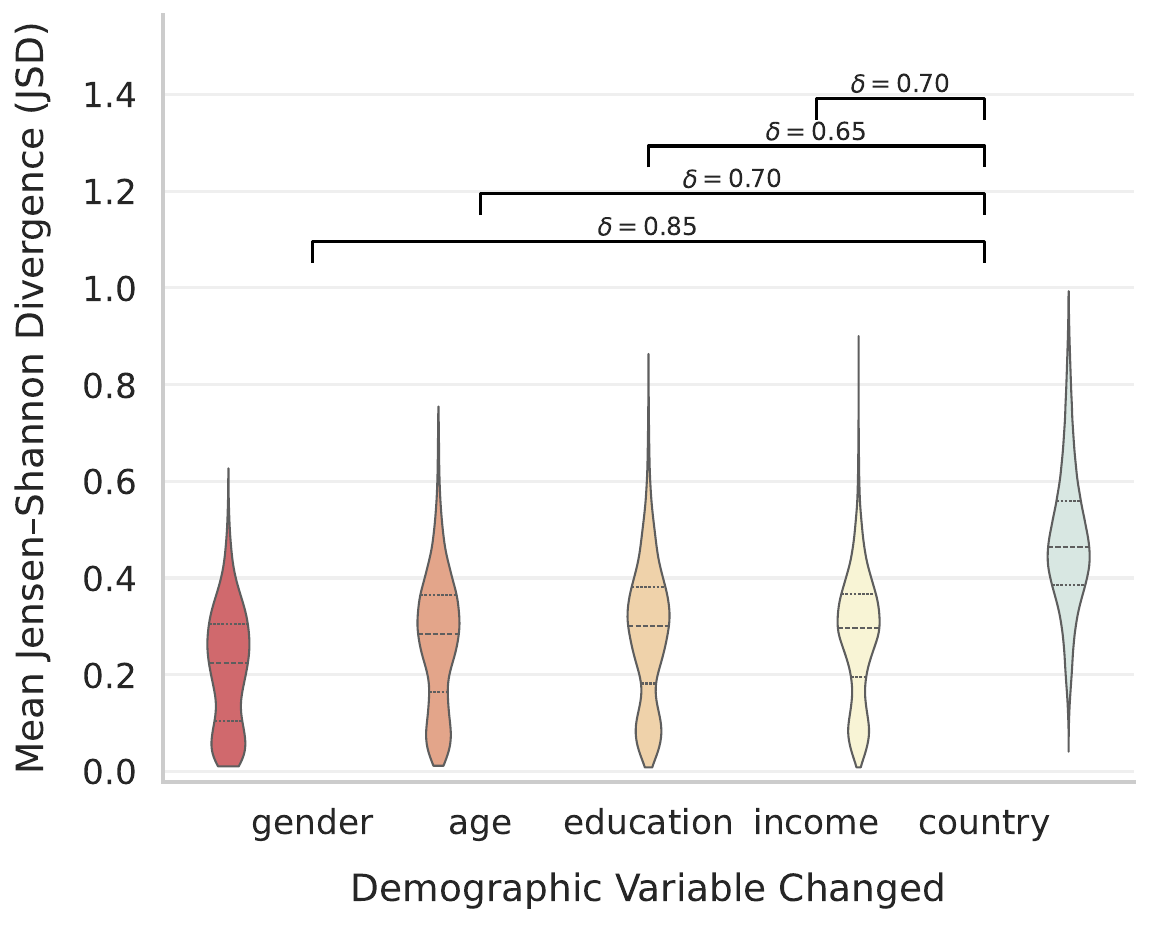}}
  \caption{Violin plot of mean JSD in real human value distributions when varying one demographic variable at a time. Annotations report descriptive Cliff's $\delta$ effect sizes comparing \textit{country} with each other variable. Pair-level observations overlap, so no significance test is applied.}
  \label{fig:demo_jsd_boxplot}
\end{figure*}

Across the observed pair distributions, Cliff's $\delta$~\citep{cliff1993dominance} for country versus gender, age, education, and income is $0.846$, $0.703$, $0.649$, and $0.700$, respectively. These positive contrasts indicate that country-pair divergences are more often larger than divergences associated with each other variable. Figure~\ref{fig:demo_jsd_boxplot} presents the distributions and the corresponding descriptive effect sizes.

Taken together, these results show that, in the real human WVS data, changing country leads to substantially larger shifts in value distributions than changing the other demographic variables considered here. This supports the use of country as a primary aggregation dimension in our simulation framework. At the same time, this result does not imply that other demographic variables are unimportant; rather, country yields the strongest cross-group differentiation among the demographic variables examined.

\subsection{Robustness: within-country response heterogeneity and simulation accuracy}
\label{appendix:heterogeneity}

A potential concern when comparing simulation accuracy across countries is whether countries with more internally diverse human response distributions are inherently harder to simulate, which would confound the cross-country comparison. To test this directly, we measure within-country response heterogeneity from real human WVS data and correlate it with country-level simulation accuracy.

Each subgroup is a fully specified combination of country and two demographic attributes (for example, women aged 18--24 in a given country). We retain the subgroups that pass the common-subgroup filter used in the main analysis. For each survey question, we group subgroups from the same country that share the same pair of demographic attributes. We average their empirical answer distributions to form a reference distribution and compute the Jensen--Shannon divergence between each subgroup and this reference. Each comparison therefore uses directly comparable subgroups answering the same question.

We aggregate these divergences into a country-level heterogeneity score by averaging across questions and comparable subgroup sets, weighted by the number of contributing subgroups. Higher scores indicate greater divergence among comparable subgroups within a country.

We estimate this association separately for each model within an experimental block. We first average each model's country accuracies across the settings retained in the block, such as the four retrieval or alignment settings. This yields one accuracy value per country for each model. We then compute its Spearman correlation with the country heterogeneity score over the available countries. Table~\ref{tab:heterogeneity_correlation} reports the median model-specific correlation and the min--max range for each block. The sample size $N$ denotes the number of countries in each model-specific correlation.

\begin{table}[htbp]
\centering
\begin{threeparttable}
\caption{Spearman association between country-level human response heterogeneity and simulation accuracy. Entries are medians of per-model country-level Spearman correlations, with the min--max range across models in brackets. $N$ is the number of countries in each per-model correlation.}
\label{tab:heterogeneity_correlation}
\begin{tabular}{lccc}
\toprule
\textbf{Evaluation setting} & \textbf{Spearman $r$} & \textbf{Models} & \textbf{$N$ (countries)} \\
\midrule
\makecell[l]{Foundational profile} & 0.132 [$-$0.402, 0.231] & 9 & 59 \\
\makecell[l]{Native-language} & 0.183 [$-$0.140, 0.311] & 8 & 27 \\
\makecell[l]{Additional-information} & $-$0.042 [$-$0.386, 0.132] & 9 & 59 \\
\makecell[l]{CPT} & $-$0.476 & 6 & 12 \\
Preference alignment & 0.075 [$-$0.327, 0.174] & 5 & 59 \\
\midrule
\makecell[l]{Pooled broad-coverage\\settings} & 0.074 [$-$0.437, 0.188] & 13 & 59 \\
\bottomrule
\end{tabular}
\end{threeparttable}
\end{table}

Across the broad-coverage blocks, model-specific correlations are centered near zero and vary in direction. The median correlation is $0.132$ for foundational profile prompting and $0.074$ for the pooled broad-coverage settings. Seven of the nine foundational models yield non-negative associations. The additional-information and preference-alignment blocks also have median correlations near zero, with ranges spanning both directions. The CPT condition shows a larger correlation, but its narrower country coverage warrants interpreting it as a small-coverage observation rather than a general pattern. Overall, within-country response heterogeneity shows no consistent association with country-level simulation accuracy across the broad-coverage analyses and does not explain the cross-country inequality pattern in the main analysis.

\section{Profile Generation Strategies}

\subsection{Full Prompt Structure and Output Format}
\label{appendix:prompt_structure}

Our evaluation uses a fixed prompt template to standardize outputs across model queries. When a model-specific chat template is available, the components are serialized as a multi-turn conversation; otherwise, the same components are concatenated in the order described below. The template contains a system profile, three few-shot format demonstrations, an optional historical memory module, and a target survey question. In each native-language condition, the question and answer options follow the corresponding WVS questionnaire. The profile, response-format instruction, and few-shot demonstrations use fixed templates in the same target language.

\textbf{System Instruction and Target Profile}. We define the target subpopulation in the system prompt using different profile formats. This anchors the model's simulated persona before the target question is presented. The main experiments use the second-person profile format; Appendix~\ref{appendix:profile_format} reports robustness checks for alternative formats.

\paragraph{Few-shot format demonstrations}
We explicitly instruct the model to respond with a single option letter (e.g., A, B, C...). This strict constraint is crucial, as it allows our pipeline to read the first generated token as the answer and extract its probability distribution. To enforce this format, we use simple, non-survey few-shot examples (e.g., ``What is 1+1?'') to avoid leaking substantive value information into the target question. Furthermore, the correct answers in these examples are randomized across different option positions to prevent the model from learning any positional bias. 
\paragraph{Historical memory}
In the additional-information settings, the final user message includes exactly three retrieved non-target WVS items for the same target subgroup, providing supplementary context about the target persona. This component is omitted in the base setting and in other settings where no additional information is provided. Each item reports the subgroup's empirical response distribution in the form ``When asked ..., I am ... likely to answer ....'' The retrieval strategy selects these three items; no memory is included in the base setting or in settings without additional information. For a target question, the module contains exactly three eligible non-target items selected by the corresponding retrieval strategy.

\paragraph{Target question and generation}
The final user message contains the historical memory when applicable, followed by the target WVS question and its answer options. The model is instructed to answer from the perspective of the specified persona using only one option letter. We use the first generated token to construct the simulated response distribution over the available options.

The complete multi-turn structure is illustrated in the example below. Square-bracketed role headers are reader-facing annotations, not literal prompt strings. They identify the system, user, and assistant turns, which are serialized through model-specific chat-template control tokens when available. The parenthetical labels ``F1'', ``F2'', and ``F3'' identify the three few-shot demonstrations; ``target turn'' identifies the final user message; and ``generated at inference'' marks the assistant position where the model produces one option letter. These labels are not sent as ordinary prompt text.

\begin{tcolorbox}[colback=black!5!white,colframe=black!55!black,width=\columnwidth,title={Illustrative Multi-turn Prompt in the Additional-information Condition},breakable]
\textbf{[System message]} \\
Your age falls into the 18--24 range. Your income level is Eight step. You are from Ethiopia.

\vspace{2mm}
\hrule
\vspace{2mm}

\textbf{[User message (F1)]} \\
You need to read the multiple-choice question carefully and only answer with ONE of the listed options.\\
Question: What is 1+1?\\
A. 3\\
B. 2\\
C. 1\\
Answer:\\
\textbf{[Assistant message (F1)]} \\
B

\vspace{1mm}
\textbf{[User message (F2)]} \\
You need to read the multiple-choice question carefully and only answer with ONE of the listed options.\\
Question: What is 1+1?\\
A. 3\\
B. 1\\
C. 2\\
Answer:\\
\textbf{[Assistant message (F2)]} \\
C

\vspace{1mm}
\textbf{[User message (F3)]} \\
You need to read the multiple-choice question carefully and only answer with ONE of the listed options.\\
Question: What is 1+1?\\
A. 2\\
B. 3\\
C. 1\\
Answer:\\
\textbf{[Assistant message (F3)]} \\
A

\vspace{2mm}
\hrule
\vspace{2mm}

\textbf{[User message (target turn)]} \\
\textit{Historical memory:}\\
\textit{When asked ``How important are friends in your life?'', I am 62.50\% likely to answer ``Very important'' and 37.50\% likely to answer ``Rather important''.}\\
\textit{When asked ``How important is religion in your life?'', I am 100.00\% likely to answer ``Very important''.}\\
\textit{When asked ``How important is work in your life?'', I am 100.00\% likely to answer ``Very important''.}\\
\vspace{1mm}
You need to read the multiple-choice question carefully and only answer with ONE of the listed options.\\
Question: How important is family in your life?\\
A. Very important\\
B. Rather important\\
C. Not very important\\
D. Not at all important\\
Answer:

\vspace{1mm}
\textbf{[Assistant message (generated at inference)]} \\
\textit{One option letter is generated here.}
\end{tcolorbox}
\label{tab:full_prompt_structure}

\subsection{Profile-format robustness check}
\label{appendix:profile_format}

To validate the robustness across different profile settings, we conducted experiments with various profile formats. In previous LLM simulation studies, researchers have employed diverse approaches to profile configuration, yet no standardized methodology has emerged. Moreover, there has been limited comparative analysis of different profile settings' effectiveness. Therefore, we synthesized existing approaches from previous studies into two representative generation strategies: template-based and LLM-generated. In the template-based examples below, double-braced fields such as \texttt{\{\{gender\}\}} denote values substituted before inference and are not literal prompt text. In the template-based strategy, a distinction is made between natural language expressions and attribute lists. We evaluate five profile formats:

\begin{center}
\begin{tcolorbox}[colback=black!5!white,colframe=black!55!black,width=0.95\columnwidth,title={Template-based profile: first-person format}]
I am a \{\{gender\}\} from \{\{country\}\}. Age-wise, I fall into the \{\{age\}\} range. My highest educational attainment is \{\{education\}\}. In terms of income, I am in the \{\{income\}\} bracket.
\end{tcolorbox}
\label{tab:prompt_profile1}
\end{center}

\begin{center}
\begin{tcolorbox}[colback=black!5!white,colframe=black!55!black,width=0.95\columnwidth,title={Template-based profile: second-person format}]
You are a \{\{gender\}\} from \{\{country\}\}. Your age falls into the \{\{age\}\} range. Your highest level of education is \{\{education\}\}. Your income level is \{\{income\}\}.
\end{tcolorbox}
\label{tab:prompt_profile2}
\end{center}

\begin{center}
\begin{tcolorbox}[colback=black!5!white,colframe=black!55!black,width=0.95\columnwidth,title={Template-based profile: third-person format}]
This person is a \{\{gender\}\} from \{\{country\}\}. This person's age falls into the \{\{age\}\} range. This person's highest level of education is \{\{education\}\}. This person's income level is \{\{income\}\}.
\end{tcolorbox}
\label{tab:prompt_profile3}
\end{center}

\begin{center}
\begin{tcolorbox}[colback=black!5!white,colframe=black!55!black,width=0.95\columnwidth,title={Template-based profile: attributes list}]
Role play with this person.\\
age: \{\{age\}\}\\
gender: \{\{gender\}\}\\
education: \{\{education\}\}\\
income: "\{\{income\}\}"\\
country: \{\{country\}\}\\
\end{tcolorbox}
\label{tab:prompt_profile4}
\end{center}

\begin{center}
\begin{tcolorbox}[colback=black!5!white,colframe=black!55!black,width=0.95\columnwidth,title={LLM-generated profile (an example for young Chinese college graduates generated by Llama-3-8B)}]
As you navigate your early twenties in China, you're likely shaped by a unique blend of traditional and modern influences. Growing up in a rapidly changing society, you've been exposed to a mix of Confucian values emphasizing social hierarchy, respect for authority, and collectivism, alongside the individualistic and meritocratic ideals of modernization. Your education, likely a Bachelor's degree or equivalent, has instilled in you a strong work ethic and a desire for personal achievement. You're likely driven to succeed, not just for personal fulfillment, but also to bring pride to your family and community. Your age and demographic background have also influenced your outlook on life, with a strong emphasis on social media and technology, and a desire for social status and recognition. You're likely drawn to careers that offer stability, security, and prestige, and you're willing to put in the effort required to achieve your goals. Despite the challenges and uncertainties of your generation, you're optimistic about the future, and you're eager to make your mark on the world.
\end{tcolorbox}
\label{tab:prompt_profile5}
\end{center}

The robustness results show that profile format has only a minor effect on overall simulation performance. Across the five formats, average simulation accuracy varies by less than 1.3\%. Moreover, the relative performance patterns across models remain stable, indicating that our main findings are not driven by a specific wording style or profile-construction strategy. Based on this robustness check, we adopt the second-person format in all main experiments for consistency and clarity.

\newpage
\raggedbottom
\section{Full Results of Representational Equality Indices}
\label{appendix:all_equality_scores}

This appendix provides the detailed results for simulation accuracy and all four indices used to measure Representational Equality. These indices offer complementary perspectives on the distribution of simulation performance across the target countries.
\begin{table}[H]
\centering
\begingroup
\renewcommand{\arraystretch}{1.60}
\begin{threeparttable}
\caption{Complete results for accuracy, equality indices, and the AE composite metric in the foundational analysis}
\label{tab:equality_summary_exp1}
{\small
\setlength{\tabcolsep}{1pt}
\begin{tabular}{
    >{\raggedright\arraybackslash}p{2.5cm}
    c
    c
    c
    c
    c
    c
}
\toprule
\textbf{Model} & \makecell{\textbf{$A_{c}$(Avg.) [95\% CI]}} & \makecell{\textbf{AE composite}\\\textbf{metric (\(\downarrow\))}} & \makecell{\textbf{Max-Min}\\\textbf{Diff (\(\downarrow\))}} & \makecell{\textbf{Min-Max}\\\textbf{Ratio (\(\uparrow\))}} & \makecell{\textbf{CV (\(\times 10^{2}\))}\\\textbf{(\(\downarrow\))}} & \makecell{\textbf{Gini (\(\times 10^{2}\))}\\\textbf{(\(\downarrow\))}} \\
\midrule
ChatGLM3-6B & 0.675 [0.670, 0.680] & 0.096 & 0.096 & 0.865 & 2.86 & 1.59 \\
\cmidrule(lr){2-7}
GLM-4-9B & 0.739 [0.730, 0.749] & 0.113 & 0.140 & 0.826 & 4.86 & 2.77 \\
\cmidrule(lr){2-7}
Llama-2-7B & 0.642 [0.635, 0.649] & 0.124 & 0.108 & 0.843 & 4.30 & 2.44 \\
\cmidrule(lr){2-7}
Llama-3-8B & 0.628 [0.621, 0.636] & 0.130 & 0.135 & 0.804 & 4.51 & 2.52 \\
\cmidrule(lr){2-7}
Llama-3-70B & 0.623 [0.616, 0.630] & 0.130 & 0.121 & 0.826 & 4.50 & 2.49 \\
\cmidrule(lr){2-7}
Mistral-7B-v0.3 & 0.627 [0.621, 0.633] & 0.116 & 0.101 & 0.848 & 3.63 & 2.05 \\
\cmidrule(lr){2-7}
Qwen2.5-3B & 0.538 [0.531, 0.545] & 0.156 & 0.111 & 0.812 & 5.26 & 3.01 \\
\cmidrule(lr){2-7}
Qwen2.5-7B & 0.539 [0.533, 0.546] & 0.143 & 0.102 & 0.830 & 4.46 & 2.50 \\
\cmidrule(lr){2-7}
Qwen2.5-72B & 0.602 [0.594, 0.611] & 0.147 & 0.122 & 0.819 & 5.46 & 3.05 \\
\bottomrule
\end{tabular}}
\begin{tablenotes}\footnotesize
    \item[*] Accuracy is reported together with a 95\% confidence interval computed over country-level accuracy scores using a t interval.
    \item[*] AE composite metric is computed as $\sqrt{(1-A_c)\times CV}$; lower values indicate better joint accuracy-equality performance.
    \item[*] The arrows (\(\uparrow\)/\(\downarrow\)) indicate the direction of improvement in equality. For indices marked with (\(\downarrow\)), a lower value represents higher equality. For indices marked with (\(\uparrow\)), a higher value represents higher equality.
    \item[*] CV and Gini coefficient values are multiplied by $10^{2}$ for display readability.
\end{tablenotes}
\end{threeparttable}
\endgroup
\end{table}

\newpage
\flushbottom

\begin{table}[H]
\centering
\resizebox{0.95\linewidth}{!}{%
\begin{threeparttable}
\caption{Complete results for accuracy, equality indices, and the AE composite metric in the contextual adaptation analysis}
\label{tab:equality_summary_exp2}
\begin{tabular}{
    >{\raggedright\arraybackslash}p{2.5cm}
    c
    c
    c
    c
    c
    c
    c
}
\toprule
\textbf{Model} & \textbf{Settings} & \makecell{\textbf{$A_{c}$(Avg.) [95\% CI]}} & \makecell{\textbf{AE composite}\\\textbf{metric (\(\downarrow\))}} & \makecell{\textbf{Max-Min}\\\textbf{Diff (\(\downarrow\))}} & \makecell{\textbf{Min-Max}\\\textbf{Ratio (\(\uparrow\))}} & \makecell{\textbf{CV (\(\times 10^{2}\))}\\\textbf{(\(\downarrow\))}} & \makecell{\textbf{Gini (\(\times 10^{2}\))}\\\textbf{(\(\downarrow\))}} \\
\midrule
\multicolumn{8}{l}{\textit{Contextual adaptation: native language}} \\
\midrule
ChatGLM3-6B & - & \textbf{0.715 [0.699, 0.730]} & 0.124 & 0.143 & 0.822 & 5.41 & 2.99 \\
\cmidrule(lr){2-8}
GLM-4-9B & - & \textbf{0.766 [0.753, 0.779]} & \textbf{0.099} & \textbf{0.112} & \textbf{0.863} & \textbf{4.21} & \textbf{2.40} \\
\cmidrule(lr){2-8}
Llama-3-8B & - & \textbf{0.640 [0.624, 0.657]} & 0.153 & 0.138 & 0.801 & 6.47 & 3.55 \\
\cmidrule(lr){2-8}
Llama-3-70B & - & \textbf{0.641 [0.632, 0.650]} & \textbf{0.112} & \textbf{0.094} & \textbf{0.862} & \textbf{3.50} & \textbf{1.93} \\
\cmidrule(lr){2-8}
Mistral-7B-v0.3 & - & \textbf{0.666 [0.660, 0.672]} & \textbf{0.088} & \textbf{0.072} & \textbf{0.899} & \textbf{2.32} & \textbf{1.24} \\
\cmidrule(lr){2-8}
Qwen2.5-3B & - & \textbf{0.576 [0.552, 0.600]} & 0.209 & 0.198 & 0.707 & 10.35 & 5.89 \\
\cmidrule(lr){2-8}
Qwen2.5-7B & - & \textbf{0.586 [0.568, 0.604]} & 0.177 & 0.153 & 0.773 & 7.58 & 4.23 \\
\cmidrule(lr){2-8}
Qwen2.5-72B & - & \textbf{0.638 [0.627, 0.650]} & \textbf{0.127} & \textbf{0.106} & \textbf{0.845} & \textbf{4.47} & \textbf{2.52} \\

\midrule

\multicolumn{8}{l}{\textit{Contextual adaptation: additional information}} \\
\midrule
\multirow{4}{*}{ChatGLM3-6B} & BM25 & \textbf{0.746 [0.742, 0.750]} & \textbf{0.075} & \textbf{0.067} & \textbf{0.914} & \textbf{2.23} & \textbf{1.27} \\
 & Embedding & \textbf{0.746 [0.742, 0.750]} & \textbf{0.074} & \textbf{0.057} & \textbf{0.926} & \textbf{2.13} & \textbf{1.22} \\
 & Expansion & \textbf{0.737 [0.733, 0.741]} & \textbf{0.074} & \textbf{0.068} & \textbf{0.912} & \textbf{2.08} & \textbf{1.17} \\
 & Theory & \textbf{0.740 [0.736, 0.744]} & \textbf{0.075} & \textbf{0.088} & \textbf{0.887} & \textbf{2.18} & \textbf{1.19} \\
\cmidrule(lr){2-8}
\multirow{4}{*}{GLM-4-9B} & BM25 & \textbf{0.768 [0.763, 0.773]} & \textbf{0.077} & \textbf{0.082} & \textbf{0.897} & \textbf{2.55} & \textbf{1.45} \\
 & Embedding & \textbf{0.754 [0.749, 0.760]} & \textbf{0.083} & \textbf{0.084} & \textbf{0.894} & \textbf{2.78} & \textbf{1.58} \\
 & Expansion & \textbf{0.774 [0.769, 0.779]} & \textbf{0.077} & \textbf{0.079} & \textbf{0.902} & \textbf{2.59} & \textbf{1.47} \\
 & Theory & \textbf{0.753 [0.746, 0.760]} & \textbf{0.092} & \textbf{0.113} & \textbf{0.860} & \textbf{3.44} & \textbf{1.96} \\
\cmidrule(lr){2-8}
\multirow{4}{*}{Llama-2-7B} & BM25 & 0.607 [0.603, 0.611] & \textbf{0.099} & \textbf{0.070} & \textbf{0.892} & \textbf{2.52} & \textbf{1.42} \\
 & Embedding & 0.614 [0.610, 0.618] & \textbf{0.098} & \textbf{0.065} & \textbf{0.901} & \textbf{2.49} & \textbf{1.42} \\
 & Expansion & 0.629 [0.625, 0.634] & \textbf{0.101} & \textbf{0.073} & \textbf{0.890} & \textbf{2.75} & \textbf{1.55} \\
 & Theory & 0.599 [0.594, 0.603] & \textbf{0.105} & \textbf{0.070} & \textbf{0.889} & \textbf{2.76} & \textbf{1.56} \\
\cmidrule(lr){2-8}
\multirow{4}{*}{Llama-3-8B} & BM25 & \textbf{0.711 [0.706, 0.715]} & \textbf{0.088} & \textbf{0.099} & \textbf{0.870} & \textbf{2.65} & \textbf{1.48} \\
 & Embedding & \textbf{0.721 [0.717, 0.726]} & \textbf{0.078} & \textbf{0.083} & \textbf{0.890} & \textbf{2.19} & \textbf{1.21} \\
 & Expansion & \textbf{0.724 [0.717, 0.730]} & \textbf{0.097} & \textbf{0.116} & \textbf{0.849} & \textbf{3.40} & \textbf{1.91} \\
 & Theory & \textbf{0.702 [0.698, 0.705]} & \textbf{0.077} & \textbf{0.075} & \textbf{0.898} & \textbf{2.01} & \textbf{1.09} \\
\cmidrule(lr){2-8}
\multirow{4}{*}{Llama-3-70B} & BM25 & \textbf{0.642 [0.636, 0.649]} & \textbf{0.117} & \textbf{0.120} & \textbf{0.831} & \textbf{3.85} & \textbf{2.14} \\
 & Embedding & \textbf{0.650 [0.644, 0.656]} & \textbf{0.108} & \textbf{0.108} & \textbf{0.848} & \textbf{3.36} & \textbf{1.90} \\
 & Expansion & \textbf{0.653 [0.647, 0.660]} & \textbf{0.112} & \textbf{0.110} & \textbf{0.845} & \textbf{3.64} & \textbf{2.07} \\
 & Theory & \textbf{0.645 [0.639, 0.650]} & \textbf{0.109} & \textbf{0.115} & \textbf{0.835} & \textbf{3.34} & \textbf{1.86} \\
\cmidrule(lr){2-8}
\multirow{4}{*}{Mistral-7B-v0.3} & BM25 & \textbf{0.710 [0.706, 0.714]} & \textbf{0.080} & \textbf{0.078} & \textbf{0.896} & \textbf{2.23} & \textbf{1.22} \\
 & Embedding & \textbf{0.703 [0.699, 0.708]} & \textbf{0.082} & \textbf{0.072} & \textbf{0.903} & \textbf{2.26} & \textbf{1.28} \\
 & Expansion & \textbf{0.703 [0.699, 0.706]} & \textbf{0.079} & \textbf{0.083} & \textbf{0.887} & \textbf{2.12} & \textbf{1.15} \\
 & Theory & \textbf{0.676 [0.671, 0.680]} & \textbf{0.088} & \textbf{0.068} & \textbf{0.904} & \textbf{2.41} & \textbf{1.36} \\
\cmidrule(lr){2-8}
\multirow{4}{*}{Qwen2.5-3B} & BM25 & \textbf{0.594 [0.588, 0.599]} & \textbf{0.118} & \textbf{0.104} & \textbf{0.837} & \textbf{3.43} & \textbf{1.89} \\
 & Embedding & \textbf{0.600 [0.595, 0.606]} & \textbf{0.117} & \textbf{0.093} & \textbf{0.853} & \textbf{3.44} & \textbf{1.92} \\
 & Expansion & \textbf{0.564 [0.558, 0.570]} & \textbf{0.138} & \textbf{0.104} & \textbf{0.830} & \textbf{4.34} & \textbf{2.46} \\
 & Theory & \textbf{0.553 [0.547, 0.559]} & \textbf{0.136} & \textbf{0.102} & \textbf{0.831} & \textbf{4.13} & \textbf{2.34} \\
\cmidrule(lr){2-8}
\multirow{4}{*}{Qwen2.5-7B} & BM25 & \textbf{0.608 [0.601, 0.614]} & \textbf{0.125} & 0.118 & 0.826 & \textbf{4.00} & \textbf{2.26} \\
 & Embedding & \textbf{0.620 [0.614, 0.626]} & \textbf{0.119} & \textbf{0.093} & \textbf{0.861} & \textbf{3.73} & \textbf{2.11} \\
 & Expansion & \textbf{0.573 [0.562, 0.583]} & 0.175 & 0.149 & 0.772 & 7.16 & 4.09 \\
 & Theory & \textbf{0.601 [0.595, 0.608]} & \textbf{0.131} & 0.122 & 0.813 & \textbf{4.29} & \textbf{2.44} \\
\cmidrule(lr){2-8}
\multirow{4}{*}{Qwen2.5-72B} & BM25 & \textbf{0.637 [0.631, 0.643]} & \textbf{0.116} & \textbf{0.110} & \textbf{0.842} & \textbf{3.72} & \textbf{2.08} \\
 & Embedding & \textbf{0.644 [0.637, 0.650]} & \textbf{0.114} & \textbf{0.109} & \textbf{0.846} & \textbf{3.64} & \textbf{2.03} \\
 & Expansion & \textbf{0.612 [0.605, 0.618]} & \textbf{0.130} & \textbf{0.121} & \textbf{0.820} & \textbf{4.33} & \textbf{2.38} \\
 & Theory & \textbf{0.612 [0.606, 0.619]} & \textbf{0.125} & \textbf{0.108} & \textbf{0.839} & \textbf{4.05} & \textbf{2.26} \\
\bottomrule
\end{tabular}
\begin{tablenotes}\footnotesize
    \item[*] Accuracy is reported together with a 95\% confidence interval computed over country-level accuracy scores using a t interval.
    \item[*] AE composite metric is computed as $\sqrt{(1-A_c)\times CV}$; lower values indicate better joint accuracy-equality performance.
    \item[*] The arrows (\(\uparrow\)/\(\downarrow\)) indicate the direction of improvement in equality. For indices marked with (\(\downarrow\)), a lower value represents higher equality. For indices marked with (\(\uparrow\)), a higher value represents higher equality.
    \item[*] If improvements in accuracy or equality are observed relative to the reference base settings, they are displayed in \textbf{bold}. For language and CPT rows, one line aggregates the union of countries covered by all sub-settings, and the reference base is recomputed on that same country union before comparison. For AE composite metric, \textbf{bold} likewise denotes a lower (better) cost than the matched base. Language rows aggregate Mandarin-, Arabic-, Spanish-, and English-native countries together.
    \item[*] CV and Gini coefficient values are multiplied by $10^{2}$ for display readability.
\end{tablenotes}
\end{threeparttable}}
\end{table}

\begin{table}[H]
\centering
\resizebox{0.94\linewidth}{!}{%
\begin{threeparttable}
\caption{Complete results for accuracy, equality indices, and the AE composite metric in the parametric modification analysis}
\label{tab:equality_summary_exp3}
\begin{tabular}{
    >{\raggedright\arraybackslash}p{1.8cm}
    c
    c
    c
    c
    c
    c
    c
}
\toprule
\textbf{Model} & \textbf{Settings} & \makecell{\textbf{$A_{c}$(Avg.) [95\% CI]}} & \makecell{\textbf{AE composite}\\\textbf{metric (\(\downarrow\))}} & \makecell{\textbf{Max-Min}\\\textbf{Diff (\(\downarrow\))}} & \makecell{\textbf{Min-Max}\\\textbf{Ratio (\(\uparrow\))}} & \makecell{\textbf{CV (\(\times 10^{2}\))}\\\textbf{(\(\downarrow\))}} & \makecell{\textbf{Gini (\(\times 10^{2}\))}\\\textbf{(\(\downarrow\))}} \\
\midrule
\multicolumn{8}{l}{\textit{Parametric modification: CPT languages}} \\
\midrule
Llama-2 & - & \textbf{0.826 [0.803, 0.849]} & 0.086 & 0.105 & 0.880 & 4.26 & 2.39 \\

\midrule

\multicolumn{8}{l}{\textit{Parametric modification: alignment sources}} \\
\midrule
\multirow{4}{*}{Llama-2-7B} & DPO-GPT & \textbf{0.825 [0.817, 0.833]} & 0.080 & 0.138 & 0.845 & 3.66 & 2.04 \\
 & DPO-Human & \textbf{0.828 [0.821, 0.836]} & 0.077 & 0.134 & 0.850 & 3.46 & 1.94 \\
 & GRPO-GPT & 0.628 [0.620, 0.637] & 0.138 & 0.136 & 0.800 & 5.13 & 2.86 \\
 & GRPO-Human & 0.597 [0.589, 0.605] & 0.140 & 0.130 & 0.802 & 4.87 & 2.73 \\
\cmidrule(lr){2-8}
\multirow{4}{*}{Llama-3-8B} & DPO-GPT & 0.717 [0.708, 0.725] & 0.112 & \textbf{0.136} & \textbf{0.823} & \textbf{4.40} & \textbf{2.47} \\
 & DPO-Human & 0.734 [0.726, 0.742] & \textbf{0.105} & \textbf{0.133} & \textbf{0.832} & \textbf{4.13} & \textbf{2.33} \\
 & GRPO-GPT & 0.536 [0.528, 0.544] & 0.164 & \textbf{0.143} & 0.760 & 5.78 & 3.20 \\
 & GRPO-Human & 0.545 [0.537, 0.553] & 0.158 & \textbf{0.134} & 0.780 & 5.51 & 3.07 \\
\cmidrule(lr){2-8}
\multirow{4}{*}{Mistral-7B-v0.3} & DPO-GPT & 0.558 [0.553, 0.564] & 0.128 & \textbf{0.085} & \textbf{0.858} & 3.72 & 2.11 \\
 & DPO-Human & 0.579 [0.573, 0.586] & 0.131 & 0.103 & 0.837 & 4.11 & 2.30 \\
 & GRPO-GPT & 0.533 [0.527, 0.539] & 0.140 & 0.102 & 0.822 & 4.22 & 2.34 \\
 & GRPO-Human & 0.533 [0.526, 0.540] & 0.156 & 0.108 & 0.816 & 5.23 & 2.97 \\
\cmidrule(lr){2-8}
\multirow{4}{*}{Qwen2.5-3B} & DPO-GPT & 0.502 [0.493, 0.511] & 0.185 & 0.135 & 0.765 & 6.86 & 3.89 \\
 & DPO-Human & 0.512 [0.503, 0.521] & 0.182 & 0.136 & 0.767 & 6.82 & 3.89 \\
 & GRPO-GPT & 0.502 [0.495, 0.509] & 0.163 & \textbf{0.109} & 0.801 & 5.33 & 3.04 \\
 & GRPO-Human & 0.511 [0.504, 0.518] & 0.157 & \textbf{0.096} & \textbf{0.826} & \textbf{5.02} & \textbf{2.88} \\
\cmidrule(lr){2-8}
\multirow{4}{*}{Qwen2.5-7B} & DPO-GPT & 0.511 [0.502, 0.520] & 0.178 & 0.158 & 0.740 & 6.46 & 3.47 \\
 & DPO-Human & 0.514 [0.506, 0.523] & 0.176 & 0.155 & 0.746 & 6.36 & 3.44 \\
 & GRPO-GPT & 0.520 [0.514, 0.526] & 0.145 & 0.110 & 0.808 & \textbf{4.37} & \textbf{2.41} \\
 & GRPO-Human & 0.529 [0.523, 0.536] & 0.150 & 0.116 & 0.805 & 4.78 & 2.62 \\
\bottomrule
\end{tabular}
\begin{tablenotes}\footnotesize
    \item[*] Accuracy is reported together with a 95\% confidence interval computed over country-level accuracy scores using a t interval.
    \item[*] AE composite metric is computed as $\sqrt{(1-A_c)\times CV}$; lower values indicate better joint accuracy-equality performance.
    \item[*] The arrows (\(\uparrow\)/\(\downarrow\)) indicate the direction of improvement in equality. For indices marked with (\(\downarrow\)), a lower value represents higher equality. For indices marked with (\(\uparrow\)), a higher value represents higher equality.
    \item[*] If improvements in accuracy or equality are observed relative to the reference base settings, they are displayed in \textbf{bold}. For language and CPT rows, one line aggregates the union of countries covered by all sub-settings, and the reference base is recomputed on that same country union before comparison. For AE composite metric, \textbf{bold} likewise denotes a lower (better) cost than the matched base.
    \item[*] CV and Gini coefficient values are multiplied by $10^{2}$ for display readability.
\end{tablenotes}
\end{threeparttable}}
\end{table}

\section{Supplementary Evaluation on Four ISSP Modules (2020--2023)}
\label{appendix:issp_benchmark}

To examine whether the main findings are specific to the WVS question inventory, we conduct a supplementary evaluation on the International Social Survey Program (ISSP) from 2020 to 2023. ISSP is a long-running cross-country survey program that conducts annual surveys on diverse social science topics. This supplementary benchmark combines four annual core modules, including Environment in 2020, Health in 2021, Gender in 2022, and Identity in 2023, for a total of 270 multiple-choice questions. After applying the same criteria of country comparability and retaining value-related multiple-choice items suitable for distributional evaluation, we restrict the comparison to the 20 sovereign countries shared across these four yearly modules.

This benchmark differs from WVS in two important respects. First, the shared-country set is substantially smaller and more regionally concentrated. More than half of the retained countries are located in Europe, and no Latin American country appears in the retained country set. Second, the four ISSP modules focus on different thematic blocks rather than a single unified questionnaire wave. For these reasons, we use ISSP as a supplementary robustness check rather than a replacement for the main WVS benchmark. The ISSP evaluation uses the same profile design, simulation protocol, and aggregation method as the main WVS analysis. Specifically, question-level accuracy is first averaged within each retained value-related domain and then averaged across domains. This makes the ISSP results comparable with the main findings at the level of evaluation procedure. WVS remains the primary dataset because it provides broader multilingual coverage and richer respondent-level demographic information, which are needed for the native-language prompting analysis and for constructing subpopulations defined by country together with other demographic attributes.

\subsection{Foundational analysis}

Under the foundational profile setting, where the model receives only the demographic profile and the survey question, the ISSP results still show clear differences in simulation accuracy across countries. Table~\ref{tab:issp_equality_summary_exp1} reports the same metrics used in the main WVS tables, including mean accuracy, the AE composite metric, the max-min difference, the min-max ratio, CV, and Gini. Averaging country-level accuracy across the nine evaluated models yields a mean accuracy of 0.598. At the model level, mean accuracy ranges from 0.470 for Qwen2.5-3B to 0.737 for GLM-4-9B. GLM-4-9B also achieves the lowest AE value of 0.105. Mistral-7B-v0.3 yields the lowest country-level accuracy dispersion ($Eq_{CV}=0.0308$, Gini $=0.0171$).

These results are consistent with the main WVS analysis in two respects. First, differences in simulation accuracy across countries remain visible on a second international survey benchmark with a different set of topics. Second, the model with the highest mean accuracy is not the model with the most even performance across countries. This confirms that average accuracy and country-level representational equality should be evaluated jointly, as also summarized by the AE composite metric.

\begin{table}[H]
\centering
\resizebox{\linewidth}{!}{%
\begin{threeparttable}
\caption{Complete results for accuracy, equality indices, and the AE composite metric in the foundational analysis on the ISSP benchmark}
\label{tab:issp_equality_summary_exp1}
\begin{tabular}{
    >{\raggedright\arraybackslash}p{2.5cm}
    c
    c
    c
    c
    c
    c
}
\toprule
\textbf{Model} & \makecell{\textbf{$A_{c}$(Avg.) [95\% CI]}} & \makecell{\textbf{AE composite}\\\textbf{metric (\(\downarrow\))}} & \makecell{\textbf{Max-Min}\\\textbf{Diff (\(\downarrow\))}} & \makecell{\textbf{Min-Max}\\\textbf{Ratio (\(\uparrow\))}} & \makecell{\textbf{CV (\(\times 10^{2}\))}\\\textbf{(\(\downarrow\))}} & \makecell{\textbf{Gini (\(\times 10^{2}\))}\\\textbf{(\(\downarrow\))}} \\
\midrule
GLM-4-9B & 0.737 [0.722, 0.752] & 0.105 & 0.116 & 0.851 & 4.16 & 2.32 \\
\cmidrule(lr){2-7}
ChatGLM3-6B & 0.652 [0.640, 0.664] & 0.116 & 0.094 & 0.863 & 3.89 & 2.14 \\
\cmidrule(lr){2-7}
Llama-2-7B & 0.640 [0.627, 0.653] & 0.124 & 0.119 & 0.828 & 4.25 & 2.34 \\
\cmidrule(lr){2-7}
Llama-3-8B & 0.632 [0.621, 0.643] & 0.115 & 0.089 & 0.866 & 3.61 & 2.01 \\
\cmidrule(lr){2-7}
Qwen2.5-72B & 0.588 [0.573, 0.603] & 0.147 & 0.110 & 0.826 & 5.25 & 2.93 \\
\cmidrule(lr){2-7}
Mistral-7B-v0.3 & 0.578 [0.570, 0.587] & 0.114 & 0.076 & 0.875 & 3.08 & 1.71 \\
\cmidrule(lr){2-7}
Llama-3-70B & 0.565 [0.554, 0.576] & 0.132 & 0.067 & 0.887 & 3.98 & 2.26 \\
\cmidrule(lr){2-7}
Qwen2.5-7B & 0.518 [0.509, 0.526] & 0.128 & 0.069 & 0.873 & 3.42 & 1.89 \\
\cmidrule(lr){2-7}
Qwen2.5-3B & 0.470 [0.461, 0.480] & 0.149 & 0.065 & 0.871 & 4.18 & 2.37 \\
\bottomrule
\end{tabular}
\begin{tablenotes}\footnotesize
    \item[*] Accuracy is reported together with a 95\% confidence interval computed over country-level accuracy scores using a t interval.
    \item[*] AE composite metric is computed as $\sqrt{(1-A_c)\times CV}$; lower values indicate better joint accuracy-equality performance.
    \item[*] The arrows (\(\uparrow\)/\(\downarrow\)) indicate the direction of improvement in equality. For indices marked with (\(\downarrow\)), a lower value represents higher equality. For indices marked with (\(\uparrow\)), a higher value represents higher equality.
    \item[*] CV and Gini coefficient values are multiplied by $10^{2}$ for display readability.
\end{tablenotes}
\end{threeparttable}}
\end{table}

\subsection{Contextual adaptation with additional information}

We replicate the additional-information experiments on the ISSP benchmark using the same four retrieval strategies evaluated under WVS: BM25, Embedding, Expansion, and Theory. Because the ISSP benchmark does not provide the multilingual questionnaire coverage needed for the native-language prompting analysis, the contextual adaptation analysis on ISSP is limited to the additional-information setting.

The ISSP results align closely with the WVS findings. Compared with the profile-only setting, all four retrieval strategies improve mean accuracy and reduce country-level accuracy dispersion. BM25 yields the strongest overall effect. Averaged across models, it raises mean accuracy to 0.648, with mean AE of 0.091 and mean $Eq_{CV}$ of 0.0238. Embedding produces a similar result, with mean accuracy of 0.645, mean AE of 0.091, and mean $Eq_{CV}$ of 0.0236. Expansion and Theory also improve both accuracy and equality, but the gains are smaller. Table~\ref{tab:issp_equality_summary_exp2} reports the corresponding results for each model in the same format as the main contextual table.

\begin{table}[H]
\centering
\resizebox{0.95\linewidth}{!}{%
\begin{threeparttable}
\caption{Complete results for accuracy, equality indices, and the AE composite metric in the contextual adaptation analysis on the ISSP benchmark}
\label{tab:issp_equality_summary_exp2}
\begin{tabular}{
    >{\raggedright\arraybackslash}p{2.5cm}
    c
    c
    c
    c
    c
    c
    c
}
\toprule
\textbf{Model} & \textbf{Settings} & \makecell{\textbf{$A_{c}$(Avg.) [95\% CI]}} & \makecell{\textbf{AE composite}\\\textbf{metric (\(\downarrow\))}} & \makecell{\textbf{Max-Min}\\\textbf{Diff (\(\downarrow\))}} & \makecell{\textbf{Min-Max}\\\textbf{Ratio (\(\uparrow\))}} & \makecell{\textbf{CV (\(\times 10^{2}\))}\\\textbf{(\(\downarrow\))}} & \makecell{\textbf{Gini (\(\times 10^{2}\))}\\\textbf{(\(\downarrow\))}} \\
\midrule
\multicolumn{8}{l}{\textit{Contextual adaptation: additional information}} \\
\midrule
\multirow{4}{*}{GLM-4-9B} & BM25 & \textbf{0.743 [0.734, 0.751]} & \textbf{0.078} & \textbf{0.069} & \textbf{0.910} & \textbf{2.37} & \textbf{1.34} \\
 & Embedding & 0.722 [0.714, 0.730] & \textbf{0.080} & \textbf{0.066} & \textbf{0.913} & \textbf{2.32} & \textbf{1.29} \\
 & Expansion & \textbf{0.745 [0.731, 0.758]} & \textbf{0.098} & \textbf{0.103} & \textbf{0.869} & \textbf{3.77} & \textbf{2.08} \\
 & Theory & 0.717 [0.708, 0.727] & \textbf{0.088} & \textbf{0.077} & \textbf{0.897} & \textbf{2.72} & \textbf{1.49} \\
\cmidrule(lr){2-8}
\multirow{4}{*}{ChatGLM3-6B} & BM25 & \textbf{0.722 [0.712, 0.732]} & \textbf{0.090} & \textbf{0.081} & \textbf{0.893} & \textbf{2.94} & \textbf{1.65} \\
 & Embedding & \textbf{0.728 [0.717, 0.738]} & \textbf{0.091} & \textbf{0.084} & \textbf{0.890} & \textbf{3.01} & \textbf{1.67} \\
 & Expansion & \textbf{0.718 [0.707, 0.728]} & \textbf{0.093} & \textbf{0.089} & \textbf{0.882} & \textbf{3.09} & \textbf{1.72} \\
 & Theory & \textbf{0.708 [0.696, 0.719]} & \textbf{0.100} & 0.098 & \textbf{0.869} & \textbf{3.43} & \textbf{1.89} \\
\cmidrule(lr){2-8}
\multirow{4}{*}{Llama-2-7B} & BM25 & \textbf{0.646 [0.637, 0.655]} & \textbf{0.102} & \textbf{0.085} & \textbf{0.878} & \textbf{2.91} & \textbf{1.56} \\
 & Embedding & \textbf{0.644 [0.634, 0.653]} & \textbf{0.104} & \textbf{0.090} & \textbf{0.870} & \textbf{3.04} & \textbf{1.65} \\
 & Expansion & 0.623 [0.614, 0.632] & \textbf{0.106} & \textbf{0.084} & \textbf{0.877} & \textbf{3.01} & \textbf{1.58} \\
 & Theory & 0.616 [0.606, 0.626] & \textbf{0.114} & \textbf{0.097} & \textbf{0.857} & \textbf{3.36} & \textbf{1.74} \\
\cmidrule(lr){2-8}
\multirow{4}{*}{Llama-3-8B} & BM25 & \textbf{0.696 [0.688, 0.703]} & \textbf{0.083} & \textbf{0.056} & \textbf{0.922} & \textbf{2.27} & \textbf{1.29} \\
 & Embedding & \textbf{0.693 [0.686, 0.700]} & \textbf{0.081} & \textbf{0.051} & \textbf{0.929} & \textbf{2.16} & \textbf{1.23} \\
 & Expansion & \textbf{0.679 [0.669, 0.689]} & \textbf{0.098} & \textbf{0.081} & \textbf{0.888} & \textbf{3.01} & \textbf{1.70} \\
 & Theory & \textbf{0.693 [0.683, 0.703]} & \textbf{0.096} & \textbf{0.080} & \textbf{0.891} & \textbf{2.99} & \textbf{1.69} \\
\cmidrule(lr){2-8}
\multirow{4}{*}{Qwen2.5-72B} & BM25 & \textbf{0.604 [0.597, 0.612]} & \textbf{0.099} & \textbf{0.066} & \textbf{0.894} & \textbf{2.50} & \textbf{1.28} \\
 & Embedding & \textbf{0.601 [0.593, 0.608]} & \textbf{0.103} & \textbf{0.072} & \textbf{0.884} & \textbf{2.67} & \textbf{1.39} \\
 & Expansion & 0.578 [0.566, 0.590] & \textbf{0.135} & \textbf{0.087} & \textbf{0.859} & \textbf{4.35} & \textbf{2.46} \\
 & Theory & \textbf{0.595 [0.586, 0.604]} & \textbf{0.116} & \textbf{0.076} & \textbf{0.877} & \textbf{3.30} & \textbf{1.82} \\
\cmidrule(lr){2-8}
\multirow{4}{*}{Mistral-7B-v0.3} & BM25 & \textbf{0.666 [0.660, 0.672]} & \textbf{0.082} & \textbf{0.051} & \textbf{0.926} & \textbf{2.00} & \textbf{1.14} \\
 & Embedding & \textbf{0.665 [0.659, 0.670]} & \textbf{0.074} & \textbf{0.037} & \textbf{0.946} & \textbf{1.64} & \textbf{0.94} \\
 & Expansion & \textbf{0.645 [0.637, 0.653]} & \textbf{0.094} & \textbf{0.061} & \textbf{0.910} & \textbf{2.49} & \textbf{1.41} \\
 & Theory & \textbf{0.643 [0.637, 0.650]} & \textbf{0.085} & \textbf{0.058} & \textbf{0.913} & \textbf{2.04} & \textbf{1.13} \\
\cmidrule(lr){2-8}
\multirow{4}{*}{Llama-3-70B} & BM25 & \textbf{0.613 [0.608, 0.618]} & \textbf{0.081} & \textbf{0.044} & \textbf{0.930} & \textbf{1.69} & \textbf{0.94} \\
 & Embedding & \textbf{0.611 [0.606, 0.617]} & \textbf{0.086} & \textbf{0.050} & \textbf{0.921} & \textbf{1.91} & \textbf{1.07} \\
 & Expansion & \textbf{0.611 [0.603, 0.619]} & \textbf{0.103} & \textbf{0.054} & \textbf{0.915} & \textbf{2.72} & \textbf{1.55} \\
 & Theory & \textbf{0.611 [0.604, 0.619]} & \textbf{0.097} & \textbf{0.051} & \textbf{0.920} & \textbf{2.43} & \textbf{1.38} \\
\cmidrule(lr){2-8}
\multirow{4}{*}{Qwen2.5-7B} & BM25 & \textbf{0.591 [0.583, 0.598]} & \textbf{0.102} & \textbf{0.058} & \textbf{0.906} & \textbf{2.56} & \textbf{1.44} \\
 & Embedding & \textbf{0.587 [0.580, 0.593]} & \textbf{0.094} & \textbf{0.053} & \textbf{0.912} & \textbf{2.15} & \textbf{1.17} \\
 & Expansion & \textbf{0.574 [0.565, 0.583]} & \textbf{0.118} & \textbf{0.064} & \textbf{0.894} & \textbf{3.27} & \textbf{1.85} \\
 & Theory & \textbf{0.575 [0.566, 0.584]} & \textbf{0.115} & \textbf{0.057} & \textbf{0.904} & \textbf{3.09} & \textbf{1.74} \\
\cmidrule(lr){2-8}
\multirow{4}{*}{Qwen2.5-3B} & BM25 & \textbf{0.554 [0.548, 0.560]} & \textbf{0.099} & \textbf{0.044} & \textbf{0.923} & \textbf{2.22} & \textbf{1.27} \\
 & Embedding & \textbf{0.553 [0.546, 0.559]} & \textbf{0.102} & \textbf{0.044} & \textbf{0.923} & \textbf{2.34} & \textbf{1.34} \\
 & Expansion & \textbf{0.530 [0.521, 0.538]} & \textbf{0.123} & \textbf{0.055} & \textbf{0.901} & \textbf{3.21} & \textbf{1.84} \\
 & Theory & \textbf{0.530 [0.522, 0.537]} & \textbf{0.117} & \textbf{0.059} & \textbf{0.894} & \textbf{2.92} & \textbf{1.65} \\
\bottomrule
\end{tabular}
\begin{tablenotes}\footnotesize
    \item[*] Accuracy is reported together with a 95\% confidence interval computed over country-level accuracy scores using a t interval.
    \item[*] AE composite metric is computed as $\sqrt{(1-A_c)\times CV}$; lower values indicate better joint accuracy-equality performance.
    \item[*] The arrows (\(\uparrow\)/\(\downarrow\)) indicate the direction of improvement in equality. For indices marked with (\(\downarrow\)), a lower value represents higher equality. For indices marked with (\(\uparrow\)), a higher value represents higher equality.
    \item[*] If improvements in accuracy or equality are observed relative to the reference base settings, they are displayed in \textbf{bold}. For the AE composite metric, \textbf{bold} likewise denotes a lower (better) cost than the matched base.
    \item[*] CV and Gini coefficient values are multiplied by $10^{2}$ for display readability.
\end{tablenotes}
\end{threeparttable}}
\end{table}

This result is consistent with the WVS contextual analysis. Providing retrieved additional information is among the more effective contextual interventions across both benchmarks. BM25 and Embedding produce the largest reductions in AE and $Eq_{CV}$ on ISSP, as they do on WVS.

\subsection{Parametric modification with alignment}

We further evaluate the alignment experiments on the ISSP dataset. Because ISSP does not provide the same multilingual coverage as WVS, the parametric modification analysis on ISSP is limited to alignment. We evaluate DPO and GRPO with human-annotated and GPT-annotated preferences on five model families.

Table~\ref{tab:issp_equality_summary_exp3} reports the ISSP alignment results in the same format as the main WVS table. We compare each alignment setting with the same model family before alignment under the profile-only setting. In this comparison, the four alignment settings do not show a stable advantage in either accuracy or representational equality. DPO with human-annotated preferences reaches a mean accuracy of 0.605, a mean AE of 0.120, and a mean $Eq_{CV}$ of 0.0375. GRPO with human-annotated preferences reaches a mean accuracy of 0.518, a mean AE of 0.147, and a mean $Eq_{CV}$ of 0.0452. Both DPO and GRPO with GPT-annotated preferences produce similar or lower accuracy and do not improve $Eq_{CV}$. This pattern is consistent with the WVS alignment analysis. Some individual model settings improve after alignment, but none of the four settings improves the overall benchmark summary on both accuracy and equality.

\begin{table}[H]
\centering
\resizebox{0.94\linewidth}{!}{%
\begin{threeparttable}
\caption{Complete results for accuracy, equality indices, and the AE composite metric in the parametric modification analysis on the ISSP benchmark}
\label{tab:issp_equality_summary_exp3}
\begin{tabular}{
    >{\raggedright\arraybackslash}p{1.8cm}
    c
    c
    c
    c
    c
    c
    c
}
\toprule
\textbf{Model} & \textbf{Settings} & \makecell{\textbf{$A_{c}$(Avg.) [95\% CI]}} & \makecell{\textbf{AE composite}\\\textbf{metric (\(\downarrow\))}} & \makecell{\textbf{Max-Min}\\\textbf{Diff (\(\downarrow\))}} & \makecell{\textbf{Min-Max}\\\textbf{Ratio (\(\uparrow\))}} & \makecell{\textbf{CV (\(\times 10^{2}\))}\\\textbf{(\(\downarrow\))}} & \makecell{\textbf{Gini (\(\times 10^{2}\))}\\\textbf{(\(\downarrow\))}} \\
\midrule
\multicolumn{8}{l}{\textit{Parametric modification: alignment sources}} \\
\midrule
\multirow{4}{*}{Llama-2-7B} & DPO-GPT & \textbf{0.796 [0.783, 0.808]} & \textbf{0.082} & 0.094 & 0.888 & 3.29 & 1.88 \\
 & DPO-Human & \textbf{0.793 [0.781, 0.805]} & \textbf{0.081} & 0.103 & 0.880 & \textbf{3.20} & \textbf{1.80} \\
 & GRPO-GPT & 0.568 [0.558, 0.579] & 0.127 & \textbf{0.084} & 0.860 & 3.77 & 2.10 \\
 & GRPO-Human & 0.545 [0.535, 0.556] & 0.136 & \textbf{0.083} & 0.855 & 4.05 & 2.23 \\
\cmidrule(lr){2-8}
\multirow{4}{*}{Llama-3-8B} & DPO-GPT & \textbf{0.761 [0.747, 0.776]} & \textbf{0.096} & 0.110 & 0.864 & \textbf{3.89} & \textbf{2.16} \\
 & DPO-Human & \textbf{0.770 [0.756, 0.784]} & \textbf{0.093} & \textbf{0.103} & \textbf{0.873} & \textbf{3.75} & \textbf{2.10} \\
 & GRPO-GPT & 0.543 [0.529, 0.558] & 0.161 & 0.109 & 0.814 & 5.66 & 3.19 \\
 & GRPO-Human & 0.535 [0.521, 0.550] & 0.162 & 0.108 & 0.814 & 5.64 & 3.19 \\
\cmidrule(lr){2-8}
\multirow{4}{*}{Mistral-7B-v0.3} & DPO-GPT & 0.489 [0.480, 0.499] & 0.143 & 0.085 & 0.840 & 4.02 & 2.25 \\
 & DPO-Human & 0.492 [0.482, 0.503] & 0.151 & 0.091 & 0.829 & 4.50 & 2.52 \\
 & GRPO-GPT & 0.515 [0.506, 0.524] & 0.134 & 0.085 & 0.845 & 3.69 & 2.03 \\
 & GRPO-Human & 0.522 [0.511, 0.533] & 0.146 & 0.096 & 0.828 & 4.47 & 2.44 \\
\cmidrule(lr){2-8}
\multirow{4}{*}{Qwen2.5-3B} & DPO-GPT & 0.460 [0.450, 0.469] & 0.153 & 0.071 & 0.854 & 4.35 & 2.48 \\
 & DPO-Human & \textbf{0.475 [0.465, 0.484]} & 0.150 & 0.069 & 0.863 & 4.31 & 2.46 \\
 & GRPO-GPT & 0.465 [0.455, 0.475] & 0.156 & 0.074 & 0.851 & 4.56 & 2.56 \\
 & GRPO-Human & 0.466 [0.456, 0.475] & 0.152 & 0.067 & 0.865 & 4.35 & 2.41 \\
\cmidrule(lr){2-8}
\multirow{4}{*}{Qwen2.5-7B} & DPO-GPT & 0.497 [0.490, 0.505] & \textbf{0.123} & \textbf{0.058} & \textbf{0.889} & \textbf{3.03} & \textbf{1.71} \\
 & DPO-Human & 0.497 [0.490, 0.504] & \textbf{0.123} & \textbf{0.058} & \textbf{0.888} & \textbf{3.00} & \textbf{1.69} \\
 & GRPO-GPT & 0.506 [0.497, 0.516] & 0.139 & 0.070 & 0.869 & 3.92 & 2.23 \\
 & GRPO-Human & \textbf{0.521 [0.510, 0.531]} & 0.140 & 0.080 & 0.855 & 4.09 & 2.27 \\
\bottomrule
\end{tabular}
\begin{tablenotes}\footnotesize
    \item[*] Accuracy is reported together with a 95\% confidence interval computed over country-level accuracy scores using a t interval.
    \item[*] AE composite metric is computed as $\sqrt{(1-A_c)\times CV}$; lower values indicate better joint accuracy-equality performance.
    \item[*] The arrows (\(\uparrow\)/\(\downarrow\)) indicate the direction of improvement in equality. For indices marked with (\(\downarrow\)), a lower value represents higher equality. For indices marked with (\(\uparrow\)), a higher value represents higher equality.
    \item[*] If improvements in accuracy or equality are observed relative to the reference base settings, they are displayed in \textbf{bold}. For AE composite metric, \textbf{bold} likewise denotes a lower (better) cost than the matched base.
    \item[*] CV and Gini coefficient values are multiplied by $10^{2}$ for display readability.
\end{tablenotes}
\end{threeparttable}}
\end{table}

\subsection{Consistency with the main WVS findings}

The supplementary ISSP analysis supports three main patterns from the WVS results. First, accuracy differences across countries remain visible under the foundational profile setting, showing that the observed cross-country inequality is not specific to the WVS question inventory. Second, retrieved additional information often improves both mean accuracy and country-level equality, with BM25 and Embedding producing the largest improvements. Third, preference alignment yields no systematic gains in representational equality, consistent with the WVS alignment results. These consistent patterns across two independent survey instruments demonstrate that the main findings are not an artifact of a single dataset.

\begin{acknowledgments}
We thank the action editor and the anonymous reviewers for their thoughtful and constructive feedback, which substantially improved this article. This work was supported by the National Natural Science Foundation of China (Grant No. 62106126).
\end{acknowledgments}


\bibliographystyle{compling}
\bibliography{main}

\end{document}